%% file: sample2.tex
\documentclass[fleqn,10pt]{wlscirep}
\usepackage[utf8]{inputenc}
\usepackage[T1]{fontenc}
\usepackage{tikz-cd}
\usepackage{float}
\usepackage[normalem]{ulem}

\title{Pedestrian Flux Analysis in a Confined University Campus Network Using Entropy and Accessibility Robustness Metrics}

\author[1]{Adamo Cerioli}
\author[4]{Barbara Caselli}
\author[4]{Lea Jeanne Marinelli}
\author[1, 2, 3, *]{Alessandro Vezzani}
\author[1, 2]{Raffaella Burioni}
\affil[1]{Department of Mathematics, Physics and 
Computer Science, University of Parma, Parco Area delle Scienze, 7/A, 43124, Parma, Italy}
\affil[2]{INFN- Istituto Nazionale di Fisica Nucleare, Gruppo Collegato di Parma, Parco Area delle Scienze 7/A, 43124 Parma, Italy}
\affil[3]{IMEM, CNR Parco Area delle Scienze 37/A, 43124, Parma, Italy.}
\affil[4]{Department of Engineering and Architecture, University of Parma, Parco Area delle Scienze, 181/A, 43124, Parma, Italy}

\affil[*]{alessandro.vezzani@unipr.it}

\begin{abstract}
When discussing urban life, pedestrian accessibility to all main services is crucial for fostering social interactions, promoting healthy lifestyles, and reducing pollution. This is especially relevant in coherent urban agglomerations like university campuses, which feature a high concentration of streets and social facilities. Using Wi-Fi data, we study pedestrian movements within a confined geometric network representing the pathways on a university campus. We estimate the level of crowding in each arc of the network and identify pedestrian flows along all possible paths, measuring the entropy and robustness of the network. In particular, we calculate the information gain achieved through the use of Wi-Fi data and we assess how pedestrian traffic redistributes within the network after the removal of individual arcs. Our results can be used to facilitate the investigation of the current state of walkability across the university campus while also testing a set of methods for analyzing urban complex networks, potentially allowing us to pinpoint areas in urgent need of road maintenance and enhancement.
\end{abstract}

\begin{document}

\maketitle

\section{Introduction}

Urban planning practices pursued after World War II led to the creation of urban areas mainly centred on the use of private motorised vehicles \cite{rossetti2020planning}. This practice has led to the proliferation of the phenomenon of urban sprawl, i.e. low-density urban expansion, instead of compact urban forms. However, this kind of urban development is no longer sustainable in all aspects (social, environmental and economic). Therefore, it is necessary to resort to urban planning centred on active mobility modes, like walking, which constitutes the primary mode of human movement, as well as being the most environmentally friendly, accessible, and affordable. Additionally, planning for pedestrian-friendly spaces and enhancing walkability not only increases environmental and health outcomes \cite{koszowski2019active} but also fosters more, attractive, livable, and inclusive cities \cite{fuller2017analysis, biazzo2019general}, where diverse transportation methods are more efficient and better integrated with each other \cite{gallotti2014anatomy}.

Another outcome of post-World War II urban planning in Europe is also the development of suburban university campuses, following the model pioneered by American cities. They derive, among other reasons, from the need for large, non-urbanised and affordable areas. Nonetheless, within these specialised urban enclaves, pedestrian movement and walkability appears to be of particular relevance. In fact, university campuses, as bustling centers of academic and social activities, represent microcosms of urban life where students primarily move by walking. However, the walkability of a campus is more than a matter of practical convenience as it profoundly influences the overall well-being of its academic community \cite{zhang2024campus}.

Similarly to other urban centers, university campuses, with their intricate layout and high concentration of streets and social facilities, are examples of complex networks, and more specifically,
transportation networks \cite{orozco2021multimodal, barthelemy2011spatial, morris2012transport, salingaros2020planning}. Therefore, many of their features can be analysed using strategies commonly used for studying complex systems, like measures of effective temperature and entropy, testing how the system reacts to perturbations or studying the network efficiency \cite{bontorin2024emergence, biazzo2020efficiency}. More specifically, studying the transport properties of a university campus translates to placing pedestrians along paths in a confined geometric network.

In this context, the utilization of Wi-Fi data emerges as a powerful tool for detecting pedestrian movements. In contrast to mobile networks data, WiFi data are easily available by public institutions and they rely on relatively economic infrastructures that are already present in public spaces. Other Wi-Fi applications \cite{pahlavan2021evolution} can be found in the context of epidemic control, which became particularly urgent in 2020 and 2021 due to the Covid-19 pandemic \cite{10.1371/journal.pone.0249839}, or for tracking people's trajectories and occupancy to study pedestrian accessibility in outdoor locations, like cities \cite{TRAUNMUELLER20184}. Similar studies in indoor locations, like large buildings \cite{pluchino2013agent}, can be relevant for reducing energy consumption \cite{JAGADEESHSIMMA2019495} or testing mixed techniques \cite{budrikis2022walkable, zhao2024unravelling}.

This research endeavors to shed light on the current state of pedestrian movements on the University Campus located in Northern Italy (Parma) and, especially, to test a set of methods for analyzing urban complex networks. To investigate the context of pedestrian movements, we first establish a realistic network for pedestrian paths in the campus. Afterward, using extensive Wi-Fi data collected at the University, we measure the average occupancy in each building and the pedestrian flux between each pair of buildings. We derive the pedestrian traffic on each arc and the overall entropy of the network, with the aim of estimating the information gain due to Wi-Fi data. Finally, we perform different tests to assess the robustness of the network, suggesting optimal strategies for reducing overcrowding and the excessive lengthening of pedestrian paths due to the removal of individual arcs. As we navigate the complex interplay of physical spaces and human dynamics, the insights gleaned from this study hold the potential to enhance walkability \cite{southworth2005designing} in urban and educational settings.

Apart from this study, in the past years, other analysis had been conducted on accessibility and walkability state of Parma Campus. For example, in 2023, the study of the pedestrian accessibility of the Parma Campus was conducted through the Space Syntax approach \cite{marinelli2023accessibility}. Furthermore, the Wi-Fi data on Parma University had been already used in other analysis, especially regarding the control of epidemics \cite{mancastroppa2022sideward, guizzo2022simplicial}. Besides Wi-Fi, other technologies have been used to track pedestrians' movements, like Bluetooth \cite{YOSHIMURA201743} and GPS \cite{YAMAGATA2020239, mizzi2018unraveling}. Since each technology has its own limitations, analyses for tracking pedestrians are often characterized by the fusion of multiple techniques \cite{6817847, lesani2018development}. In this way, it is also possible to compare directly different technologies and approaches.

The paper is organized as follows: the next chapter outlines the strategies and models we adopted for distributing pedestrians on the possible paths, and ultimately, on the arcs of a realistic pedestrian network of the Campus of Parma. In Chapter 3, we present our results of the distribution of pedestrian traffic on the network. Chapter 4 is dedicated to the study of the entropy of the network and the information gain provided by Wi-Fi data. In Chapter 5, we analyse the participation ratio of the arcs of the network in terms of the contributions from the pedestrian fluxes. In Chapter 6, considering different types of measures, we test how much our network is susceptible to changes like the removal of single arcs. The final chapter provides a summary of our findings, presents our conclusions and possible future applications of our analysis.   

\section{Methods}

Pedestrian movements within the university campus will be described by first introducing a campus walking network. In this network, arcs represent pedestrian infrastructures, i.e. pavements, other footpaths and pedestrian crossings, with their physical length; but also main lanes of parking areas and roadsides (in absence of dedicated pavements), as pedestrians tend to occupy the road space as well as walkways. Nodes represent either the junctions between these infrastructures or the main access to university buildings interconnected by the network. Within these buildings, pedestrians can connect their devices to the university Wi-Fi system. This network was created using Geographic Information Systems (GIS) data available trough the municipality GIS platform, reprocessed and implemented as needed through additional sources and on-site observations, to accurately capture real pedestrian movements on the campus. Further details on the network construction can be found in \textbf{Section A} of \textbf{Supplementary material}.

The first goal of our study is to estimate the pedestrian traffic $p_i$ in each arc of the network $i$ by using the data obtained from the Wi-Fi connection of the pedestrians devices. As we will show  in \textbf{Section} \ref{sec:Flux_occupation} the Wi-Fi dataset can be efficiently used to observe not only the occupation of the university buildings but also the travels between them. These can be quantified in terms of an average daily flux $\Psi_{\alpha \beta}$ between buildings $\alpha$ and $\beta$. In \textbf{Section} \ref{sec:Flux_occupation} we also discuss some basic characteristic of the building occupancy as a function of the building location in the campus and of the different part of the working day.

In \textbf{Section} \ref{sec:Pedestrian_traffic} we outline a method to distribute the pedestrian fluxes between buildings $\Psi_{\alpha \beta}$  within the campus walking network to obtain the arc pedestrian traffic, $p_i$. This method prioritizes the shortest paths by introducing an effective inverse temperature related to the length of the path \cite{omodei2014physics}. It not only provides a reliable description of movement within the campus but also allows us to estimate the network's response to perturbations, such as adding or removing pedestrian infrastructure or changing the occupancy of a specific building.

\subsection{Population density occupancy and fluxes between buildings}
\label{sec:Flux_occupation}

The Wi-Fi have been automatically collected by the University of Parma’s IT staff \cite{ guizzo2022simplicial}.
The structure of the dataset is described in details \textbf{Section B} of \textbf{Supplementary material}. Instead, In \textbf{section C} we discuss some operations needed to clean up errors in the dataset due to occasional malfunctions of the Wi-Fi systems, and to deal reliably with data relevant to an agent using multiple Wi-Fi devices.
After completing this preliminary process, we determine whether Wi-Fi users were present on campus at any given time and in which building. Whenever a Wi-Fi user connects to at least two buildings in a day, we can extrapolate a virtual trajectory of the relevant user. However, Wi-Fi data displays limitations, which include the unpredictability of periodic time checks for new connections, gaps in Wi-Fi coverage in certain areas of the campus, limited Wi-Fi usage and a limited precision in the location of the Wi-Fi device connected to a certain antenna. Such limitations do not allow us to extrapolate the precise path associated to each movement and one can obtain with a certain precision only the flux between couples of buildings. For this reason, in order to obtain an estimate of the pedestrian traffic in each arc of the network we redistribute such fluxes into all paths that connect the relevant buildings according to a rule the we will discuss in the next section.

It is important to stress that approximately one third of the campus population connects to the Wi-Fi network. However, assuming a uniform likelihood of connection across individuals, this does not affect the structure of the inferred mobility patterns and only results in a global rescaling of the measured fluxes. Therefore, this limitation introduces a systematic bias in the absolute values of traffic, but it is largely irrelevant for the relative comparisons and structural analyses performed in this study.

Our analysis covers three months, with 41 working days from October 9, 2023, to December 7, 2023. This period was selected for its regular daily activities. To calculate pedestrian fluxes, we count each time a person's device connects to a different building's Wi-Fi network, and we then determine the average number of people moving between the two buildings per day. This approach identifies buildings where users connect to Wi-Fi as common starting points or destinations within the campus environment.

\begin{figure}
\centering
\includegraphics[width=0.85\linewidth]{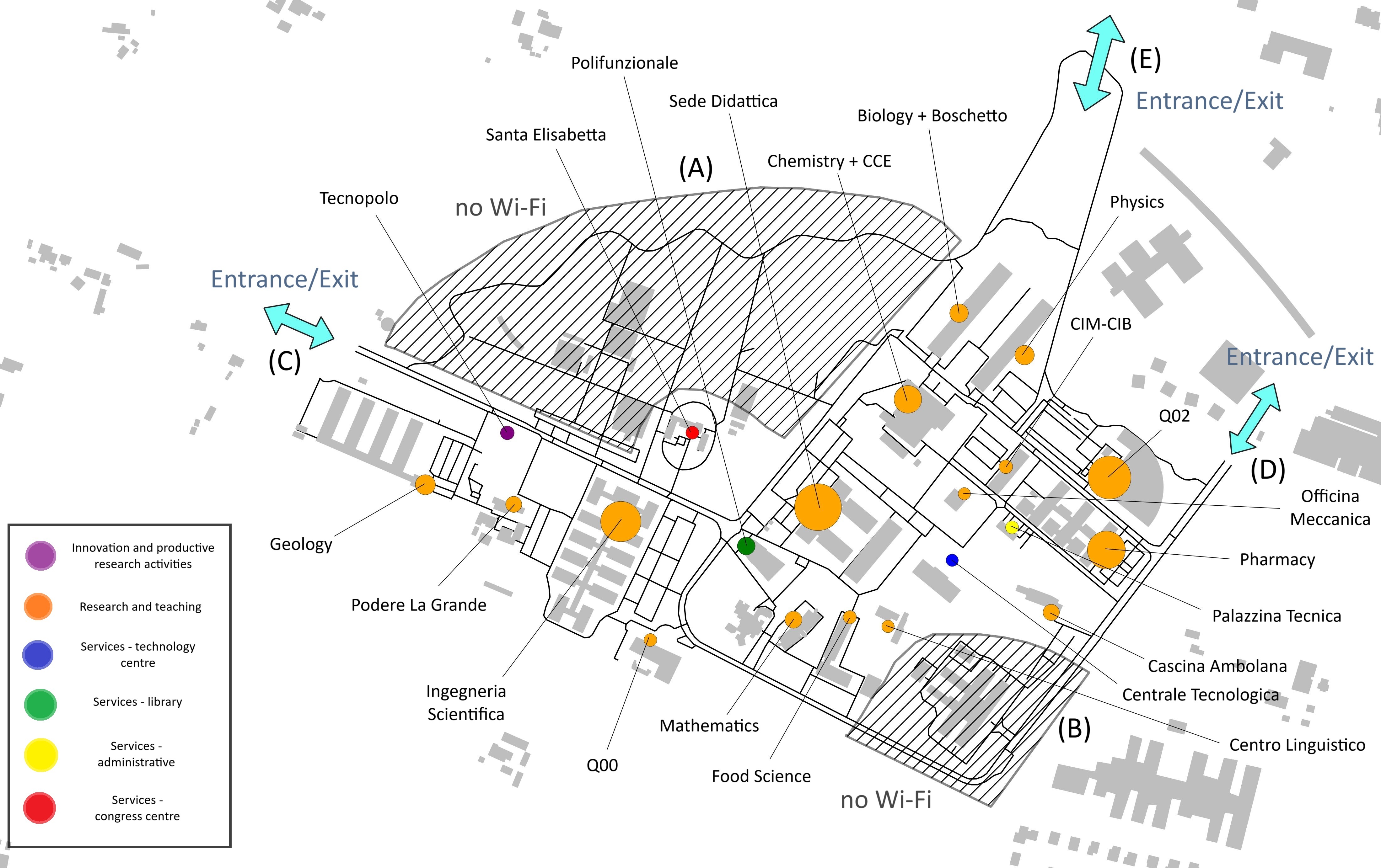}
\caption{\label{fig:frog4} Analysis of building occupancy. We realised a scheme where buildings are highlighted as dots, which dimension is proportional to their average population occupancy. The buildings with the highest occupancy are "Ingegneria Scientifica" and "Sede Didattica", primarily used for engineering and architecture courses, as well as "Q02" and the "Pharmacy" building. The black lines represent the pedestrian paths that constitute the network. The buildings are colored according to their function. The classification used is shown in the legend at the bottom left. Moreover, we highlighted the regions that are not covered by Wi-Fi connection in grey and the accesses of the campus in lightblue.}
\end{figure}

Specifically, we measure the average number of connected people in 30-minute intervals between 8:00 a.m. and 7:00 p.m. for each building, as this time frame encompasses all academic activities. The results of occupancy are shown in Figure \ref{fig:frog4}, where each building is depicted as a dot which size is proportional to its average population density. As anticipated, the Figure illustrates that the majority of people who remain indoors occupy only a few university buildings. The physical layout of all campus buildings, including those not covered by Wi-Fi or located in peripheral areas, is shown in grey in the background of the pedestrian network.

In order to better understand these results, and especially those regarding pedestrian traffic which will be discussed later, it is important to know that certain regions within campus are not covered by Wi-Fi connection. These are represented in grey in Figure \ref{fig:frog4} and are the sports center (A) and the IMEM-CNR building (B). Furthermore, Wi-Fi connection is also absent in the accesses of the campus in blue. The main one (D) is served by car lanes flanked by pavements on both sides; the other car access (C) is only served by car lanes; the third access (E), can only be traveled on foot or by bicycle.

\subsection{Pedestrian traffic on campus network}
\label{sec:Pedestrian_traffic}

To assign a pedestrian traffic to each arc, based on pedestrian fluxes between buidings, as we cannot directly determine a pedestrian trajectory, we make assumptions based on common human behavior \cite{omodei2014physics, barbosa2018human, gonzalez2008understanding}. For instance, people typically choose the shortest path when moving from one place to another, although this is not always the case \cite{zhu2015people}. As a matter of fact, secondary factors \cite{seneviratne1985analysis, ubaldi2021heterogeneity} influencing path choice, some of which are highly subjective, include the number and intensity of angular deviations, perceived safety, the beauty of the surroundings, and weather conditions.

In the context of pedestrian decision-making, defining a rigorous framework is challenging due to the multitude of influencing factors \cite{tong2022principles}. Among the factors influencing pedestrian behavior, one of the most extensively researched and inherently intriguing phenomena is the impact of crowding on movement patterns and decision-making processes \cite{corbetta2018physics, garcimartin2018redefining}.

Considering the factors we mentioned, the most straightforward approach for distributing pedestrian flows between each pair of buildings is to allocate them among all suitable paths, i.e. sequences of arcs connected by nodes that do not form loops. The distribution of weights depends on the lengths of the paths, with an intuitive expectation of an exponential decay pattern. Specifically, we opted to use the following law to assign the weight for each path:
\begin{equation}
    W(\gamma_{\alpha \beta}) = \dfrac{\textit{N}}{1+ \exp \left( k \ \dfrac{D_{\gamma_{\alpha \beta}}-D_{\min}}{D_{\gamma_{\alpha \beta}}}\right)} \label{eq:Venice}
\end{equation}
where $\gamma_{\alpha \beta}$ is a selected path, between the buildings $\alpha$ and $\beta$, $D_{\gamma_{\alpha \beta}}$ is the length of the path $\gamma_{\alpha \beta}$, $D_{\min}$ is the length of the shortest path, $\textit{N}$ is a normalization factor, evaluated numerically, and $k$ is a control parameter. This law had been formulated in the context of algorithmic modeling and simulations based on Markov theory applied to the study of pedestrian dynamics in Venice \cite{omodei2014physics}.

The value of the parameter \( k \) can be roughly estimated based on typical pedestrian behavior. Since we do not have access to the actual trajectories chosen by pedestrians, it is essential to test the effect of varying \( k \) within a realistic range. In particular, we considered three values: \( 5 \), \( 20 \), and \( 50 \). To provide an intuition behind these choices, we calculated---according to equation~\ref{eq:Venice}---how much more likely pedestrians are to choose the shortest path over one that is \( 20\% \) longer. For the three values of \( k \), the corresponding probability ratios are approximately \( 1.86 \), \( 27.8 \), and \( 11,\!014 \), respectively. In the first case, pedestrians are encouraged to also explore paths which are considerably longer than the shortest path even if they are always less likely. In the latter case, pedestrians almost always choose the shortest path available, or at most slightly longer alternatives, as significantly longer paths become extremely unlikely.

This parameter plays a role analogous to the inverse temperature in statistical systems, as it controls the intensity of fluctuations or perturbations in the system. In the limit \( k \to \infty \) (corresponding to zero temperature), the system becomes "frozen" in the configuration where pedestrians exclusively follow the shortest paths. Conversely, when \( k \to 0 \) (corresponding to infinite temperature), all paths become equally likely, and the system behaves as if it were in a highly excited and strongly disordered state. A clear visualization of the effect of the parameter $k$ on the pedestrian traffic is reported in Section G of the Supplementary Material by focusing on the flux between two single buildings.

In order to apply the previous formula, we needed to know all paths that connect each pair of buildings. However, depending on the complexity of the network, finding the shortest paths can be computationally demanding. The literature offers various strategies for solving the shortest path problem \cite{magzhan2013review, ridel2018literature}.

In our case, we accepted to find only the majority (or at least the shortest ones) of paths that connect each pair of buildings by using random walks. Specifically, for each pair of buildings, we generated a list of all possible trajectories by allowing self-avoiding random walks to explore possible paths between the two nodes. A complete introduction to self-avoiding random walks \cite{lawler1980self} can be found in literature, along with many applications in urban mobility \cite{gallotti2016stochastic}. Moreover, an explanation, in which the role of self-avoiding random walks is contextualized in our project, can be found in \textbf{Section D} of \textbf{Supplementary material}.

The procedure for the assignment of the pedestrian weight $p_{j}$ to each arc $j$ can be summarized by the following expression:
\begin{equation}
    p_{j} =  \sum_{(\alpha , \beta), \ \alpha \neq  \beta}  \Psi_{\alpha \beta } \sum_{\gamma_{\alpha \beta} \ni j}W(\gamma_{\alpha \beta}) = \sum_{(\alpha , \beta), \ \alpha \neq  \beta} p^{\alpha \beta }_{j} \label{eq:traffic}
\end{equation}
where the first and second sum runs over all building couples and the paths $\gamma_{\alpha \beta}$, between the buildings $\alpha$ and $\beta$, that passes through the arc $j$, respectively. Furthermore, $\Psi_{\alpha \beta }$ is the pedestrian fluxes between buildings $\alpha$ and $\beta$ and $W(\gamma_{\alpha \beta})$ is the weight associated to the path $\gamma_{\alpha \beta}$.

In order to significantly decrease the computational cost of equations \ref{eq:Venice} and \ref{eq:traffic}, 
we neglect very long unlikely paths form the summation in \ref{eq:traffic}. 
In particular, we keep only paths that satisfy the following condition:
\begin{equation}
    D_{\gamma_{\alpha \beta}} < \left(1+\frac{A}{k}\right) D_{\min}
    \label{eq:soglia}
\end{equation}
where $D_{\gamma_{\alpha \beta}}$ and $D_{\min}$ are the length of the generic and shortest path that connects a fixed pair of buildings. By fixing $A=10$ we skip paths with weight $11014$ times smaller than the one assigned to the shortest path. 
In \textbf{Section E} of \textbf{Supplementary material}, we show a test of robustness regarding the choice of this threshold.
\section{Results on pedestrian movements on the network}
\subsection{Pedestrian traffic on arcs and the k dependency}

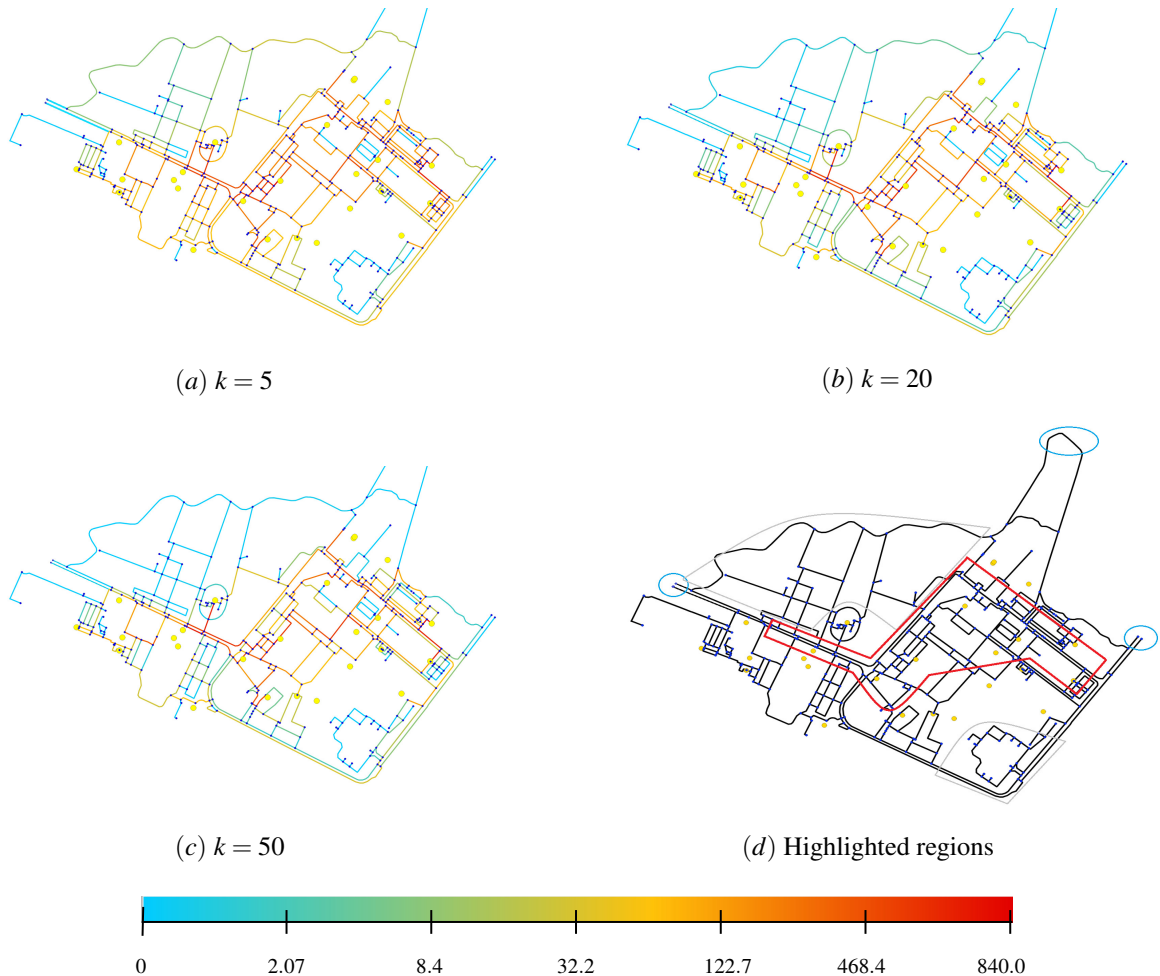
\begin{figure}
  \centering
  \input{scala_1.tikz}
  \caption{\label{fig:scala_1} Analysis of pedestrian traffic. We added a chromatic scale on arcs of the network. The color indicates the daily average pedestrian traffic evaluated with our model. As colors go from blue to red the estimated traffic increases. As we can see from the chromatic representation at the bottom of the figure, the six chromatic intervals are not associated with equally long traffic intervals, as we utilized a logarithmic scale. This scale is the same in all networks and goes from $0$ to the highest traffic found, $840.0$. The networks were obtained by using a parameter $k$ for the probability distribution equals to $5$, $20$ and $50$, respectively. Moreover, the maximum intensities of daily pedestrian traffic among all arcs are $781.5$, $794.3$ and $840.0$, respectively. In the fourth figure, we highlighted the regions where most of the pedestrian flows are distributed, where there is no Wi-Fi connection and where the entrances-exits of the campus are located in red, grey and blue, respectively. In the last figure, we show the frequency distributions of pedestrian traffic for different values of $k$.}
\end{figure}

The results on pedestrian traffic are shown in Figure \ref{fig:scala_1} which contains three networks associated to values of $k$ equal to $5$, $20$ and $50$, respectively. As a reference, in \textbf{Section F} of \textbf{Supplementary material}, we reported the list of the highest pedestrian fluxes we detected.

We removed from the representations all arcs leading to building entrances, as they can be considered extensions of the buildings themselves. Moreover, we used a logarithmic scale for pedestrian traffic, since traffic values are highly concentrated on a small number of frequently used arcs, while the majority are significantly less used. 

Upon examining the main features common to all networks, in figure \ref{fig:scala_1}d, we highlight in red the region on network that delineates most of the pedestrian flows; as expected, the buildings with the highest occupancy are located  in this area. Instead, as a reminder, the grey and blue regions (entrances) indicate where there is no Wi-Fi connection.

In figure \ref{fig:scala_1}e, we show the frequency distributions of pedestrian traffic for different values of $k$. It can be observed that decreasing $k$ effectively imposes an upper cutoff on the pedestrian traffic: the highest traffic values on the arcs decrease, and simultaneously, the number of arcs with very low traffic is reduced. In other words, decreasing $k$ leads to a more even redistribution of pedestrian traffic, with flow being distributed more regularly across the network arcs.
The histograms of pedestrian traffic across the arcs are reported in Section G of the Supplementary Material.

Regarding the dependency on $k$, as this parameter increases, the distribution of pedestrian traffic on arcs becomes steeper and steeper along specific paths. In fact, in figure \ref{fig:scala_1}a, which represents a scenario where pedestrians are less influenced by path lengths, new arcs obtain a slightly higher traffic. On the contrary, figure \ref{fig:scala_1}c represents a scenario where pedestrians choose almost exclusively the shortest paths available. As a consequence, the distribution of pedestrian traffic becomes even steeper with few arcs standing out from the others. 
Although it is not possible to directly estimate the value of \( k \) from the available data, we showed that the resulting distribution of pedestrian traffic across the network remains qualitatively similar over a broad range of realistic \( k \) values. Therefore, we fix \( k = 20 \) for all subsequent analyses, as it provides a reasonable balance and aligns with common assumptions about human movement patterns.

Furthermore, we assessed the robustness of our results with respect to measurement uncertainties in the pedestrian flows. In particular, we verified that the most trafficked paths remain stable when accounting for statistical fluctuations in the estimated fluxes. A detailed estimation of the statistical error on inter-building flows, and its propagation to other measures such as arc traffic, is provided in Section H of the Supplementary Material.

\subsection{Time distributions of pedestrian traffic on the arcs}

We also conducted a study on how pedestrian traffic on the arcs varies at different times of the day by identifying five temporal phases based on the variations in average building occupancy throughout a typical workday that are represented in figure \ref{fig:scala_2}f. The details of this preliminary study can be found in \textbf{Section G} of \textbf{Supplementary material}.

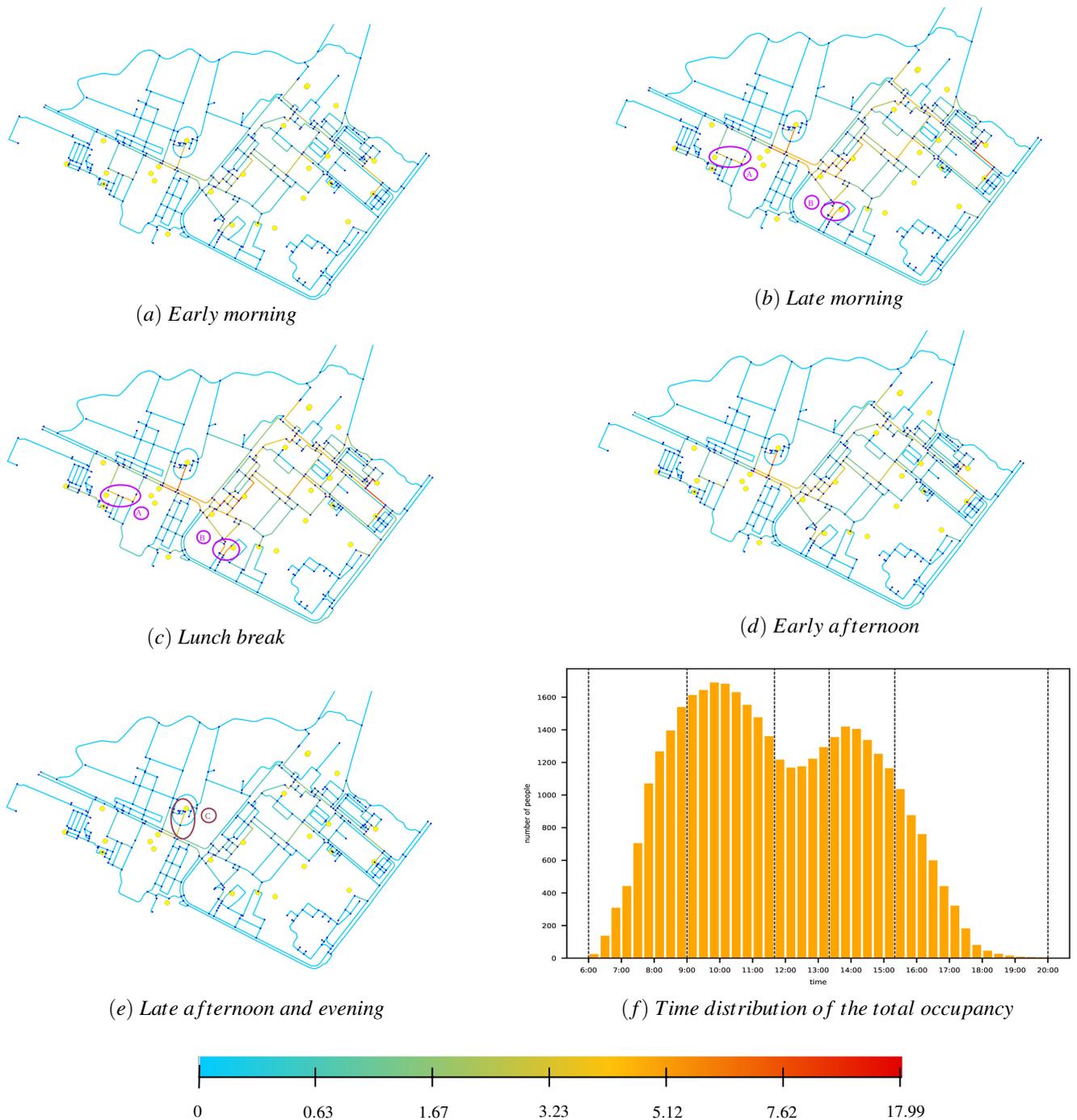
\begin{figure}
  \centering
  \input{scala_2.tikz}
        \caption{\label{fig:scala_2} Analysis of pedestrian traffic during temporal phases and (f) time distribution of the total occupancy of buildings. The represented networks were obtained using the same procedure explained before and with $k=20$. Indeed, the color indicates the average pedestrian traffic in a 20-minute interval. In particular, the logarithmic scale is upper-bounded by the maximum pedestrian traffic we found, which corresponds to the "Lunch break" phase. In particular, the maximum intensities of daily pedestrian traffic among all arcs are 5.57, 13.48, 18.02, 11.53 and 4.34, respectively.}
        \label{fig:time}
\end{figure}

The five phases are: the "early morning" from 6:00 a.m. to 9:00 a.m., the "late morning" from 9:00 a.m. to 11:40 a.m., the "lunch break" from 11:40 a.m. to 1:20 p.m., the "early afternoon" from 1:20 p.m. to 3:20 p.m. and finally, the "late afternoon and evening" period from 3:20 p.m. to 8:00 p.m. These are associated to the networks in Figure \ref{fig:scala_2}, where, within each of these intervals, the pedestrian traffic on the arcs were normalized to be averaged among 20-minutes time intervals.

In figure \ref{fig:scala_2}f, we represented the total average occupancy calculated in each 20-minutes time interval. In particular, for each of these time intervals, we detected the number of WiFi users connected in each building and summed these across all buildings to obtain the total average occupancy. Each of the five phases, delimited by vertical dashed lines, reflects different occupancy behaviors that align with the expected frequency on campus.

Firstly, we notice that the pedestrian fluxes are significantly low in the first and last phases of the working day. In fact, as far as the movement of people is concerned, the first and last intervals are characterized by a gradual arrival at and departure from the campus, respectively and some of these movements are underestimated due to the lack of Wi-Fi connection at the entrance of the campus. However, these movements typically occurs by public transport or by private cars, while most of the movements within the campus are performed by walking. During the second and fourth phases, there are peaks in occupancy due to class schedules, which commonly occur during these time intervals. The third phase shows a local minimum in occupancy because of the lunch break.

Aside from the previous general considerations, it is worth discussing specific traffic increases which are highlighted in Figure \ref{fig:scala_2}. In particular, figures \ref{fig:scala_2}b and \ref{fig:scala_2}c show two regions highlighted in purple that experience high traffic, which are not so significant in the other phases. These two regions correspond to the two university dining halls, with (B) being the most frequented one. If we consider instead figure \ref{fig:scala_2}e, we can notice that the region (C) highlighted in brown is the one characterized by the highest traffic despite not being particularly relevant in the other phases. This can be explained by the fact that the north-wester portion of the campus consists of the sporting area where the majority of activities is carried out during the evening time.

As an additional analysis, we quantify the overall people occupancy and pedestrian flows in all five phases. The main results are that "Lunch break" and "Late afternoon and evening" are the two phases in which there is the largest intensity of pedestrian fluxes compared with the "stable" people occupancy of buildings. Furthermore, the phase "Early afternoon" is the one during which there are the fewest relative pedestrian movements. The numerical results on occupancy and pedestrian flows are reported in \textbf{Section G} of \textbf{ Supplementary material}.

\section{Entropy and information of the traffic network}

The pedestrian traffic reconstructed in the previous sections was obtained from two types of information: the analysis of the pedestrian infrastructures connecting the buildings, with the arcs of the network with their length, and the Wi-Fi data providing the fluxes between the different buildings. It is now interesting to analyze how much information about pedestrian traffic comes from the two contributions.
In this perspective, we replicate our previous study without utilizing Wi-Fi data. Specifically, instead of using Wi-Fi measurements to associate the flux between each building pair, we assigned a constant flux that is independent of the chosen pair while maintaining the overall average number of moving pedestrians, so that the flux in each link should be related only to its centrality in the network \cite{barrat2008dynamical, porta2009street}. This approach allows us to isolate the impact of the network topology on pedestrian traffic, separate from the influence of the Wi-Fi data. Our results are reported in \textbf{Section H} of \textbf{Supplementary material}. We observed that, without using Wi-Fi data, pedestrians are more equally distributed among all arcs.

We now quantify how much information we gained from the pedestrian infrastructure and form the Wi-Fi data respectively, by evaluating the Shannon entropy of the pedestrian network fluxes \cite{omar2020survey, NAZARNIA201932, cabral2013entropy} which is defined as: 
\begin{equation}
    S = - \sum_{j=1}^{n}\tilde{p}_{j} \log \left(\tilde{p}_{j}\right) \ \ \ \ \ \ \  \ \  \ \ \ \tilde{p}_{j} = \dfrac{p_{j}}{\sum_{i=1}^{n} p_{i}}
\end{equation}
where $n=521$ is the number of all arcs that compose the network and $\tilde p_{j}$ is the probability that a generic walker is 
crossing the arc $j$. The maximum entropy is obtained when all the arcs are crossed with the same probability i.e. $\tilde p_{j}=1/n$ and  $S_{\max}=\log(n) \simeq 6.256$, while the entropy is vanishing if all the pedestrian traffic occurs on a single arc. The natural logarithm is used for calculating the entropy.

\begin{figure}[H]
\centering
\includegraphics[width=0.7\linewidth]{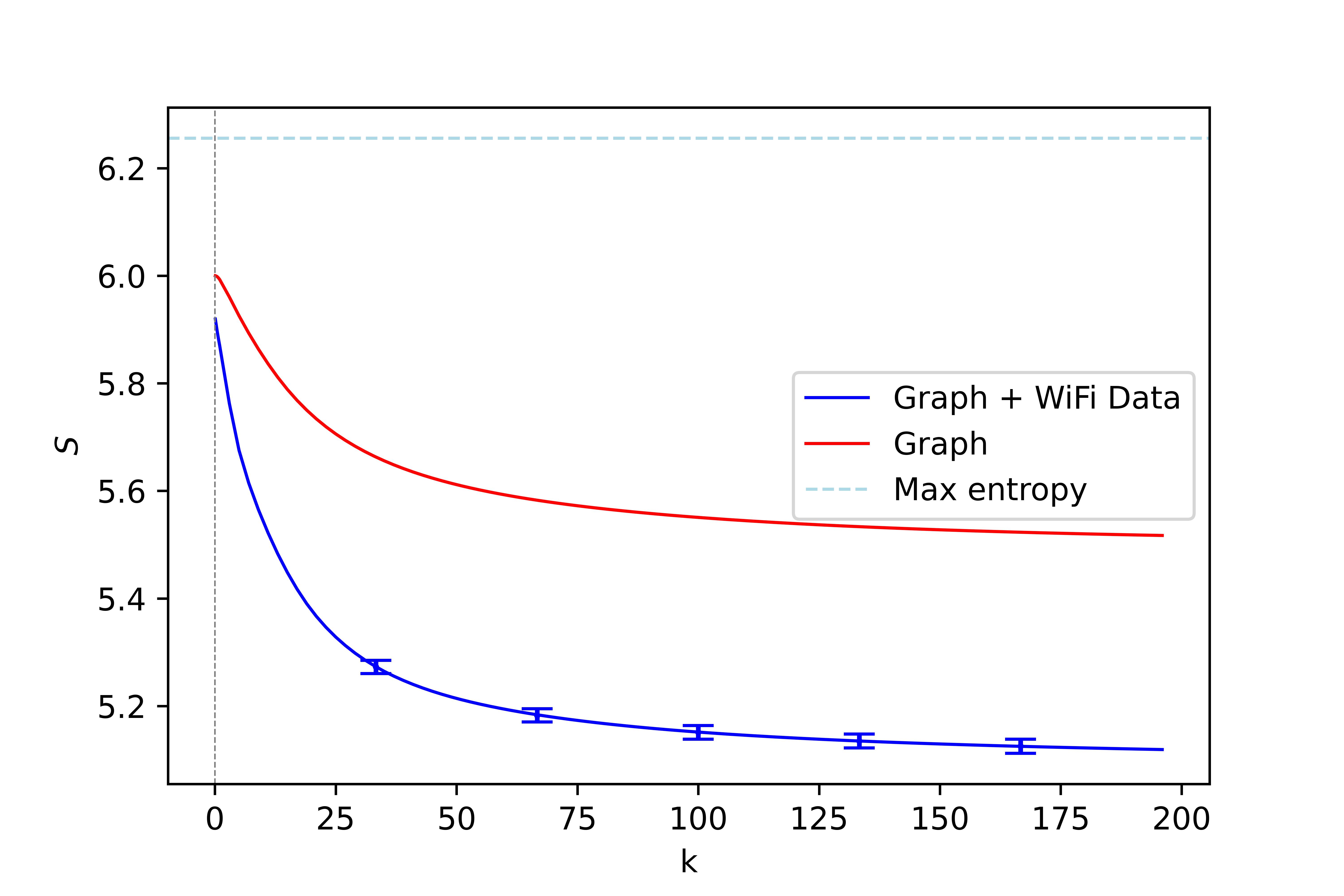}
\caption{\label{fig:entropy_k} Entropy of the network. We represented the entropy as a function of $k$ both with (blue line) and without (red line) the Wi-Fi data. We also represented the maximum value of entropy (dotted light-blue line). The error bars represent the standard deviation, calculated based on repeated simulations in which pedestrian flows between buildings were varied according to the expected statistical error. Further details are provided in Section H of the Supplementary Material.}
\end{figure}

We calculate the entropy first by using only the pedestrian network and then by introducing also the Wi-Fi data. The results are plotted in Figure \ref{fig:entropy_k}, as a function of  $k$ . 

The information gain in the two cases (difference with respect $S_{\max}$) is comparable. Therefore both the network structure and the Wi-Fi data provide significant information to our estimate of the pedestrian traffic. As expected, the information gain is minimal for $k=0$, which corresponds to the configuration in which pedestrians choose equally among all possible paths. Instead as $k \rightarrow  \infty$, which corresponds to the case in which pedestrians choose exclusively the shortest path, the entropy variation asymptotically reaches a maximum. In particular, for our case of interest $k=20$, the entropy difference is reasonably close to its value at $k \rightarrow  \infty$. 

\section{Participation ratios in the traffic network}

Based on the analysis from the previous section, some arcs in the networks experience heavy traffic primarily due to their central location, while other high-traffic arcs are involved in transporting between buildings with significant flux, independent of their position.

In the first scenario, numerous pedestrian fluxes contribute to the overall traffic of an arc, whereas, in the second scenario, a single flux tends to dominate. Distinguishing between these two cases is essential for predicting which paths will be most affected by changes in a single flux between two buildings. For instance, it is valuable to determine whether traffic in a particular arc can be reduced by limiting the movement from just one building.

In order to quantify these different behaviors, we introduce the participation ratio of the traffic of arc $j$. This is a typical tool used in the physics of localization \cite{evers2008anderson} and it provides an estimate of the number of couples of buildings that contribute to the traffic of the considered arc. In particular we define:
\begin{equation}
    L_{j} =\dfrac{\left(\sum_{\alpha \neq \beta} p_{j}^{\alpha \beta}\right)^{2}}{\sum_{\alpha \neq \beta} \left( p_{j}^{\alpha \beta} \right)^{2}}
    \label{eq:Loc}
\end{equation}
where according to Eq. \eqref{eq:traffic}, $p_{j}^{\alpha \beta}$ is the portion of pedestrian flux between buildings $\alpha$ and $\beta$ that is assigned to arc $j$. The sums run over all building couples. The participation ratio $L_{j}$ varies from $1$ to the number of building pairs $\mathfrak{N}=210$.
\begin{figure}[H]
  \centering
  \input{scala_3.tikz}
  \caption{\label{fig:scala_loc} Analysis of the participation ratio. We added a chromatic logarithmic scale on arcs network for $k=20$. The color indicates the participation ratio, which corresponds to a measure of the number of building pairs that contribute to the traffic.}
\end{figure}
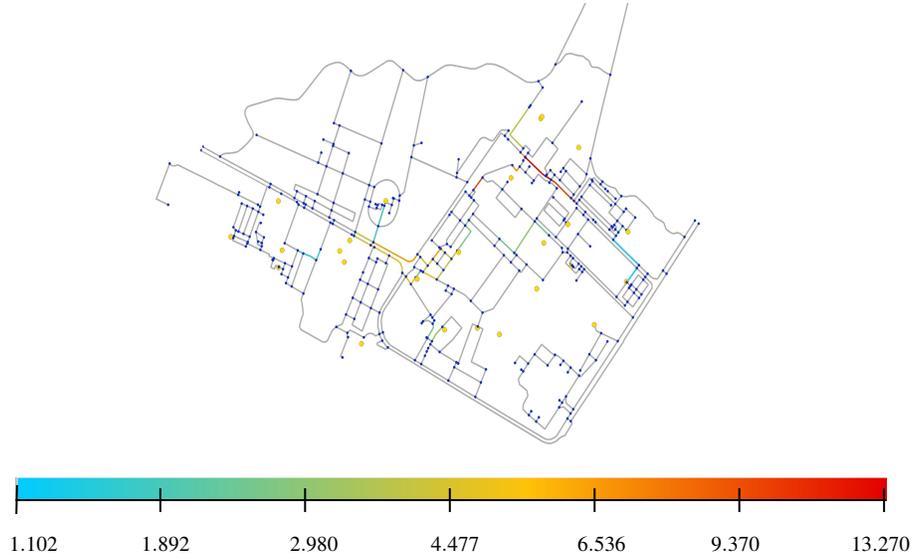

In Figure \ref{fig:scala_loc}, we depict the participation ratios of the 60 arcs with the highest pedestrian traffic, as it is most pertinent to examine the participation ratios for heavily trafficked arcs. The other arcs are shown in grey to clarify the network's structure.

We observe that these high-traffic arcs exhibit a range of participation ratios, from those influenced by a single flux to arcs where pedestrian traffic originates from more than 10 different building pairs. Generally, the arcs with the highest participation ratios are found in the central areas of the network. In contrast, arcs with the lowest participation ratios usually connect buildings near the network's boundaries, despite having significant pedestrian traffic.

\section{Measures of robustness of the network}

Whenever a network is involved for modelling a system, the network robustness \cite{barrat2008dynamical} can often be a concern. This is certainly true in the case of transportation networks, such as air routes \cite{lordan2014robustness} and road networks \cite{zhou2017robustness}, which must ensure good connectivity even after network damage. This is also true for pedestrian networks and therefore, we studied how the network responds to perturbations such as the removal of single arcs.

\begin{figure}[H]
  \centering
  \input{scala_robust.tikz}
  \caption{\label{fig:scala_robust} First analysis of the robustness of the network subjected to the removal of single arcs. The colors in the first linear scale indicate the total pedestrian traffic variation, and as they go from blue to red the effect of removing an arc increases. The networks were obtained by using the parameter $k$ equals to $20$. The two bottom networks represent two opposite situations, (a) a localized and (b) a widespread traffic redistribution. The colors in the scale indicate the pedestrian traffic due to the removal of the arc $u$ highlighted in black. The traffic originally assigned to the removed arc (a) is $p = 419.2$ and to the removed arc (b) is $p = 103.7$.}
\end{figure}
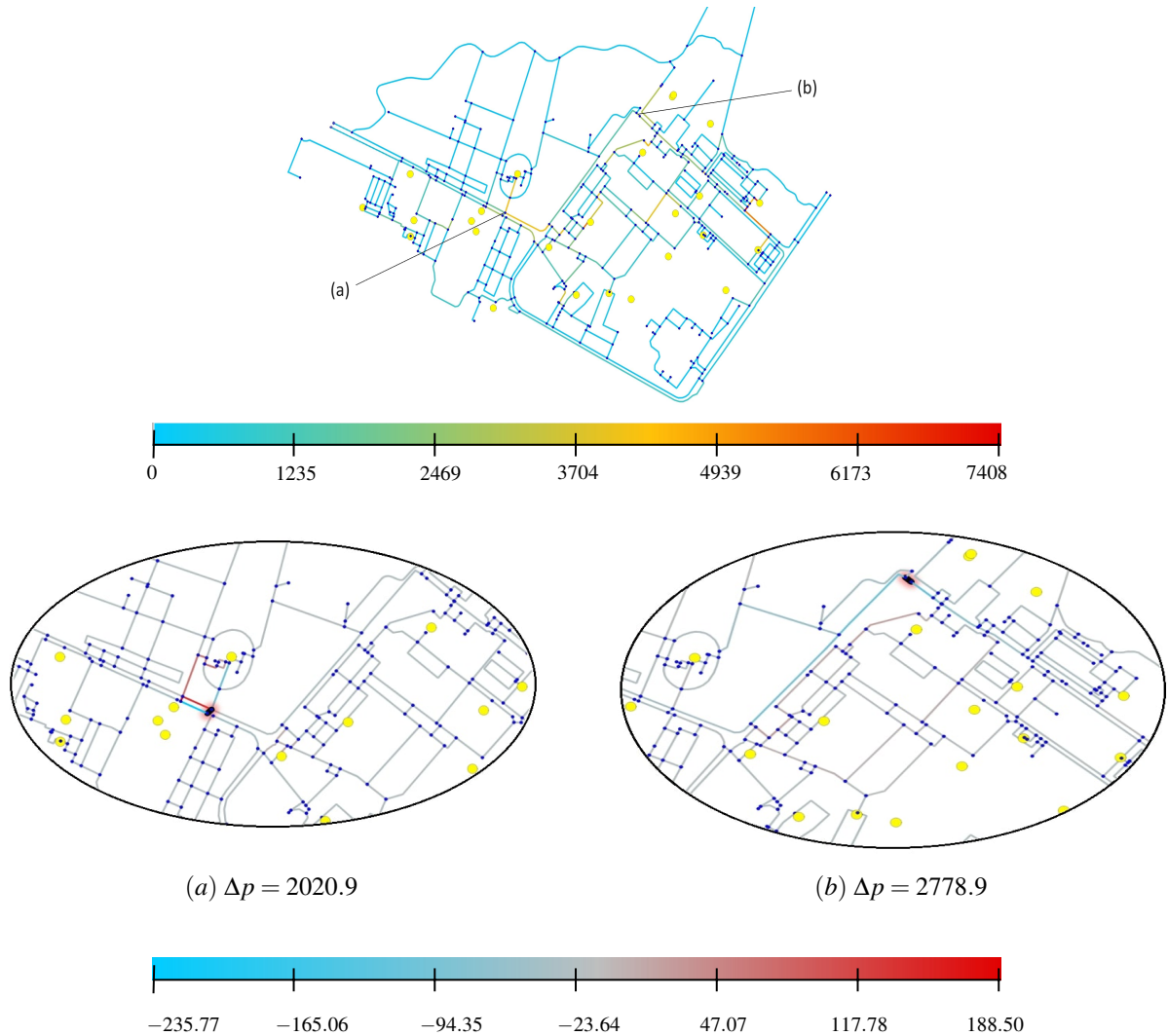

As a first measure of network response to the removal of a single arc, we chose to sum the absolute values of all traffic variations in all arcs except the removed one; i.e.:
\begin{equation}
\Delta p_{u} = \sum_{j \neq u} \left| p_{j | u} - p_{j} \right|
\end{equation}
where $p_{j | u}$ and $p_{j}$ represent the pedestrian traffic on the arc $j$ after and before the removal of arc $u$, respectively. The values of $\Delta p_{u}$ are represented in the top network of Figure \ref{fig:scala_robust}.

Despite being useful for gaining insights into the importance of each arc in the network, this global measure cannot discriminate between cases where traffic shifts to many arcs versus only a few arcs. However, in terms of network impact, these differences are extremely relevant. Although they have similar values of $\Delta p$, two examples representing the opposite situations are shown in Figure \ref{fig:scala_robust}.

In particular, in figure \ref{fig:scala_robust}a, the removal of an arc shows an high increase in traffic in some nearby arcs, which can lead to problems like overcrowding in those regions. In contrast, in figure \ref{fig:scala_robust}b, the traffic increase is almost not visible using the same scale since the flux variation is dispersed in several arcs.

Therefore, if we are interested in predicting when large traffic increases may appear due to the closure of a arc of the network, it is better to define another quantity. In order to investigate the robustness of the pedestrian network, and specifically to predict possible crowding effects, we used the measure of the maximum traffic variation caused by the removal of a arc:
\begin{equation}
\Delta p^{\max}_{u} = \max_{j \neq u} \left(   p_{j | u} - p_{j}  \right)
\end{equation}

\begin{figure}[H]
  \centering
  \input{scala_robust_2.tikz}
  \caption{\label{fig:scala_robust_2} Second analysis of the robustness of the network subjected to the removal of single arcs. The colors in the first linear scale indicate the maximum pedestrian traffic variation, and as they go from blue to red the effect of removing an arc increases. The networks were obtained by using $k=20$. The bottom figures represent examples of high traffic redistribution. In particular, they show the change in traffic after removing the two arcs whose removal cause the highest maximum pedestrian traffic variation. The removed arcs $u$ are marked in black. The chromatic scale is based on the largest variations that we found. The traffic variations of figure (a) fall within the range $[-443.2 , 793.8]$, while those of figure (b) fall within the range $[-327.1 , 640.2]$.}
\end{figure}
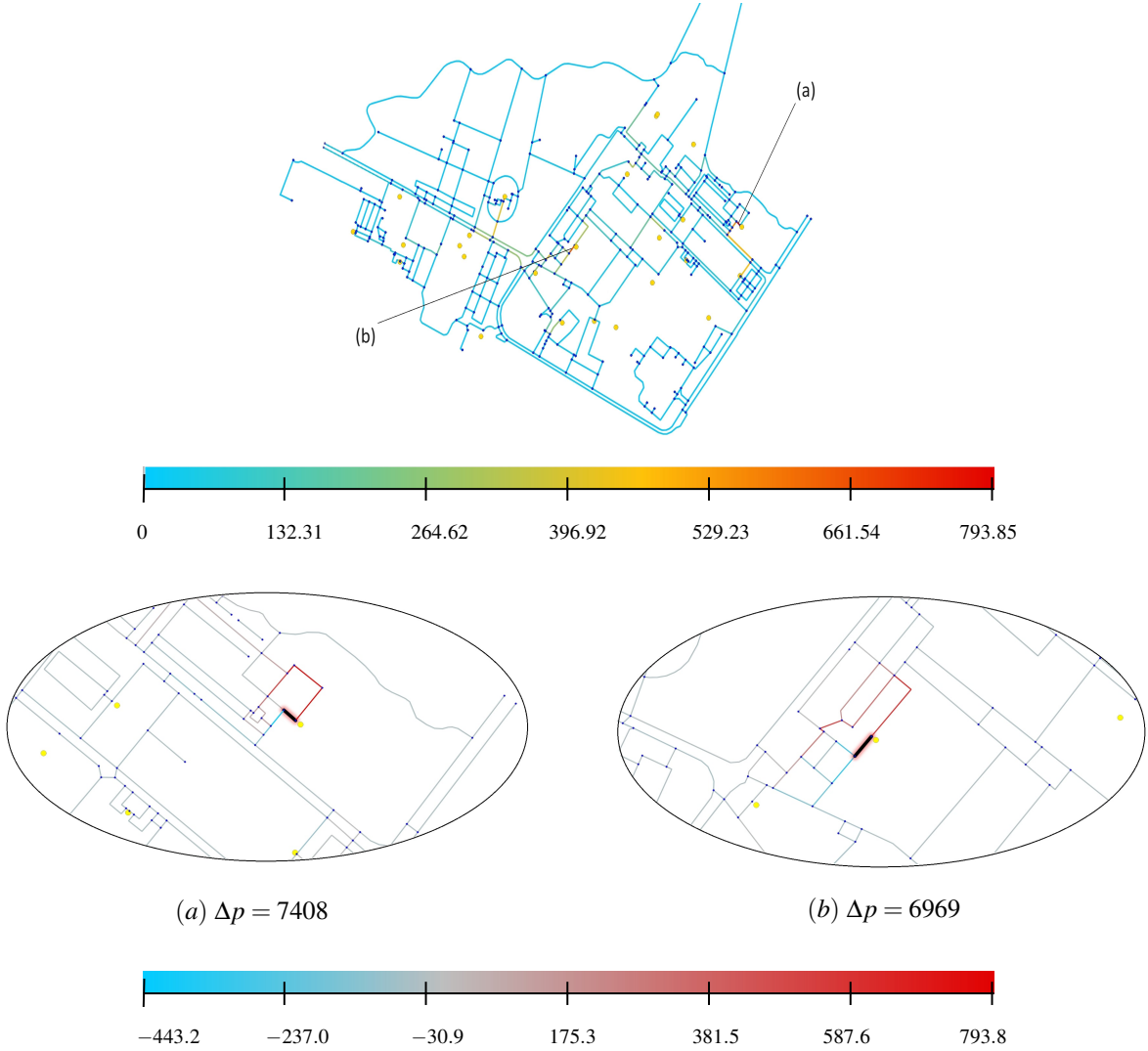

These results are represented in the top network of Figure \ref{fig:scala_robust_2}. From the latest figure, it is evident that some of the arcs adjacent to highly frequented buildings experience the greatest increase in traffic. Specifically, pedestrians redirect towards other directions immediately after leaving the buildings. This effect is shown in Figure \ref{fig:scala_robust_2} which represents the traffic variation resulting from the removal of the two arcs that cause the highest maximum pedestrian traffic variation.

The previous measures of robustness are mainly focused on the network structure and possible overcrowding. In this framework, we show that some arc removal may disperse the variation of pedestrian fluxes on several arcs; in this case it is possible that some pedestrians may face a significant lengthening of their path which is a different issue of network stability under arc removal. This new measure can be viewed in terms of travel cost, specifically as the additional time required for the journey, as has been also explored in studies related to vehicular traffic \cite{riccardo2012towards}. In this context we evaluate how many meters on the average the displaced people need to travel after the removal of a arc $u$; i.e.:
\begin{equation}
\Delta L_{u} = \frac{1}{p_{u}}\sum_{j \neq u} D_{j} (p_{j | u} - p_{j}) 
\end{equation}
where $D_{j}$ is the length of the arc $j$ and $p_{u}$ according to Eq. \eqref{eq:traffic} is the number of displaced walkers when removing the arc $u$ (i.e. the flux on the arc).

\begin{figure}[H]
  \centering
  \input{scala_walk.tikz}
  \caption{\label{fig:scala_walk} Analysis of the additional meters needed per person by the displaced people after the removal of single arcs. The colors in the linear scale indicate the length variation, in meters, of the new chosen paths for $k=20$. The three arcs with the highest values are highlighted in order.}
\end{figure}
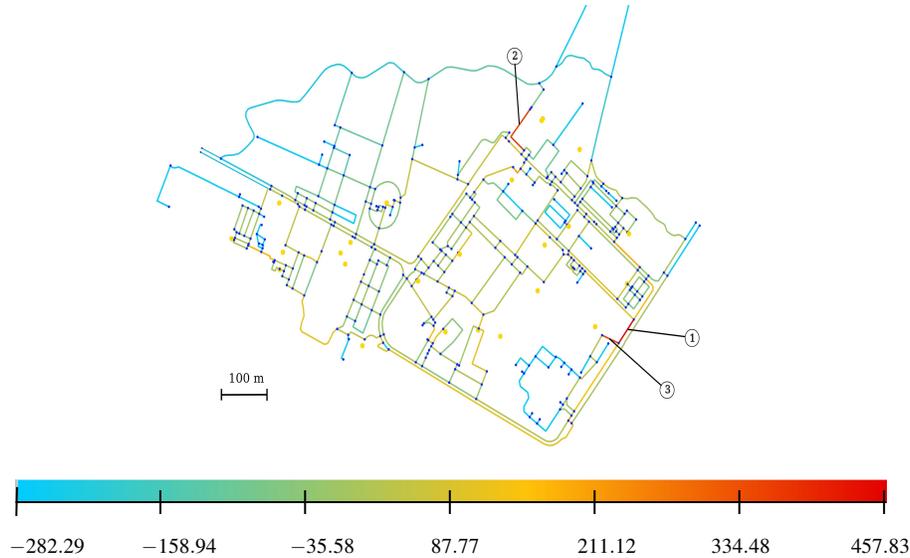

In Figure \ref{fig:scala_walk} we represent the $\Delta L_{u}$ for each arc of the network.

Contrary to what one might initially expect, many arcs have a negative $\Delta L_{u}$. This is because, for a not excessively large value of $k$, such as $k=20$, a non-negligible fraction of pedestrians are distributed on paths that are not the absolute shortest. In fact, we found that as k increases, all the negative $\Delta L_{u}$ converge to $0$.

We highlighted the three arcs whose removal cause the longest alternative paths for the displaced people. The additional meters $\Delta L_{u}$ displaced people should cross to reach their destinations and the pedestrian traffic $p_{u}$ of the three highlighted arcs in Figure \ref{fig:scala_walk} are: $\Delta L_{1}=457.8$, $p_{1}=74.4$, $\Delta L_{2}=364.3$, $p_{2}=192.8$, $\Delta L_{3}=358.1$ and $p_{3}=81.5$. All other arcs have $\Delta L_{u} < 250$.

The most critical arcs are placed close to a building so that the removal significantly elongate the path connecting this building to the rest of the network. Clearly this may have a relatively small impact on the variation of the pedestrian traffic if the relevant building is not involved in large fluxes. 

\section{Conclusions}

We studied pedestrian movements on a university campus using Wi-Fi data, which proved to be a valuable source of insight into pedestrian dynamics. Wi-Fi data are easily available through public institutions, and they rely on relatively economic infrastructures that are typically already present in public spaces. 

As we distributed pedestrian flows across all different paths, we employed a law (Formula \ref{eq:Venice}) characterized by an exponential decay with respect to path length and a free parameter $k$. We assumed that this "temperature" parameter remains constant across all regions of the network. However, this assumption is generally not true, as $k$ governs how people tend to choose one path over the other, and therefore, it may vary depending on the environment.

Considering these factors, the parameter $k$ could be promoted into a vector, with each component associated with an observable of a given path. In addition to length, an example of a relevant observable is the angular variation of a path. However, the Formula \ref{eq:Venice} may also be influenced by factors that are either unrelated to the paths themselves or highly subjective. For instance, weather conditions and decreased visibility during nighttime can significantly impact pedestrian choices. Generally, the perceived aesthetic quality of the surroundings of a path can also influence pedestrian choices.

Besides the study on building occupancy and the time distribution of pedestrian flows, which aid in understanding which buildings serve primarily as pass-through structures and how pedestrian flows vary throughout a typical working day on campus, one of the primary outcomes of our research include the establishment of a pedestrian traffic to assign to each arc of the network. These values allow us to define a list of medium and high traffic arcs that need more attention in the walkability assessment to ensure an overall improvement of pedestrian accessibility to campus spaces and buildings. The results of the analysis identified the central area of the pedestrian network, delineated by the red perimeter in figure \ref{fig:scala_1}d, as the zone with the highest concentration of pedestrian flows. This trend remains consistent across all temporal phases of the day, even though overall values decrease. The predominance of flows in this area aligns with the spatial distribution of the buildings exhibiting the highest occupancy rates. Additionally, this portion of the network includes the primary stops of the local public transport service connecting the campus to the city centre and the railway station. These insights suggest that this area should be prioritized in future interventions aimed at enhancing walkability conditions, ensuring that infrastructure quality and accessibility standards are adequately met. A careful walkability analysis assessing the quality of urban environments and infrastructures for pedestrians, combined with the flow analysis reported in this study, could effectively support the identification of intervention priorities along the most critical segments of the network. Moreover, this approach could prove useful in monitoring the effects of temporary adaptations to the pedestrian infrastructure—such as maintenance work involving construction sites that alter the layout of the walking network—and in anticipating the resulting redistribution of flows, while ensuring that safe and accessible routes are always available for all pedestrians, including those with disabilities.

By taking inspiration from other fields, we also studied the participation ratio of the most trafficked arcs. This quantity allows to highlight the arcs where pedestrian traffic is most susceptible to the closure of a building.

Using the evaluation of entropy, we were able to estimate the amount of information gained from Wi-Fi data. Furthermore, the analysis of the effect of removing an arc from the network could serve as a fundamental tool for predicting new pedestrian traffic patterns whenever a region of the campus is closed. This study was conducted both in terms of maximum traffic variation, considering the potential crowding effects that could arise, and in terms of the additional meters displaced people would need to travel to reach their destinations, focusing on the direct impact on individual pedestrians.

Finally, as previously reported, this work can be further extended by correlating the pedestrian traffic with quality assessments of footpaths. Indeed, there are numerous features of a footpath that could be considered when evaluating its overall quality. Some examples include the footpaths practicability and inclusivity, assessing whether regulatory requirements guaranteeing accessibility for all are met; safety, evaluating the protection from vehicular traffic; and comfort resulting from the suitability of the flooring, the attractiveness of the urban environment and the correct climatic design of spaces \cite{rossetti2024sumps}. These quality assessments can be performed either in situ or by 3D mapping of the campus and, eventually, utilizing image recognition. In fact, the use of big data, and especially, deep learning is becoming more and more relevant in walkability studies \cite{YANG2024102087, LI2022104140, blevcic2018towards}.

\bibliography{sample2}

\section*{Acknowledgements}

This work was supported within the framework of the “Sustainable Mobility Centre” (CNMS) Spoke 9, WP3, Task 3.2.
Funder: Project funded under the National Recovery and Resilience Plan (NRRP), Mission 4 Component 2 Investment 1.4 - Call for tender No. 3138 of 16/12/2021 of Italian Ministry of University and Research funded by the European Union – NextGenerationEU. Award Number: Project code CN00000023, Concession Decree No. 1033 of 17/06/2022 adopted by the Italian Ministry of University and Research, CUP D93C22000400001, “Sustainable Mobility Center” (CNMS).
This manuscript reflects only the authors’ views and opinions, neither the European Union nor the European Commission can be considered responsible for them.

\section*{Data availability statement}
The raw Wi-Fi data that support the findings of this study are available from the University of Parma, but restrictions apply to the availability of these data, which were used under licence for the current study and so are not publicly available. The data relevant to the population density occupancy, the average fluxes and the walking network of the pedestrian infrastructures in the Campus are available upon reasonable request from R.B (raffaella.burioni@unipr.it).
\section*{Author contributions statement}

All authors conceived the research.  L.J.M. and B.C. mapped the network of pedestrian infrastructures. A.C. conducted the numerical analysis of Wi-Fi data, calculated the relevant quantities and produced the figures.  All authors analyzed the results and reviewed the manuscript. 

\newpage

\section*{Supplementary material}
\appendix

\section{Construction of the pedestrian path network}
\label{appendix:graph}
We realized a pedestrian path network of Parma Campus which covers an area of about $77$ hectares located to the south of Parma city. Large portions of this area, approximately $10$ hectares each, are dedicated to buildings, parking areas, and sports facilities.

The arc-node network, represented in Supplementary Fig. \ref{fig:graph2}, depicts each pedestrian path segment as an arc, interrupted whenever it intersects other paths. Nodes are placed at the intersection points. The Parma Campus comprises 23 out of 60 buildings of the University of Parma. In particular, the Campus holds all scientific degree programs that are offered by the University of Parma. For practical reasons, we reduced this count to 22 buildings due to the spatial proximity of two of them: the "CCE" building, formerly the center of electronic calculus, was relocated and integrated into the "Chemistry" building. Therefore, for pedestrian flow calculations, these two buildings are treated as one.

Moreover, every building entrance becomes a node of the network for identification purposes and is represented with yellow dots slightly larger than the blue dots that represent all other nodes. Most buildings are connected to the pedestrian network (arc-node network) via the main entrance only, while the “Ingegneria Scientifica” complex, a large building, features three nodes corresponding to its main and most frequently used entrances, similarly to the building "Podere La Grande", which features two entrances.

The network encompasses pedestrian paths which typically feature dedicated infrastructure, but also other urban spaces without dedicated infrastructure (e.g., parking areas) commonly used by pedestrians. In particular, arcs inside parking areas are included if they provide continuity to pedestrian network facilitating building-to-building movement. Conversely, dead ends paths are not mapped to simplify the network.

The decision to incorporate additional arcs into the pedestrian network arises from two primary considerations: the need to create a network with highly connected arcs to evaluate all possible combinations effectively, and the aim to simulate people’s natural movements as realistically as possible. This objective cannot be achieved by limiting the network composition to paths with dedicated pedestrian infrastructure alone. These decisions are informed not only by on-site observations but also by the authors' firsthand experiences as frequent campus users.

Following this reasoning, concerning pedestrian crossings, the authors opted to include two additional arcs within the network, situated in the southeast near the "Pharmacy” complex and in the northwest between the "Mathematics” and “Q00” buildings. Other instances of this phenomenon can be observed, such as the paths along the campus boundary in its eastern and northeastern sections. In the eastern area, a bicycle path is present and commonly used by pedestrians due to the lack of other options, while in the northeastern part, there is no infrastructure, leading users to walk on the roadside or in the adjacent green space.

Furthermore, as mentioned before, certain buildings have been grouped together due to their spatial proximity. Specifically, the "Chemistry" and "CCE" buildings share the same physical structure. Therefore, there is no need to separate them in the network. However, a slightly different situation involves the "Biology" and "Boschetto" buildings. In particular, even though they are adjacent to each other, we opted to represent them separately in the network. However, for the study of pedestrian occupancy in the buildings, we considered them as a single building, resulting in a more visually elegant representation.

\renewcommand{\figurename}{Supplementary Figure}
\renewcommand{\thefigure}{S\arabic{figure}}
\renewcommand{\tablename}{Supplementary Table}
\renewcommand{\thetable}{S\arabic{table}}

\begin{figure}[H]
\centering
\includegraphics[width=1.04\linewidth]{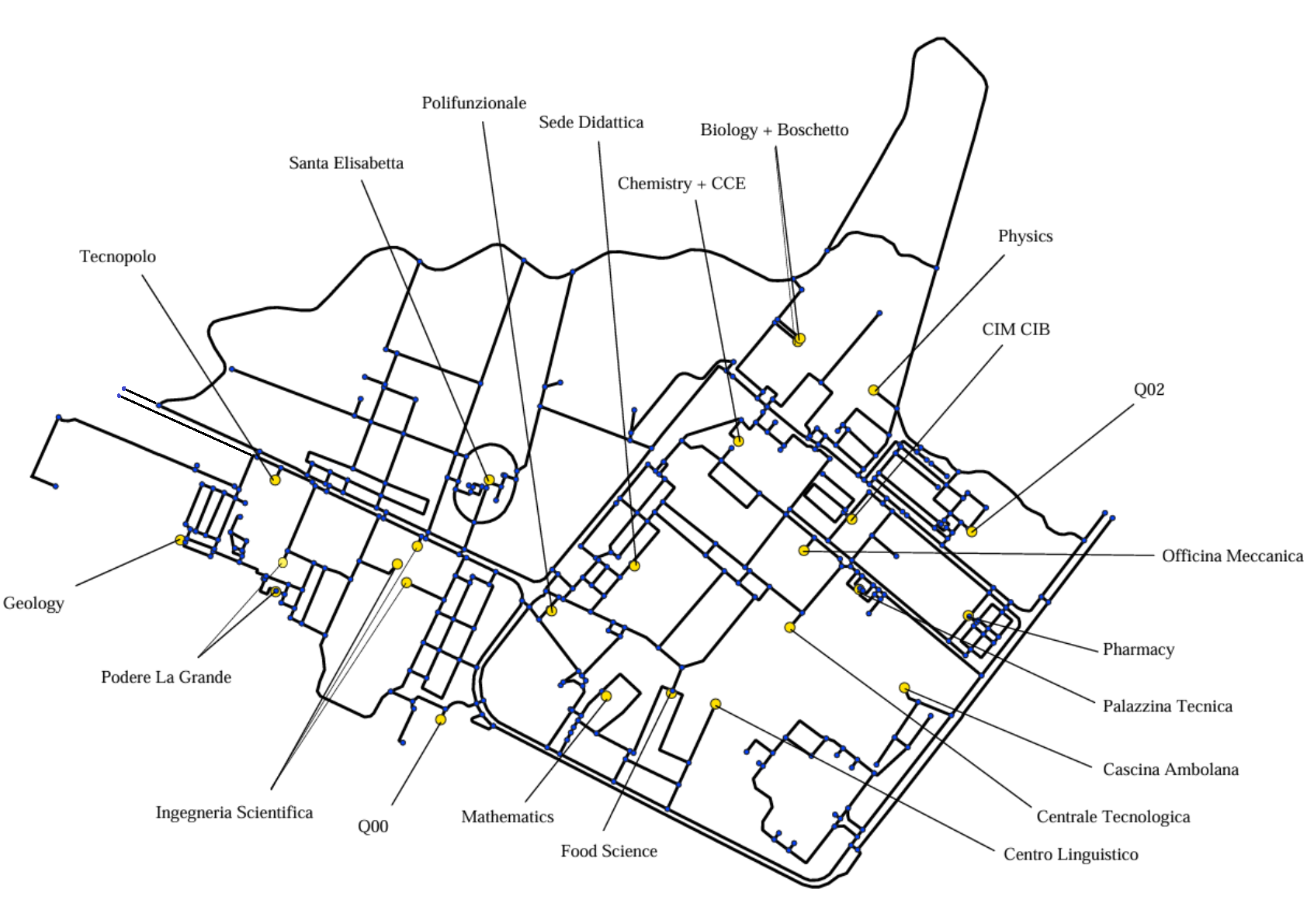}
\caption{\label{fig:graph2} Pedestrian network of Parma Campus. The links were chosen to reflect insights into the real movements of pedestrians on campus. The yellow nodes represent the entrances of the buildings, with all names provided for future reference.}
\end{figure}

\section{Structure of data}
\label{appendix:data}

The Wi-Fi data we used is semi-structured and automatically collected by the University of Parma's IT staff. The data is organised in CSV files that contain anonymous information about the Wi-Fi connections. All this information is fundamental to the IT staff for supervising the use of the university Wi-Fi and locations, such as rooms and hallways. This data is stored using AWS, a various set of cloud computing services. This strongly simplifies the management and handling of the data. Every day, the data is organised in one file and its dimension is approximately $100 MB$ for workdays and $3 MB$ for the other days. Each row of the CSV file contains several fields. Aside from all technical fields regarding the supervision of the Wi-Fi network, we have selected seven fields that are crucial for our analysis: 

\begin{itemize}
    \item a unique anonymized label ("user name") associated to each person. The label is composed of 32 alphanumeric characters and is conserved among days.
    \item a label ("acct status type") defining the event type. There are three possibilities: "start", "stop" or "intermediate update". The "start" event occurs whenever a new connection takes place. The "stop" event is associated to the ending of a connection. While the "intermediate update" event gives us information about the state of the Wi-Fi connection, immediately after a periodic check performed by the IT system. This happens with a frequency of few minutes.
    \item a number ("called station id") that identifies the access point (AP) to which the user is connected. The AP is generally associated with a room or a hallway.
    \item the MAC address ("calling station id") of the connected device.
    \item a label ("acct session id") that uniquely identifies a Wi-Fi connection. This field is fundamental for searching a chain of events ("start" - "intermediate update" - ... - "intermediate update" - "stop"). Indeed, events belonging to the same chain, or rather to the same connection, must have the same "acct session id". 
    \item a timestamp string ("timestamp") which contains information about the time of the event.
    \item a string ("domain") that indicates the domain of the account email used during the connection. This field allows to distinguish between students, staff members and foreign participants. 
\end{itemize}
Besides the Wi-Fi data, we analysed data that contains information about 
the buildings of Parma University. In particular, this additional CSV file contains a list of all access points associated with each building, identified by complete names and GPS coordinates. All of this information are crucial for predicting the trajectories of Wi-Fi users.

Regarding an estimate of frequency of Wi-Fi usage, our preliminary analysis revealed significant variations across degree programs and buildings. Specifically, we observed higher Wi-Fi usage among students located in the University Campus compared to those studying in the center of Parma. For example, the percentage of students on campus who regularly use Wi-Fi is close to $50 \%$, while in many courses outside the campus, it does not reach $25 \%$.

\section{Preliminary analysis on Wi-Fi data}
\label{appendix:WiFi_analysis}

To extract all information regarding the occupancy of access points (APs) in terms of Wi-Fi connection, the data needs to undergo certain processes. The overall needed treatment can be divided into the following steps:

\begin{itemize}
    \item We removed all rows with at least one attribute "Not a Number" (NaN) or with "acct status type" which is neither "start" nor "stop" nor even "intermediate update".
    \item We ordered in ascending sense rows according to the value of "timestamp".
    \item Using the field "acct session id", we merged all events involved in the same connection, with the same "called station id" and successive to each other. In order to be considered eligible, a sequence of events must begin with either an event "start" or with an event "intermediate update" after a previous sequence (this can be either eligible or not) and must stop with either an event "end" or with an event "intermediate update" followed by a successive sequence (this can be either eligible or not). In this way, we obtain new rows of data. Aside from the fields "user name" and "called station id", each row contains six new fields. The first two are "begin" and "end". In timestamp format, these respectively indicate the begin of the connection to that specific AP by a specific person ("user name") and the end of the same connection. The other four fields are called "domain begin", "domain end", "calling station id begin" and "called station id end". These last fields contain information regarding the domain and device of the user in the events associated to the begin and end of the connection. Essentially, the new rows contain information about the time intervals during which APs were occupied by users.
    \item We removed rows when "domain begin" and "domain end" do not coincide, or "calling station id begin" and "called station id end" do not coincide. Then, we created the unique fields "domain" and "called station id". Afterward, we proceeded to remove the other four fields.
    \item We merged the intervals associated with the same "user name" and "called station id" which overlap or are closer than three minutes.
    \item For each person, we detected the device "calling station id" that is used for the higher number of Wi-Fi connections. Then, we kept only the rows associated with the most used device. Successively, the field "calling station id" was also removed.
    \item We deleted the remaining "bilocation" using an algorithm that always favours new connections. "Bilocation" is defined as the ambiguity in assigning a position to someone due to their ostensible presence in two, or more, different locations simultaneously.
\end{itemize}

All these preliminary steps are essential for removing anomalous data and ultimately obtaining time intervals characterized by the "user name", "called station id" and "domain". Additionally, some strategies were adopted with human behaviours in mind. For example, the identification of the most used device for each person is based on the idea that usually people are connected to Wi-Fi with at most two devices, typically a smartphone and a laptop. Since we aim to identify a person and it is generally more likely that a smartphone reflects a person's real movement rather than a laptop, it is important to differentiate between devices. All these assumptions were tested using some examples in our data. In particular, the results of the analysis regarding the distribution of devices for user are shown in Supplementary Fig. \ref{fig:Distribuzione_devices}. These plots were obtained by analysing the data of a particularly active Tuesday, that is 7 November, 2023. As expected, the statistics of the data is dominated by the student activity, and the vast majority of individuals have at most two devices.

Moreover, the final step needed to address the few remaining cases of "bilocation" also takes human behaviour into account. A representation of the strategy used in these cases is shown in Supplementary Fig. \ref{fig:linee}. While the case illustrated is fictitious, it helps to convey the main idea of this approach. This strategy was necessary only for some number of cases, which are due to the variable delays in connection checks performed by the system. As illustrated in Supplementary Fig. \ref{fig:Distribuzione_Delta_time_update}, these updates are not performed at regular intervals. The key observation is that the majority of time intervals between Wi-Fi connection updates are less than ten minutes. Therefore, for future analyses, we must ensure that the smallest time interval we choose for our analysis is at least ten minutes. This will help to disregard spurious effects resulting from time delays in the Wi-Fi system.

\begin{figure}[H]
\centering
\includegraphics[width=0.80\linewidth]{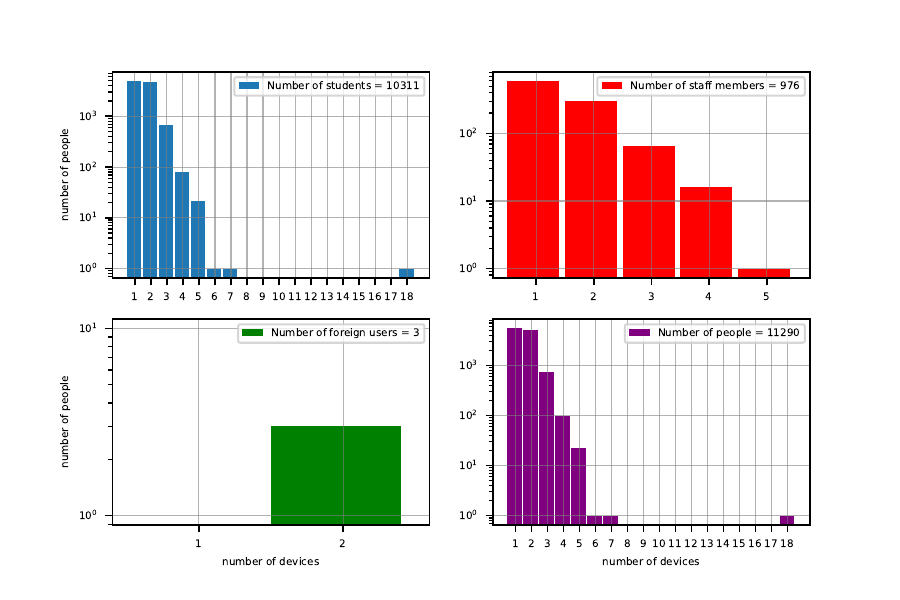}
\caption{\label{fig:Distribuzione_devices} Distribution of the number of devices per person. We used the Wi-Fi data collected on 7 November, 2023. The blue, red and green histograms were obtained using the data associated to students, staff members and foreign users, respectively. The purple plot collects all the previous data. In the upper-right corner of each graph, we can see the number of people who used Wi-Fi.}
\end{figure}
\begin{figure}[H]
\centering
\includegraphics[width=0.60\linewidth]{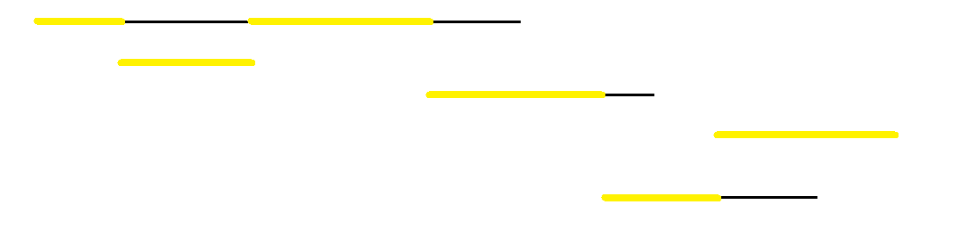}
\caption{\label{fig:linee} Scheme showing how the remaining cases of "bilocation" were removed. In particular, imagining an horizontal time axis, the segments are hypothetical time intervals of Wi-Fi connections performed by the same user. The segments that are highlighted in yellow coincide with the resulting ones. As we can notice, the priority was always given to new connections.}
\end{figure}
\begin{figure}[H]
\centering
\includegraphics[width=0.65\linewidth]{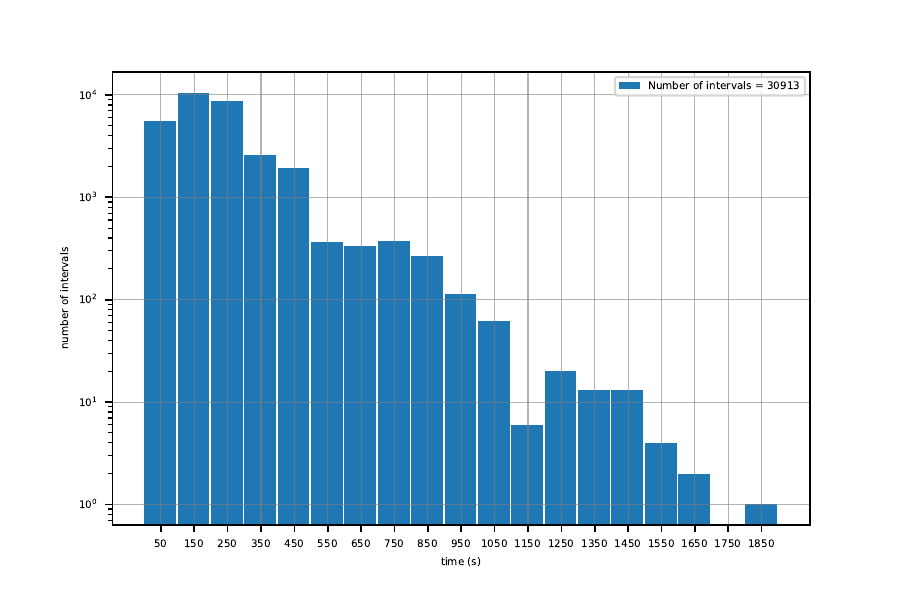}

\caption{\label{fig:Distribuzione_Delta_time_update} Distribution of time intervals between updates of the Wi-Fi data collection system performed on 7 November, 2023. On the horizontal axis the time is in indicated in seconds. These time intervals separate two consecutive events that belong to the same sequence. Moreover, one of the two events must be a "intermediate update". The total number of time intervals, defined as above, is indicated in the upper-right corner.}
\end{figure}

\section{Self-avoiding random walks}
\label{appendix:RW}
To efficiently find all, or at least nearly all, possible paths connecting each pair of buildings in the network, we opted to utilize self-avoiding random walks as our computational strategy. The use of random walks, also known as drunkard's walks, represents a statistical process in which a sequence of random steps defines a path. There are several types of random walks. To be precise, in our case, we employed discrete self-avoiding random walks. They are discrete because we are working on a network, and self-avoiding because they are not allowed to revisit nodes or cross links that have been encountered previously.

We chose to use self-avoiding random walks because we are seeking realistic pedestrian paths, and paths characterized by loops are not acceptable. Furthermore, for our project, it is important to find, for each pair of buildings, the initial portion of the list of possible paths in ascending order of total length. Only paths that are not significantly longer than the shortest one are considered realistic in the context of pedestrian movements. However, we ensured that we repeated the search for paths using random walks a sufficient number of times to also capture a range of longer paths. This approach allowed us to accurately test cases where we might have overestimated pedestrians' willingness to explore the campus network.

We tested the robustness of our analysis by changing the number of paths found by using random walks. The results are shown in Supplementary Fig. \ref{fig:RW} where we represented the entropy of the network as a function of $k$, the inverse of the effective temperature, and for different sets of found paths. In particular, in our analysis we used the largest set, consisting of $N=403{,}487$ paths. The entropy is defined as the Shannon entropy using the pedestrian traffic on each arc as probabilities. 

We performed a similar analysis for two different flow assignment strategies: one using the empirically observed pedestrian flows between buildings from the Wi-Fi data (Graph + WiFi Data), and one assuming uniform flows across all building pairs (Graph). In the latter case, the variations in entropy due to the number of detected paths are more pronounced. This is because, when Wi-Fi data are not used, flows between distant buildings—which are much harder to connect via self-avoiding random walks—become equally important. As a result, the error introduced by the incomplete sampling of trajectories is significantly larger in this case. Nonetheless, it is worth emphasizing that the difference between the entropies computed with and without Wi-Fi data remains substantially larger than the estimated uncertainty associated with the random-walk-based path sampling.

In the case of the entropy computed using the WiFi-detected pedestrian fluxes, the largest deviation from the blue line, as expected, is observed with the red line, which represents the entropy calculated using the smallest number of paths. On the other hand, if we focus on the blue and green lines, we can see they are very close, especially in our interval of interest, $k \in \left[ 5, 50 \right]$. Slightly higher deviations can be seen for small and large values of $k$.  In the first case, the deviation is due to the relevance of longer paths, which are the majority of paths that differ between the two samples. In the second case, the deviation is due to the fact that fewer paths remain relevant, only the shortest ones, and missing even a few of these paths causes the entropy to change. However, the maximum entropy deviation from the blue and green lines is around $0.02$.

\begin{figure}[H]
\centering
\includegraphics[width=0.80\linewidth]{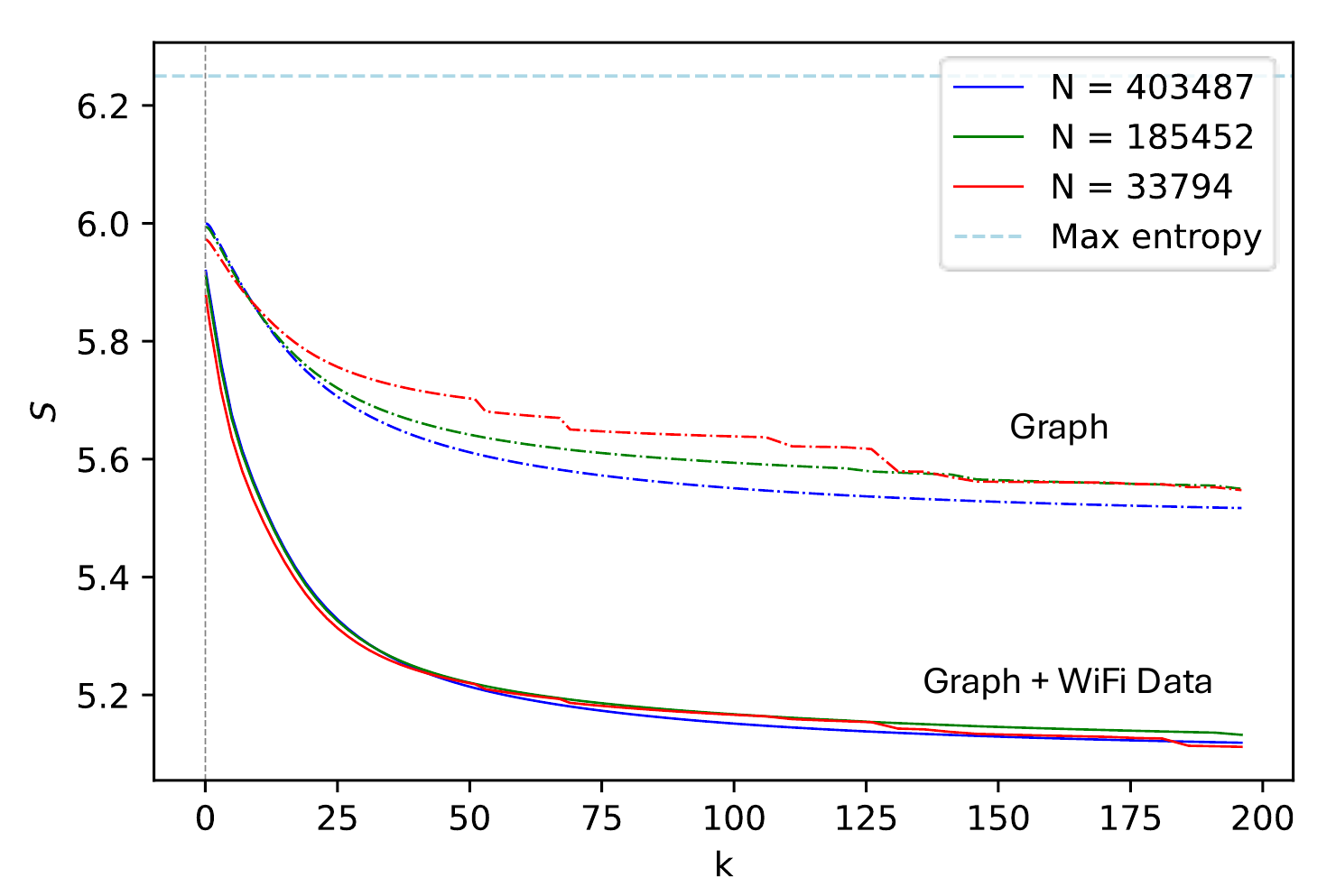}
\caption{\label{fig:RW} Robustness of the entropy as the used paths change. We represented the entropy of the network as a function of $k$, the parameter that defines how little people are willing to explore longer routes, for different numbers of paths used for our method. Solid lines correspond to the case where pedestrian fluxes between buildings are derived from WiFi data (Graph + WiFi Data), while dashed lines represent the case where equal flux is assumed between all pairs of buildings (Graph only). We also represented the maximum value of entropy.}
\end{figure}

To further assess the robustness of our results with respect to the number of trajectories identified through self-avoiding random walks, we also compared the top 30 arcs with the highest traffic values across two reduced sets of paths, corresponding to \( N = 185{,}452 \) and \( N = 33{,}794 \), respectively. In the first case, 29 out of the original top 30 arcs remained in the top 30, while in the second case, 28 did. Moreover, the arcs that dropped out of the top 30 in these two reduced sets originally occupied the lowest positions in the original top-30 ranking (i.e., positions 30 and 30, 29), and shifted only slightly downward to ranks 31 and 34, 38, respectively. These findings indicate that the changes are marginal and confirm the stability of the high-traffic arc ranking, further supporting the robustness of the overall traffic assignment and structural analysis.

\section{Robustness on path threshold}
\label{appendix:PT}
Depending on the parameter $k$, we kept all paths below a certain length for each pair of buildings in our analysis. In other words, in each case, we defined a threshold above which it is extremely unlikely for pedestrians to choose paths. In fact, for paths above this threshold the weight is at least $11014$ times smaller than the weight associated to the shortest path.

The threshold has no physical meaning but it is useful for decreasing the computation time of all analysis especially when $k$ is large. As needed, we performed a test of robustness regarding the used threshold. The results are shown in Supplementary Fig. \ref{fig:cutoff} which represent the entropy similarly to the previous figure but in this case we changed the threshold $A$. In particular, for each $k$ and $A$, we kept only paths that satisfy the condition:
\begin{equation}
    D_{\gamma_{\alpha \beta}} < \left(1+\frac{A}{k}\right) D_{\min}
    \label{eq:soglia}
\end{equation}
where $D_{\gamma_{\alpha \beta}}$ and $D_{\min}$ are the length of the generic and shortest path that connects a fixed pair of buildings.

These results show how robust the entropy of the network is to changes in the parameter $A$ (we used $A=10$ in our analysis). All lines are extremely close except for the red one, which corresponds to the fewest number of acceptable paths.
\begin{figure}[H]
\centering
\includegraphics[width=0.80\linewidth]{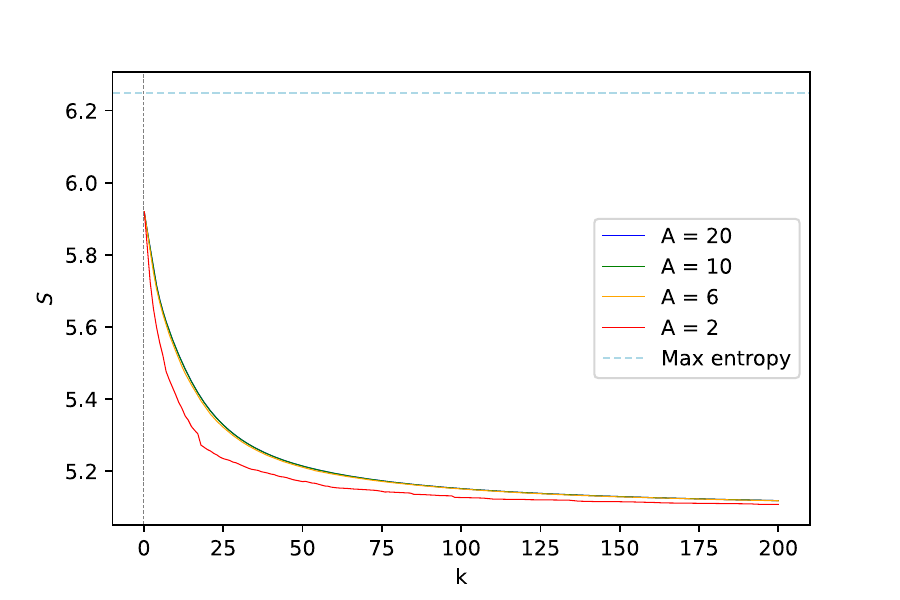}
\caption{\label{fig:cutoff} Robustness of the entropy as the path threshold changes. We represented the entropy of the network as a function of $k$, the parameter that defines how little people are willing to explore longer routes, for different path thresholds. We also represented the maximum value of entropy.}
\end{figure}

\section{Detected pedestrian fluxes}
\label{appendix:PF}
As a reference, we have reported the Supplementary Table \ref{tab:table2} which contains the highest pedestrian fluxes that we detected using the Wi-Fi data. These values were used to calculate the pedestrian traffic to associate to each arc.

In particular, the streets that connect "Q02" to "Pharmacy" are far from the central region of campus. However, this region is heavily trafficked by pedestrians because its centrality within the campus ensures that many different paths between buildings overlap. Indeed, it is important to remember that we do not have Wi-Fi data outside Campus and therefore, as we get closer to its exits, the pedestrian traffic is more and more underestimated. For example, we lose some of the information related to pedestrians that move from Campus to the superstore and vice versa.
\begin{table}[H]
  \begin{center}
    \caption{Wi-Fi pedestrian fluxes among building couples}
    \label{tab:table2}
    \begin{tabular}{c|c|c} 
      \textbf{Building 1} & \textbf{Building 2} & \textbf{Pedestrian flux} \\
      \hline
      Q02 & Pharmacy & 445.06 \\
      Ingegneria Scientifica & Santa Elisabetta & 376.22 \\
      Sede Didattica & Polifunzionale & 206.71 \\
      Ingegneria Scientifica & Sede Didattica & 200.18 \\
      Sede Didattica & Officina Meccanica & 194.39 \\
      Ingegneria Scientifica & Podere La Grande & 142.08 \\
      Sede Didattica & Mathematics & 129.39 \\
      Biology & Chemistry + CCE & 86.28 \\
      Sede Didattica & Q02 & 84.24 \\
      Ingegneria Scientifica & Polifunzionale & 78.16 \\
      Chemistry + CCE & Officina Meccanica & 65.39 \\
      Chemistry + CCE & Q02 & 63.16 \\
      Physics & Q02 & 62.12 
    \end{tabular}
  \end{center}
\end{table}

\section{Distributions of pedestrian traffic across arcs}
\label{appendix:distributions}

\begin{figure}[H]
\centering
\includegraphics[width=0.9\linewidth]{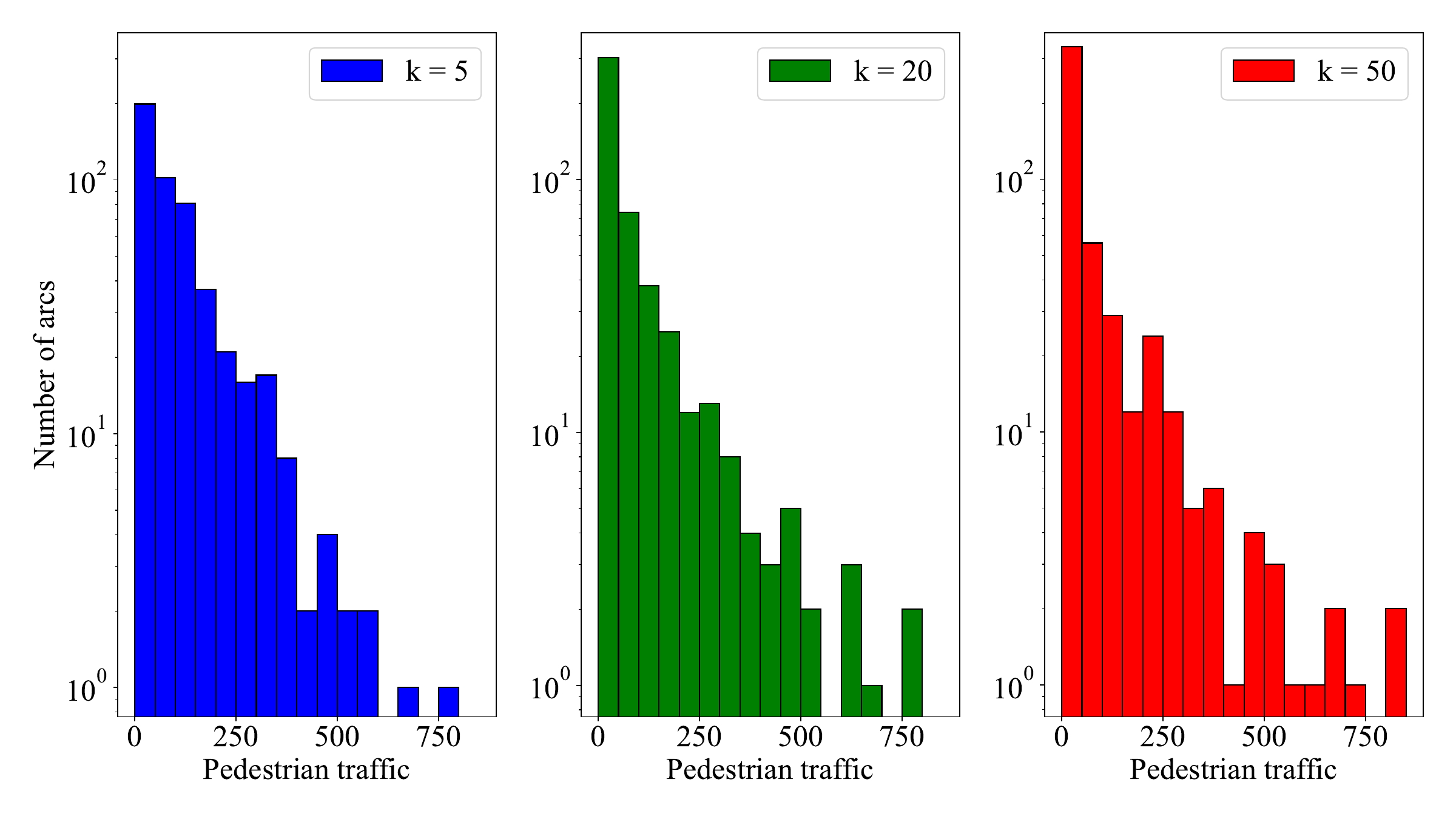}
\caption{\label{fig:Distribuzione_traffic} Distribution of pedestrian traffic values assigned to the arcs of the network, for different values of the parameter k. The three histograms correspond to k = 5 (blue), k = 20 (green), and k = 50 (red). For the three plot, the y-axis is shown on a logarithmic scale.}
\end{figure}

This section illustrates how pedestrian traffic is distributed across the network arcs (Figure \ref{fig:Distribuzione_traffic}), providing insights into its profile and how it changes according to the three selected values of parameter $k$. We observe that as $k$ increases, it becomes increasingly evident that a small number of arcs carry most of the pedestrian traffic, while the majority of arcs are barely used.

To illustrate the effect of the parameter $k$ on the distribution of pedestrian traffic across network edges, we present test cases where, for the sake of clarity, the entire pedestrian flow is assumed to occur between two buildings. We then show how this flow is distributed among the different edges (Figure \ref{fig:scala_k_WiFi}).

\begin{figure}[H]
  \centering
  \input{scala_k_WiFi.tikz}
  \caption{\label{fig:scala_k_WiFi} 
  Analysis of pedestrian traffic obtained by considering only the flow between a single pair of buildings. The two buildings are represented as green dots. A logarithmic scale was applied to the edges, with colors ranging from blue to red to indicate increasing estimated fractions of traffic. The networks were obtained by using a parameter $k$ equals to $5$, $20$ and $50$, respectively.}
\end{figure}

\section{Quantifying uncertainty in pedestrian traffic values}
\label{appendix:uncertainty}

In this section, we describe how we quantified the uncertainty associated with the measured pedestrian flow values between building pairs (origin-destination flows), and how these uncertainties propagate through the network when assigning pedestrian traffic to individual arcs. This analysis allows us to better assess the reliability of the estimated arc-level traffic and to understand the sensitivity of the network to input data variability.

We assign an uncertainty to each measured pedestrian flow value $\Psi_{\alpha \beta}$ between building pairs, proportional to $\sqrt{\Psi_{\alpha \beta}}$, assuming a Poisson distribution for the counting process. This assumption reflects the typical statistical behavior of count data, where the variance is expected to scale with the mean.

To evaluate how this uncertainty propagates to the arc-level traffic data, we generated alternative realizations of the pedestrian flows for each building pair by sampling from a Poisson distribution centered at the observed value $\Psi_{\alpha \beta}$. This procedure reflects the statistical nature of count-based data and introduces variability consistent with the assumed uncertainty model. We repeated this sampling process twenty times, thereby obtaining twenty alternative origin-destination flow matrices that differ from the original one. For each of these, we computed the corresponding arc-level pedestrian traffic by applying the same assignment model used for the observed data.

The resulting traffic distributions across arcs were remarkably similar to the original ones, demonstrating the robustness of the network assignment model with respect to fluctuations in the WiFi-based building-level flow measurements. Specifically, when comparing the top 30 arcs with the highest traffic values across the five alternative flow assignments, we found that the overlap with the original top-30 set was 30, 30, 29, 28, 28, 30, 29, 29, 30, 28, 28, 28, 29, 29, 28, 30, 30, 29, 28 and 28 arcs respectively, with average $28.85$.

To further quantify the similarity, we compared the rankings of arc-level traffic for the six lists that completely overlap with the original using the Kendall tau distance, which counts the number of adjacent swaps needed to transform one ranking into another. The normalized similarity scores (ranging from 0 for complete reversed rankings to 1 for identical rankings) were 0.9701, 0.9494, 0.959, 0.982, 0.963, and 0.968, corresponding to 13, 22, 18, 8, 16, and 14 inversions, respectively, out of a maximum of 435 (i.e., the number of inversions in a completely reversed list). These results confirm that the assignment model produces stable and consistent arc-level traffic patterns, even under plausible variations in the input flow data.

Moreover, when considering the elements that are no longer among the top 30 most-trafficked arcs, we observe that they typically occupy the lowest positions within the corresponding sub-ranking of 30 elements (e.g., positions 30, 28, 30, 30, 29, 30, 30, 30, 30, 29, 29, 29, 30, 28, 29, 30, 30, 30, 28, 29). In addition, these elements often appear just below the cutoff in the full rankings, at positions such as 31, 36, 31, 31, 33, 32, 32, 31, 33, 32, 33, 33, 35, 32, 35, 32, 33, 31, 33, 32, 32, 35. These observations suggest that the elements dropping out of the top 30 do so marginally, reinforcing the robustness of the ranking structure.

To assess how measurement uncertainty propagates to the entropy, we computed the entropy values for the same twenty alternative arc-level traffic assignments previously generated. This allowed us to estimate the variability of the entropy induced by statistical noise in the origin-destination flow data. The resulting spread in entropy values provided a measure of the uncertainty associated with the entropy itself. Our analysis shows that the reduction in entropy—i.e., the gain in information—derived from the WiFi-based data remains significantly larger than the estimated uncertainty. This indicates that the observed differences in entropy are not merely the result of measurement noise, but rather reflect meaningful and robust structure in the pedestrian traffic patterns.

\section{Temporal phases}
\label{appendix:TP}
We studied the population occupancy on campus over a larger time interval from 6:00 a.m. to 8:00 p.m., with 20-minute intervals for a more detailed analysis. Unlike the previous analysis, which only provided a general overview of averaged occupancy across buildings, this approach allows for data-driven decisions about how to segment the working day into phases. First, we extracted the time distribution of occupancy aggregated across all 21 buildings. The distribution is shown in Supplementary Fig. \ref{fig:Distribuzione_times_final}, where the vertical lines indicate five phases into which the working day can be divided, reflecting different occupancy behaviors.

The five phases are:
\begin{itemize}
    \item Phase 1 (6:00 a.m. - 9:00 a.m.): The majority of people arrives at campus during the first hours of morning. In fact, the first lessons usually start at 9:00 a.m.
    \item Phase 2 (9:00 a.m. - 11:40 a.m.): There is a peak of occupancy and the number of people remains approximately constant.
    \item Phase 3 (11:40 a.m. - 1:20 p.m.): Occupancy decreases due to lunch break.
    \item Phase 4 (1:20 p.m. - 3:20 p.m.): There is a secondary peak of occupancy and the number of people remains approximately constant.
    \item Phase 5 (3:20 p.m. - 8:00 p.m.): All the people gradually leaves the campus. In fact, the last lessons usually end at 6:00 p.m.
\end{itemize}

To attribute the presence of a Wi-Fi user to a specific building within each temporal interval, we required that the connection to that building last at least three-quarters of the interval. This is also true for the analysis of building attendance during the entire working day.

\begin{figure}[H]
\centering
\includegraphics[width=0.9\linewidth]{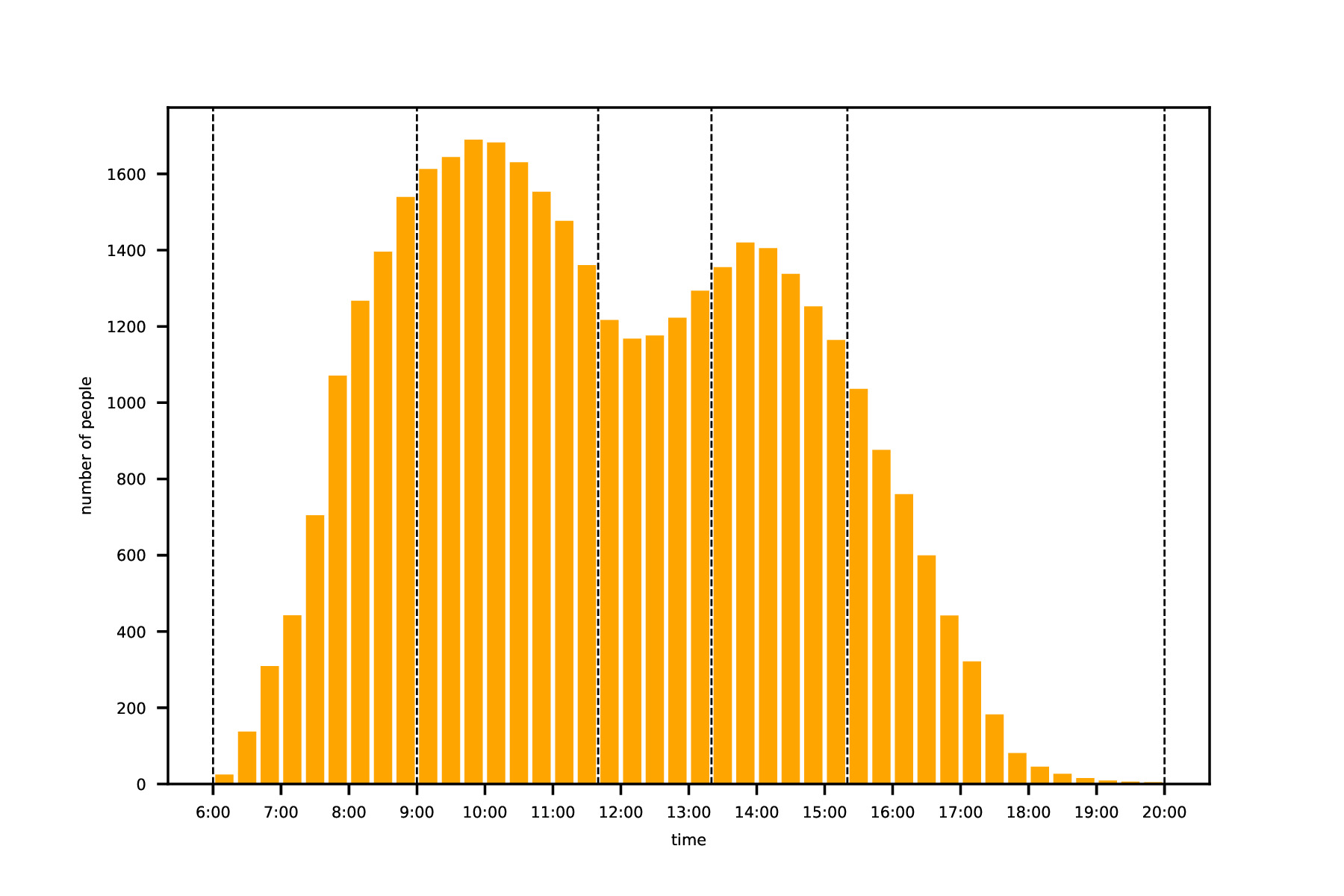}
\caption{\label{fig:Distribuzione_times_final} Time distribution of the total occupancy of buildings on Parma Campus, averaged over 41 working days from October 9, 2023, to December 7, 2023. Vertical dotted lines mark the time boundaries of the five phases into which we divided a typical working day. The first phase starts at 6:00 a.m., and the last ends at 8:00 p.m.}
\end{figure}

For each phase identified, we calculated the population density for each building, averaged over all 41 working days and discrete 20-minute intervals. Before presenting the results, it is important to describe some features of the topology of the Parma Campus, as these are essential for understanding the time variation in occupancy related to certain buildings.

To interpret the campus in its context, it is essential to consider some key features. Firstly, Parma Campus has two main streets serving as entrances and exits: firstly, Parma's Campus presents a road network consisting prevalently of a ring road accessible through two gated entrances. A first entrance located to the north-east of the study area, representing the most frequented access point, also due to its connection to the a superstore, a cinema, and some of the students residences of the University of Parma. The second entrance is located instead in the western portion of the Campus. A third access point to the Campus premises, through a pedestrian and cycling road, can be also identified in the northern area of the Campus. This segment grants access to the Campus through a green area functioning as a park and is not fully represented as the path exits the Campuses premises by intercepting Parma's ring road.

Additionally, it is important to note that there are two dining halls on the campus. The main dining hall is located next to the Mathematics building, while the other is near the building known as 'Podere La Grande.' Both dining halls typically attract a lot of people during lunchtime.

The precise values of some of the highest population densities for all phases mentioned earlier are reported in Supplementary Table \ref{tab:table1}. As expected, the majority of people on campus are present in the late morning (Phase 2) and first afternoon (Phase 4). The general trend shows a decrease in people density during the lunch break.
\begin{table}[H]
  \begin{center}
    \caption{Averaged people densities among buildings}
    \label{tab:table1}
    \begin{tabular}{c|c|c|c|c|c} 
      \textbf{Building} & \textbf{Phase 1} & \textbf{Phase 2} & \textbf{Phase 3} & \textbf{Phase 4} & \textbf{Phase 5} \\
      \hline
      Sede Didattica & 168.77 & 303.76 & 223.42 & 253.19 & 58.44 \\
      Q02 & 162.81 & 301.80 & 208.44 & 198.79 & 37.94 \\
      Pharmacy & 111.96 & 217.31 & 163.73 & 186.07 & 37.38 \\
      Sede Scientifica & 78.79 & 212.24 & 185.57 & 220.79 & 64.04 \\
      Chemistry + CCE & 69.24 & 125.16 & 87.79 & 101.62 & 29.16 \\
      Polifunzionale & 32.73 & 60.07 & 43.74 & 44.13 & 4.06\\
      Physics & 27.87 & 52.29 & 41.56 & 56.23 & 18.95 \\
      Geology & 24.89 & 70.24 & 55.58 & 59.32 & 9.78 \\
      Biology + Boschetto & 19.16 & 54.38 & 44.37 & 43.39 & 11.19 \\
      Mathematics & 17.16 & 40.15 & 31.90 & 35.74 & 6.95\\
      Cascina Ambolana & 14.80 & 36.41 & 25.08 & 28.78 & 7.14\\
      Podere La Grande & 8.76 & 28.40 & 23.21 & 30.72 & 7.88
    \end{tabular}
  \end{center}
\end{table}

In particular, for each phase, we summed all occupancies and pedestrian flows. The results are summarized in Supplementary Table \ref{tab:table2}. As expected, "Lunch break" and "Late afternoon and evening" are the two phases in which there is the largest intensity of pedestrian fluxes compared with the "stable" people occupancy of buildings. Instead, the phase "Early afternoon" is the one during which there are the fewest relative pedestrian movements.

\begin{table}[H]
  \begin{center}
    \caption{Averaged people occupancies and flows}
    \label{tab:table2}
    \begin{tabular}{c|c|c|c|c|c} 
      \textbf{Observable} & \textbf{Phase 1} & \textbf{Phase 2} & \textbf{Phase 3} & \textbf{Phase 4} & \textbf{Phase 5} \\
      \hline
      Pedestrian flows & 24.87 & 57.42 & 70.77 & 38.12 & 16.69 \\
      People occupancy & 756.03 & 1551.41 & 1177.95 & 1300.95 & 305.33 \\
      Pedestrian flows/People occupancy & 0.0329 & 0.0370 & 0.0601 & 0.0293 & 0.0547 
    \end{tabular}
  \end{center}
\end{table}

\section{Pedestrian traffic without Wi-Fi data}
\label{appendix:noWiFi}
We replicated the study on pedestrian traffic distribution without using the Wi-Fi data. By doing so, we isolated the contribution from topology, i.e. the network structure and location of buildings, from the empirical data on pedestrian traffic obtained through Wi-Fi connections. In particular, between each building pair, instead of associating the flux detected by Wi-Fi measurements, we assigned a constant flux independent of the chosen pair while we keep the total average number of moving pedestrians constant. These results are shown in Supplementary Fig. \ref{fig:scala_no_WiFi} which represent three choices for $k$. What we observed is that, without using Wi-Fi data, pedestrians are more equally distributed among all arcs.

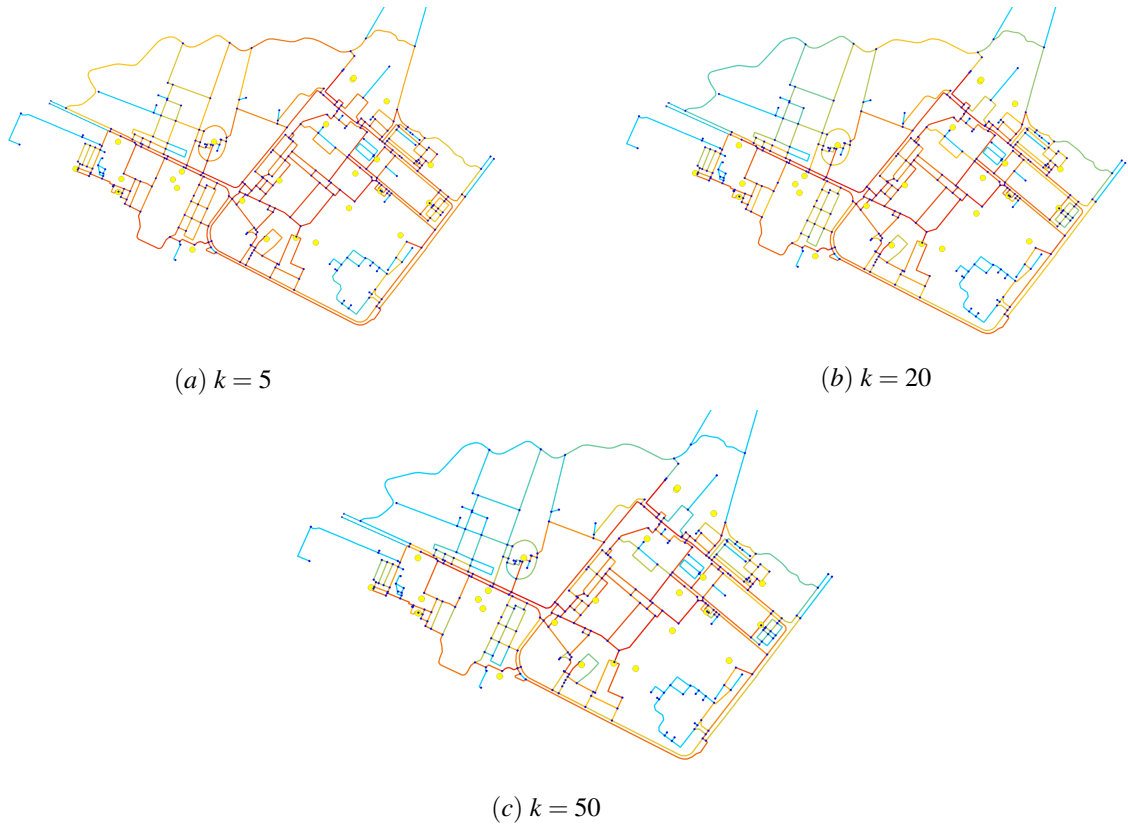
\begin{figure}[H]
  \centering
  \input{scala_no_WiFi.tikz}
  \caption{\label{fig:scala_no_WiFi} Analysis of pedestrian traffic without using Wi-Fi data. We added a logarithmic scale on arcs. As colors go from blue to red the estimated traffic increases. This scale goes from $0$ to the highest pedestrian traffic found, $669.04$. The networks were obtained by using a parameter $k$ equals to $5$, $20$ and $50$, respectively. Moreover, the maximum intensities of expected daily pedestrian traffic among all arcs are $509.3$, $594.9$ and $669.04$, respectively.}
\end{figure}

\end{document}

%% file: scala_1.tikz
\tikzset{every picture/.style={line width=0.75pt}} 

\begin{tikzpicture}[x=0.75pt,y=0.75pt,yscale=-1,xscale=1]

\draw (348.36,469.0) node  {\includegraphics[width=329.31pt,height=9.75pt]{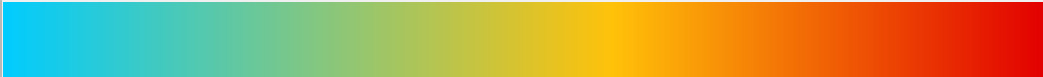}};
\draw (128.82,475.1) -- (567.91,475.1) 
      (201.82,469.1) -- (201.82,481.1)
      (274.82,469.1) -- (274.82,481.1)
      (347.82,469.1) -- (347.82,481.1)
      (420.82,469.1) -- (420.82,481.1)
      (493.82,469.1) -- (493.82,481.1)
      (566.82,469.1) -- (566.82,481.1);
\draw (129.55,467.5) -- (129.2,482.1);
\draw (245,445.6) node [anchor=north west][inner sep=0.75pt, font=\footnotesize] {Pedestrian traffic (daily average number of people)};

\draw (192.7,88.94) node  {\includegraphics[width=215.25pt,height=127.27pt]{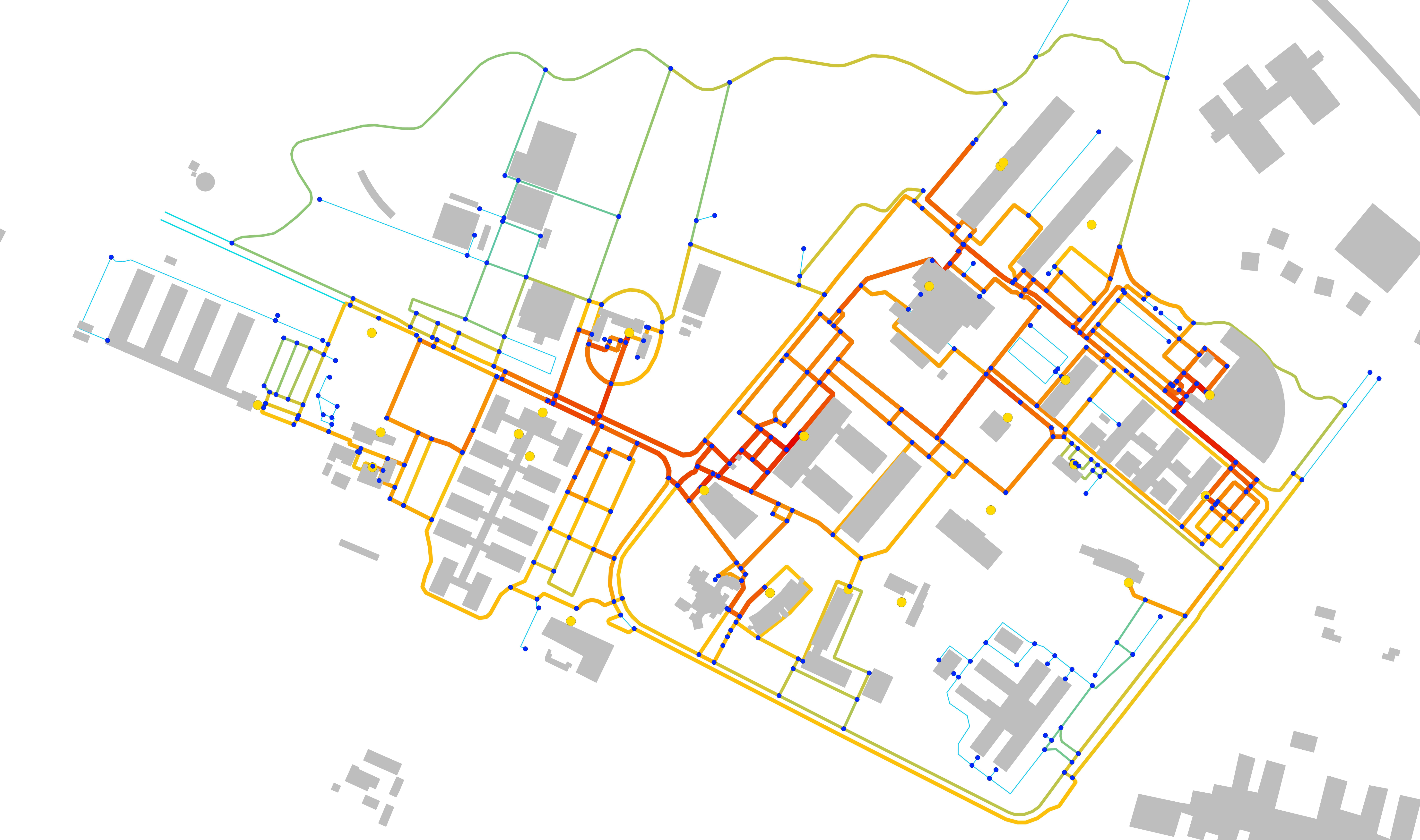}};
\draw (507.7,91.31) node  {\includegraphics[width=221.25pt,height=130.81pt]{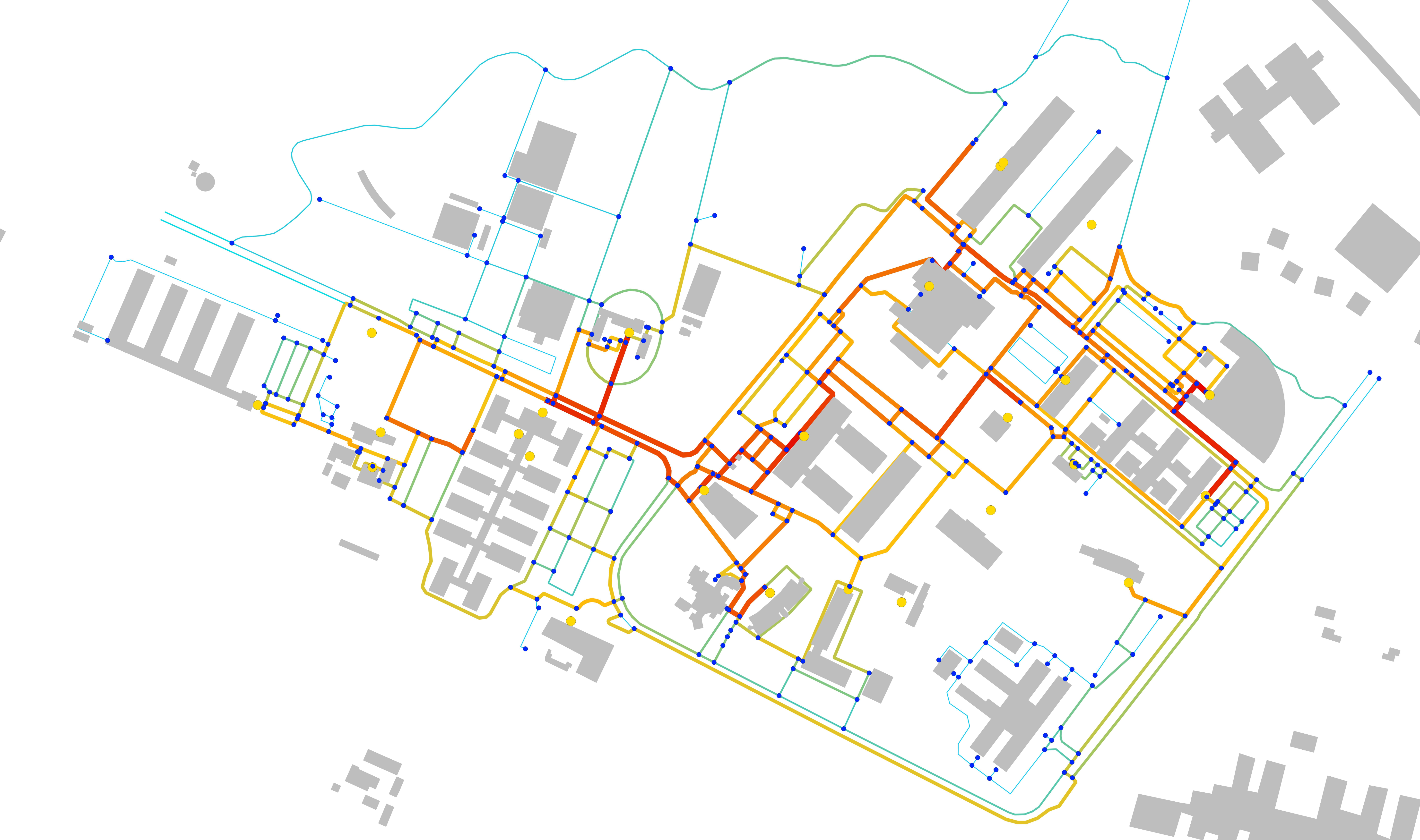}};
\draw (192.7,320) node  {\includegraphics[width=215.25pt,height=127.27pt]{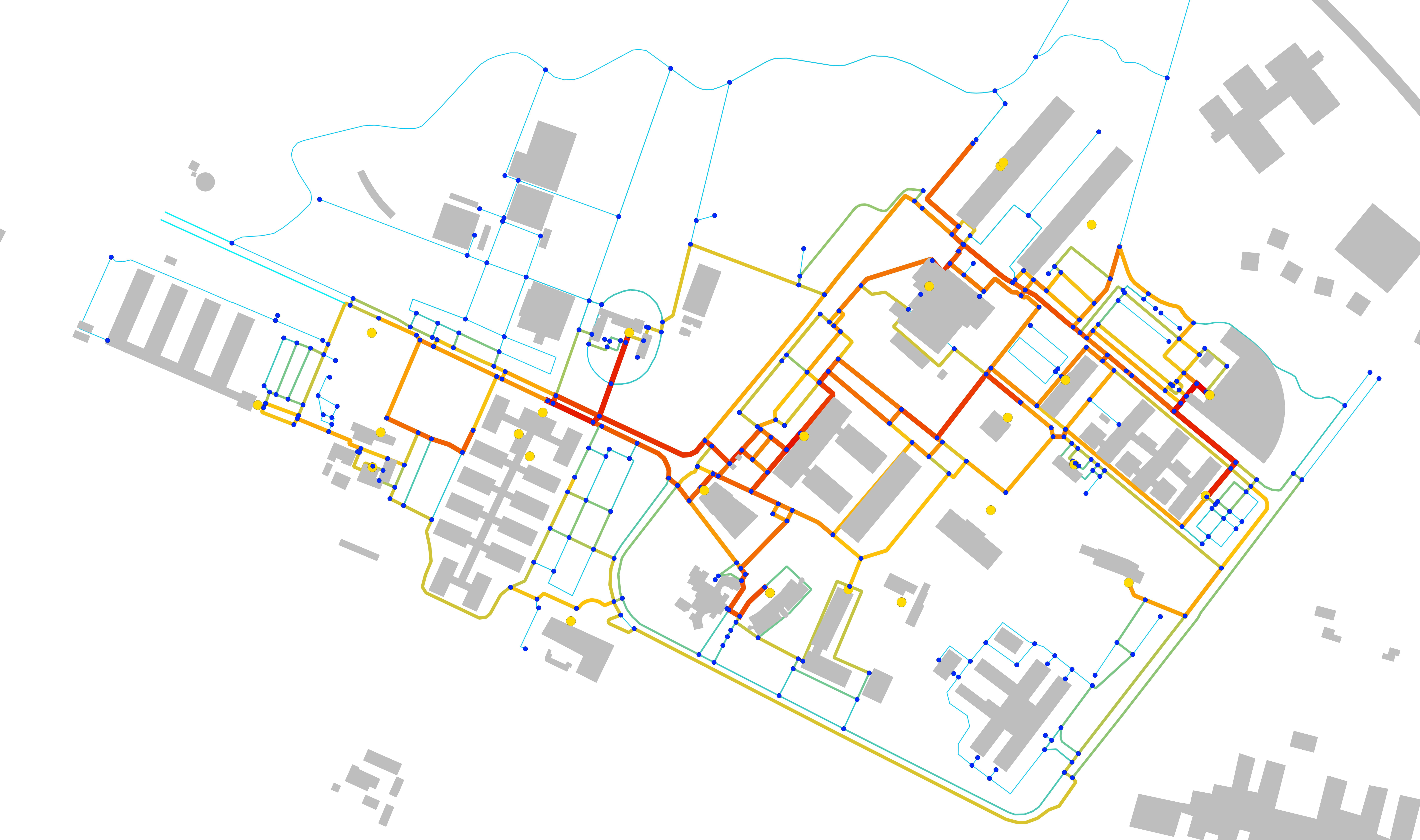}};
\draw (507.7,310) node  {\includegraphics[width=206.25pt,height=145.81pt]{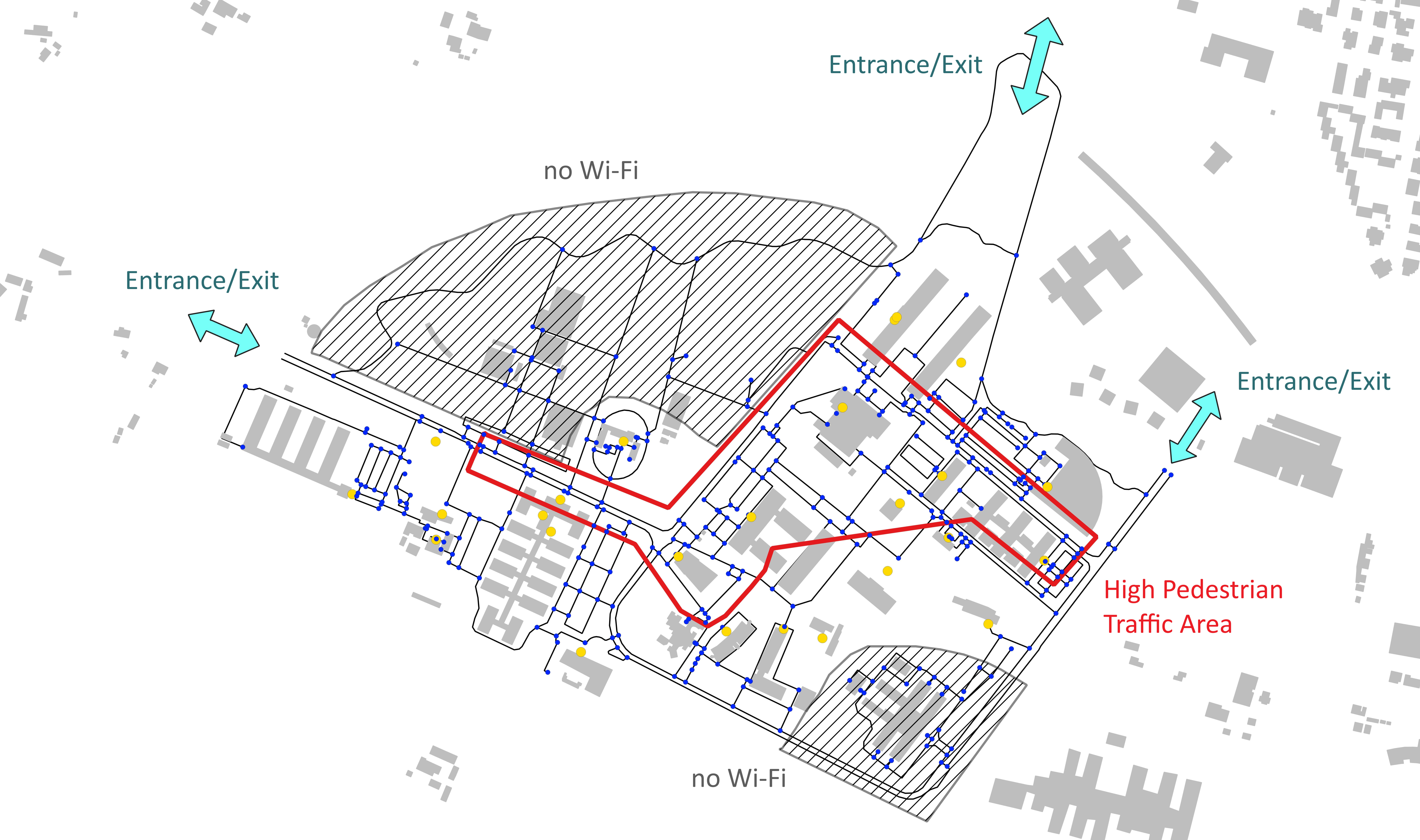}};

\draw (190.83,492.07) node [anchor=north west][inner sep=0.75pt, font=\footnotesize] {$2.07$};
\draw (124.27,492.07) node [anchor=north west][inner sep=0.75pt, font=\footnotesize] {$0$};
\draw (265.7,492.07)  node [anchor=north west][inner sep=0.75pt, font=\footnotesize] {$8.4$};
\draw (336.88,492.07) node [anchor=north west][inner sep=0.75pt, font=\footnotesize] {$32.2$};
\draw (410.78,492.07) node [anchor=north west][inner sep=0.75pt, font=\footnotesize] {$122.7$};
\draw (478.23,492.07) node [anchor=north west][inner sep=0.75pt, font=\footnotesize] {$468.4$};
\draw (548.69,492.07) node [anchor=north west][inner sep=0.75pt, font=\footnotesize] {$840.0$};

\draw (144,184.6) node [anchor=north west][inner sep=0.75pt] {$( a) \ k=5$};
\draw (470,183.6) node [anchor=north west][inner sep=0.75pt] {$( b) \ k=20$};
\draw (144,419.6) node [anchor=north west][inner sep=0.75pt] {$( c) \ k=50$};
\draw (430,419.6) node [anchor=north west][inner sep=0.75pt] {$( d)$ Highlighted regions};

\draw (348,610) node {\includegraphics[width=280pt]{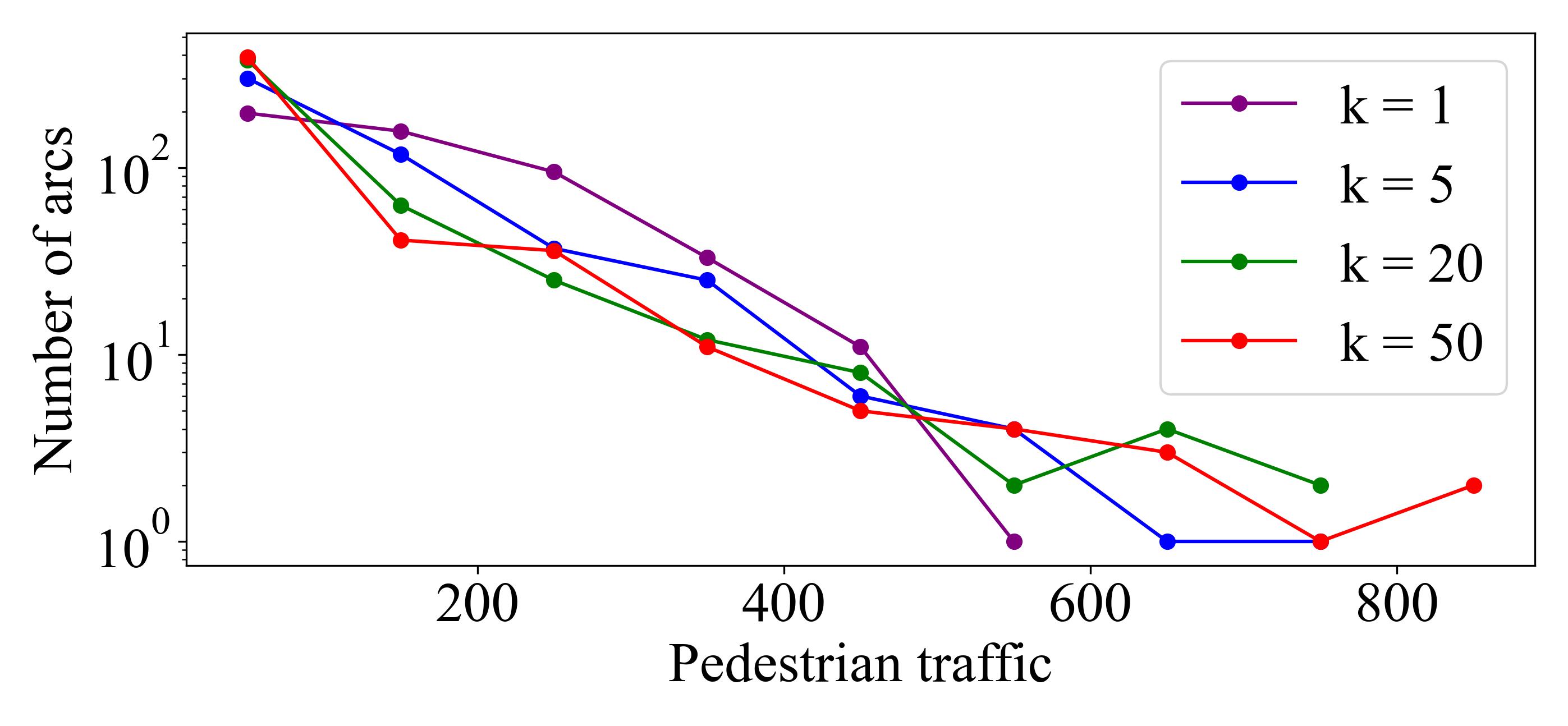}};
\draw (270,700) node [anchor=north west][inner sep=0.75pt] {$( e)$ Distribution of pedestrian traffic};

\end{tikzpicture}

%% file: scala_2.tikz
\tikzset{every picture/.style={line width=0.75pt}} 

\begin{tikzpicture}[x=0.75pt,y=0.75pt,yscale=-1,xscale=1] 

\draw (327.36,592.74) node  {\includegraphics[width=329.31pt,height=9.75pt]{scala_cromatica.jpeg}};

\draw    (107.82,599.24) -- (546.91,599.24) 
         (180.82,593.24) -- (180.82,605.24)
         (253.82,593.24) -- (253.82,605.24)
         (326.82,593.24) -- (326.82,605.24)
         (399.82,593.24) -- (399.82,605.24)
         (472.82,593.24) -- (472.82,605.24)
         (545.82,593.24) -- (545.82,605.24);

\draw    (108.55,591.64) -- (108.2,606.24);

\draw (181,567.24) node [anchor=north west][inner sep=0.75pt]  [font=\footnotesize]  {Pedestrian traffic (average number of people in a 20-minutes interval)};

\draw (171.83,615.67) node [anchor=north west][inner sep=0.75pt]  [font=\footnotesize]  {$0.63$};
\draw (103.27,615.67) node [anchor=north west][inner sep=0.75pt]  [font=\footnotesize]  {$0$};
\draw (243.7,615.67) node [anchor=north west][inner sep=0.75pt]  [font=\footnotesize]  {$1.67$};
\draw (318.88,615.67) node [anchor=north west][inner sep=0.75pt]  [font=\footnotesize]  {$3.23$};
\draw (389.78,615.67) node [anchor=north west][inner sep=0.75pt]  [font=\footnotesize]  {$5.12$};
\draw (461.23,615.67) node [anchor=north west][inner sep=0.75pt]  [font=\footnotesize]  {$7.62$};
\draw (535.69,615.67) node [anchor=north west][inner sep=0.75pt]  [font=\footnotesize]  {$17.99$};(108.2,596.24) ;

\draw (129.7,224.04) node  {\includegraphics[width=233.25pt,height=137.91pt]{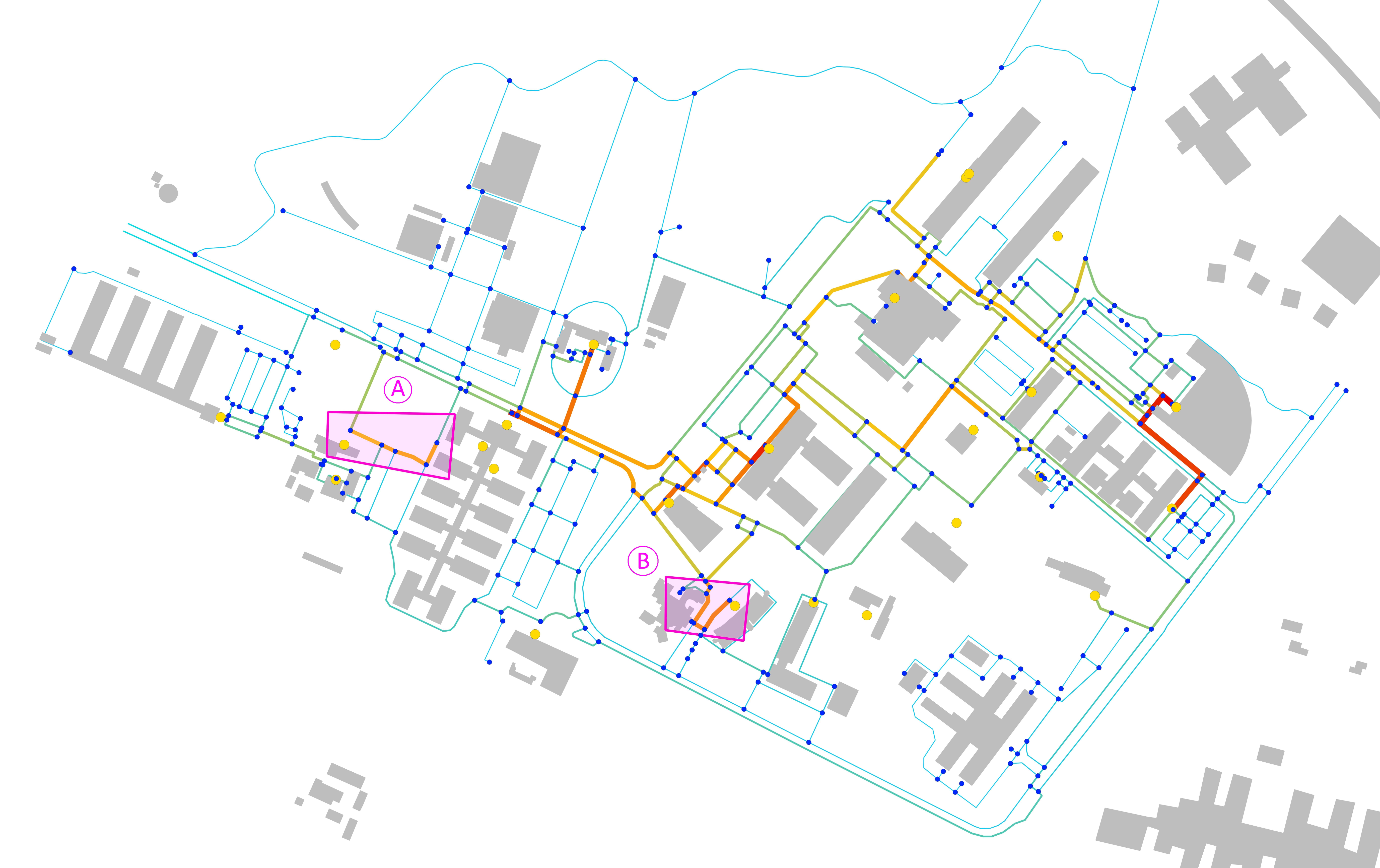}};
\draw (498.7,215.04) node  {\includegraphics[width=233.25pt,height=137.91pt]{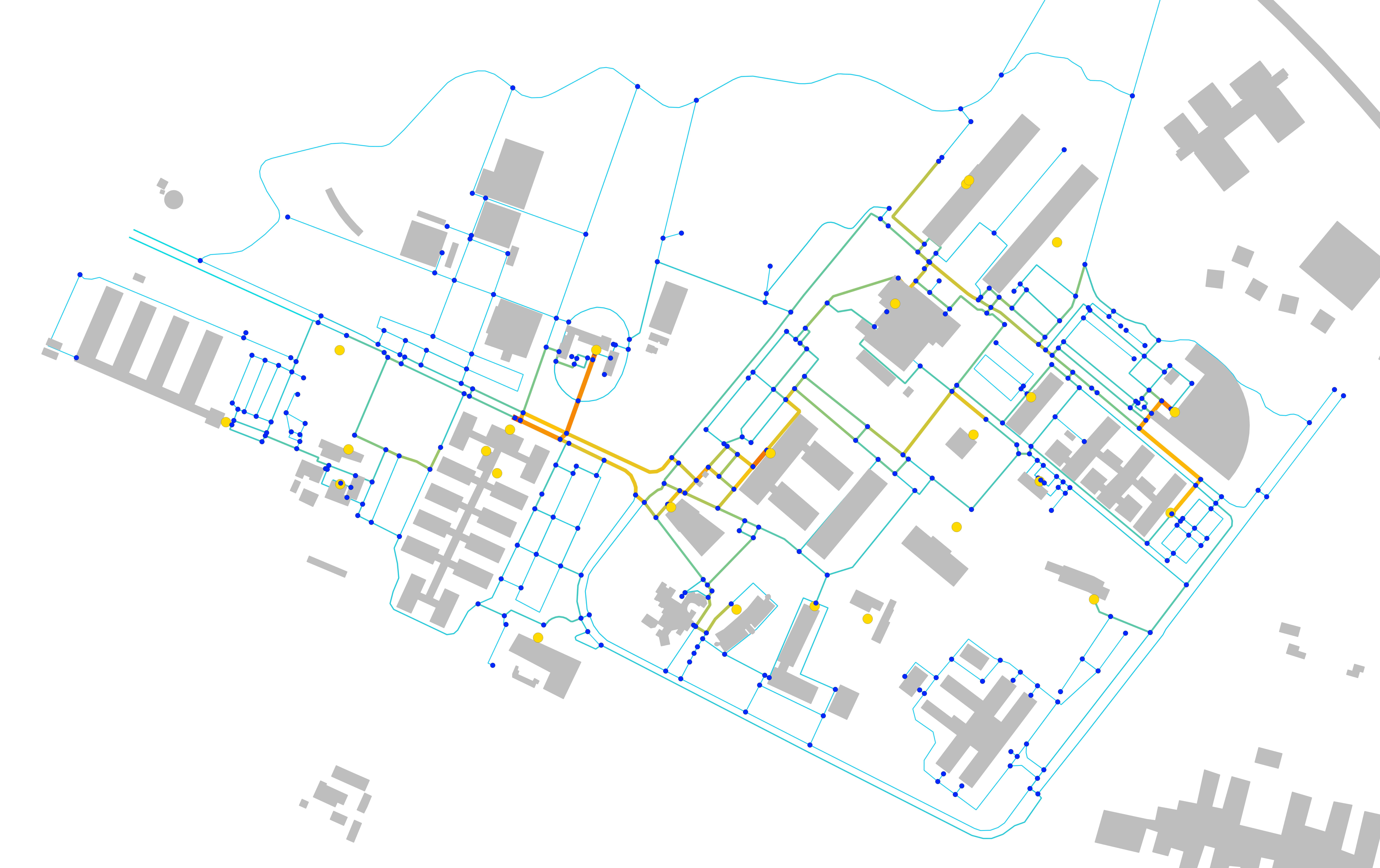}};
\draw (489.7,425.24) node  {\includegraphics[width=303.225pt,height=179.283pt]{Distribuzione_times_final.jpeg}};
\draw (509.7,13.04) node  {\includegraphics[width=233.25pt,height=137.91pt]{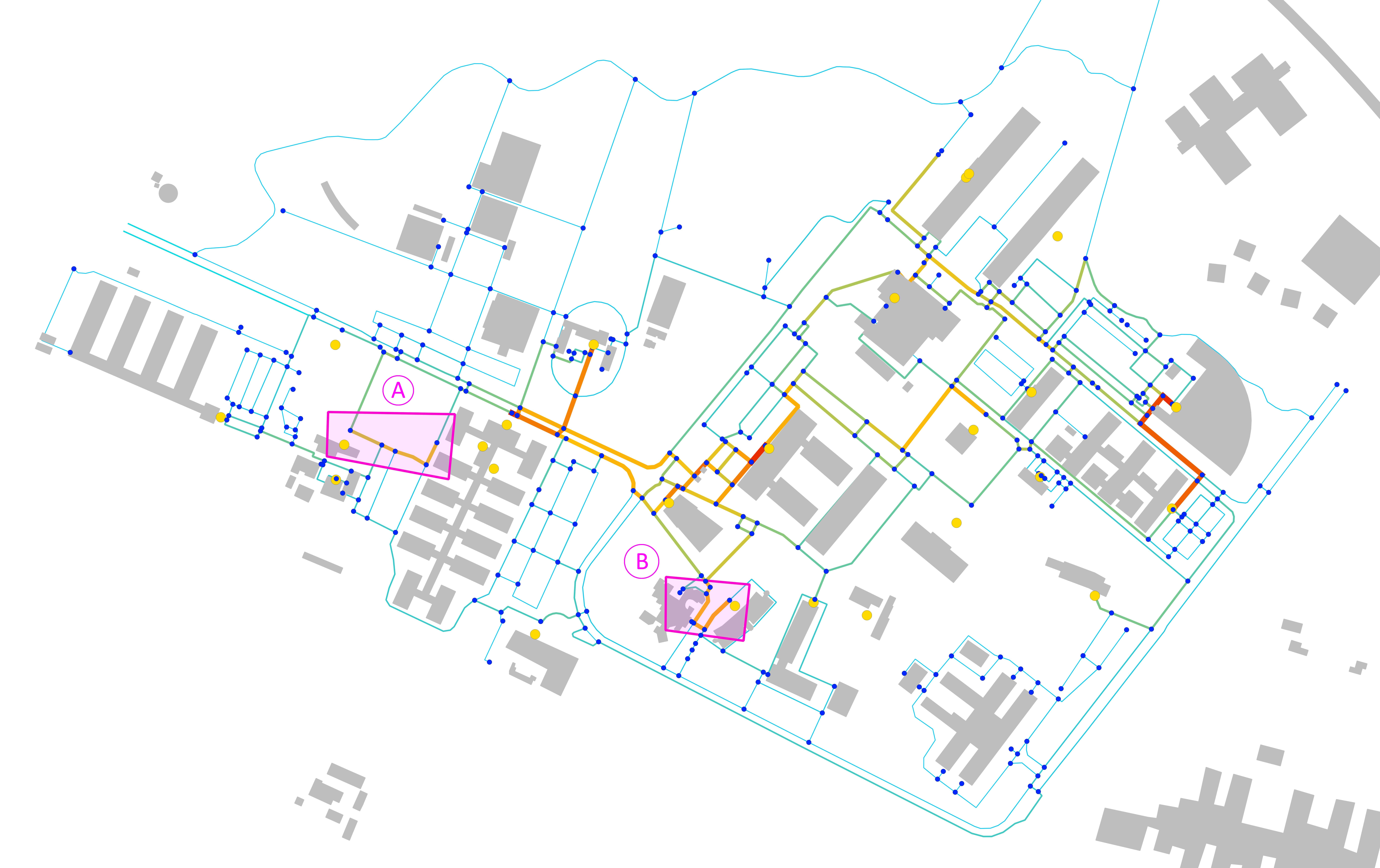}};
\draw (129,23.06) node  {\includegraphics[width=231pt,height=136.58pt]{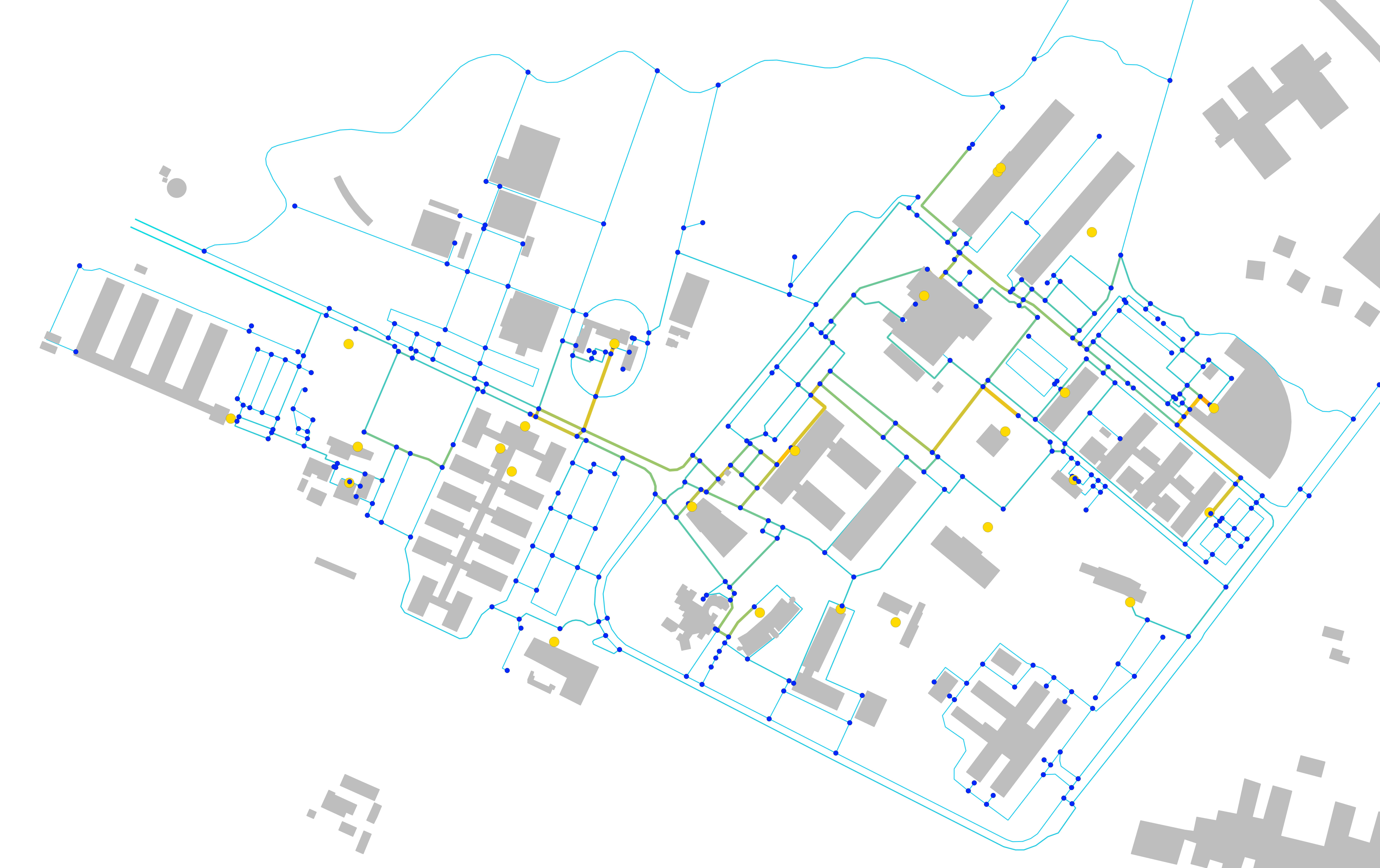}};
\draw (129,440.24) node  {\includegraphics[width=233.25pt,height=137.91pt]{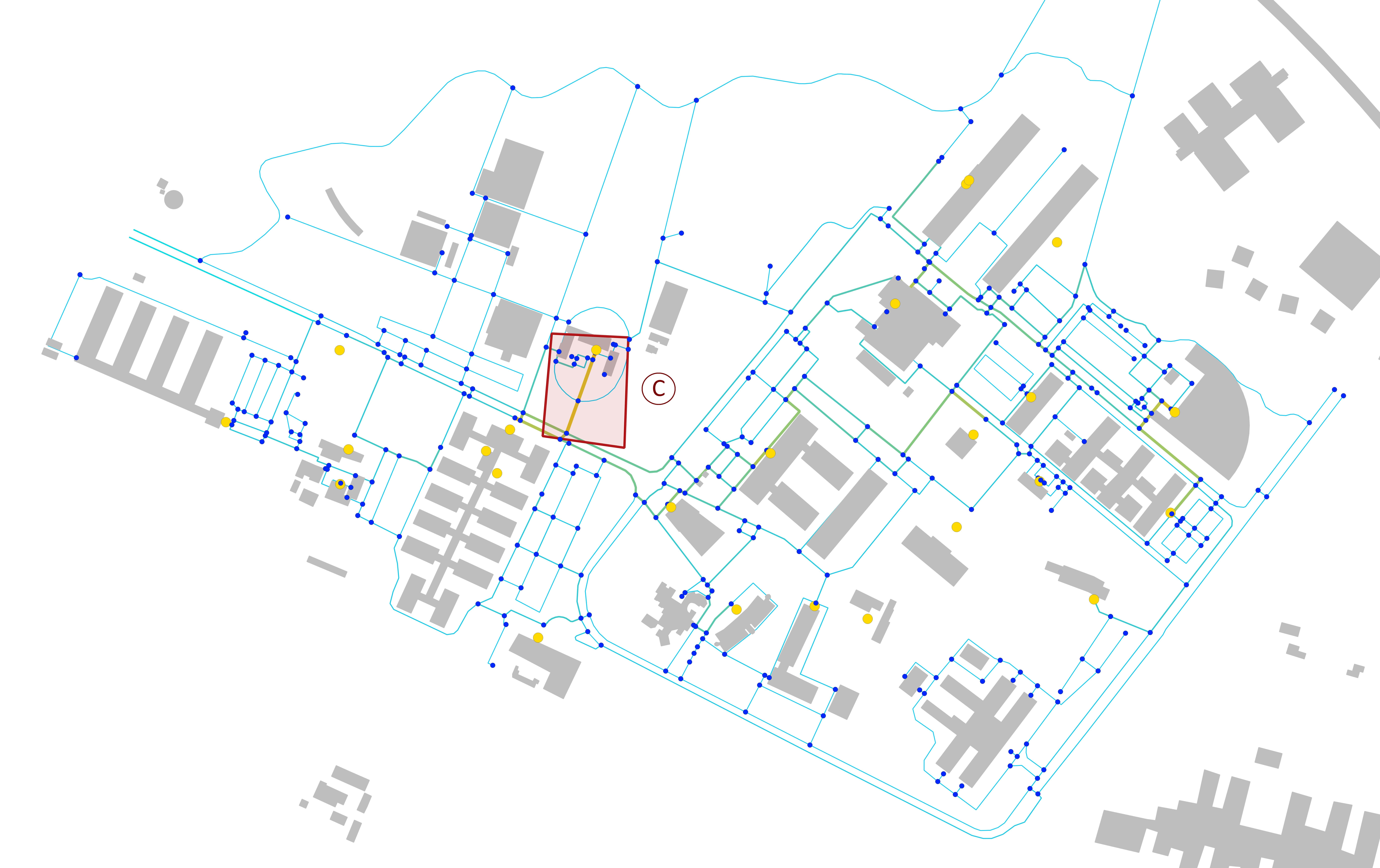}};

\draw (67,105) node [anchor=north west][inner sep=0.75pt]    {$( a) \ Early\ morning$};
\draw (453,95) node [anchor=north west][inner sep=0.75pt]    {$( b) \ Late\ morning$};
\draw (74,307) node [anchor=north west][inner sep=0.75pt]    {$( c) \ Lunch\ break$};
\draw (444,300) node [anchor=north west][inner sep=0.75pt]    {$( d) \ Early\ afternoon$};
\draw (50,539) node [anchor=north west][inner sep=0.75pt]    {$( e) \ Late\ afternoon\ and\ evening$}; 
\draw (370,539) node [anchor=north west][inner sep=0.75pt]    {$( f) \ Time\ distribution\ of\ the\ total\ occupancy$};

\end{tikzpicture}

%% file: scala_3.tikz
\tikzset{every picture/.style={line width=0.75pt}} 

\begin{tikzpicture}[x=0.75pt,y=0.75pt,yscale=-1,xscale=1]

\draw (348.36,330) node  {\includegraphics[width=246.98pt,height=173.06pt]{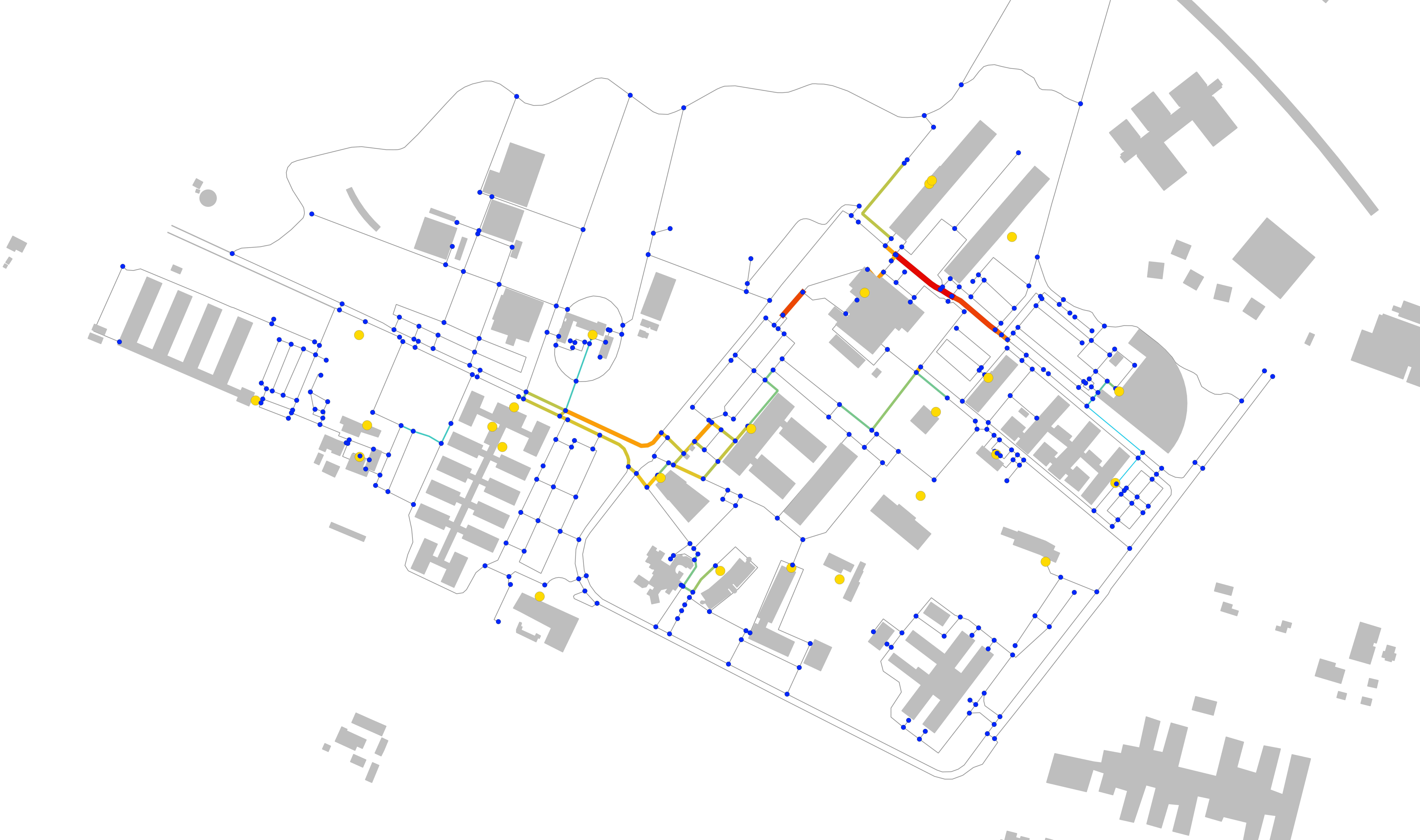}};

\draw (193,454) node [anchor=north west][inner sep=0.75pt, font=\footnotesize] {$L_{j}$: Participation ratio (number of building pairs contributing to an arc traffic)};

\draw (348.36,475.6) node  {\includegraphics[width=329.31pt,height=8.75pt]{scala_cromatica.jpeg}};

\draw    (128.82,481.1) -- (567.91,481.1) 
          (201.82,475.1) -- (201.82,487.1)
          (274.82,475.1) -- (274.82,487.1)
          (347.82,475.1) -- (347.82,487.1)
          (420.82,475.1) -- (420.82,487.1)
          (493.82,475.1) -- (493.82,487.1)
          (566.82,475.1) -- (566.82,487.1) ;

\draw    (129.55,473.5) -- (129.2,488.1) ;

\draw (190.83,498.07) node [anchor=north west][inner sep=0.75pt, font=\footnotesize]  {$1.892$};
\draw (124.27,498.07) node [anchor=north west][inner sep=0.75pt, font=\footnotesize]  {$1.102$};
\draw (265.7,498.07)  node [anchor=north west][inner sep=0.75pt, font=\footnotesize]  {$2.980$};
\draw (336.88,498.07) node [anchor=north west][inner sep=0.75pt, font=\footnotesize]  {$4.477$};
\draw (410.78,498.07) node [anchor=north west][inner sep=0.75pt, font=\footnotesize]  {$6.536$};
\draw (478.23,498.07) node [anchor=north west][inner sep=0.75pt, font=\footnotesize]  {$9.370$};
\draw (548.69,498.07) node [anchor=north west][inner sep=0.75pt, font=\footnotesize]  {$13.270$};

\end{tikzpicture}

%% file: scala_robust.tikz
\tikzset{every picture/.style={line width=0.75pt}} 

\begin{tikzpicture}[x=0.75pt,y=0.75pt,yscale=-1,xscale=1]

\draw (348.36,330) node  {\includegraphics[width=246.98pt,height=162.67pt]{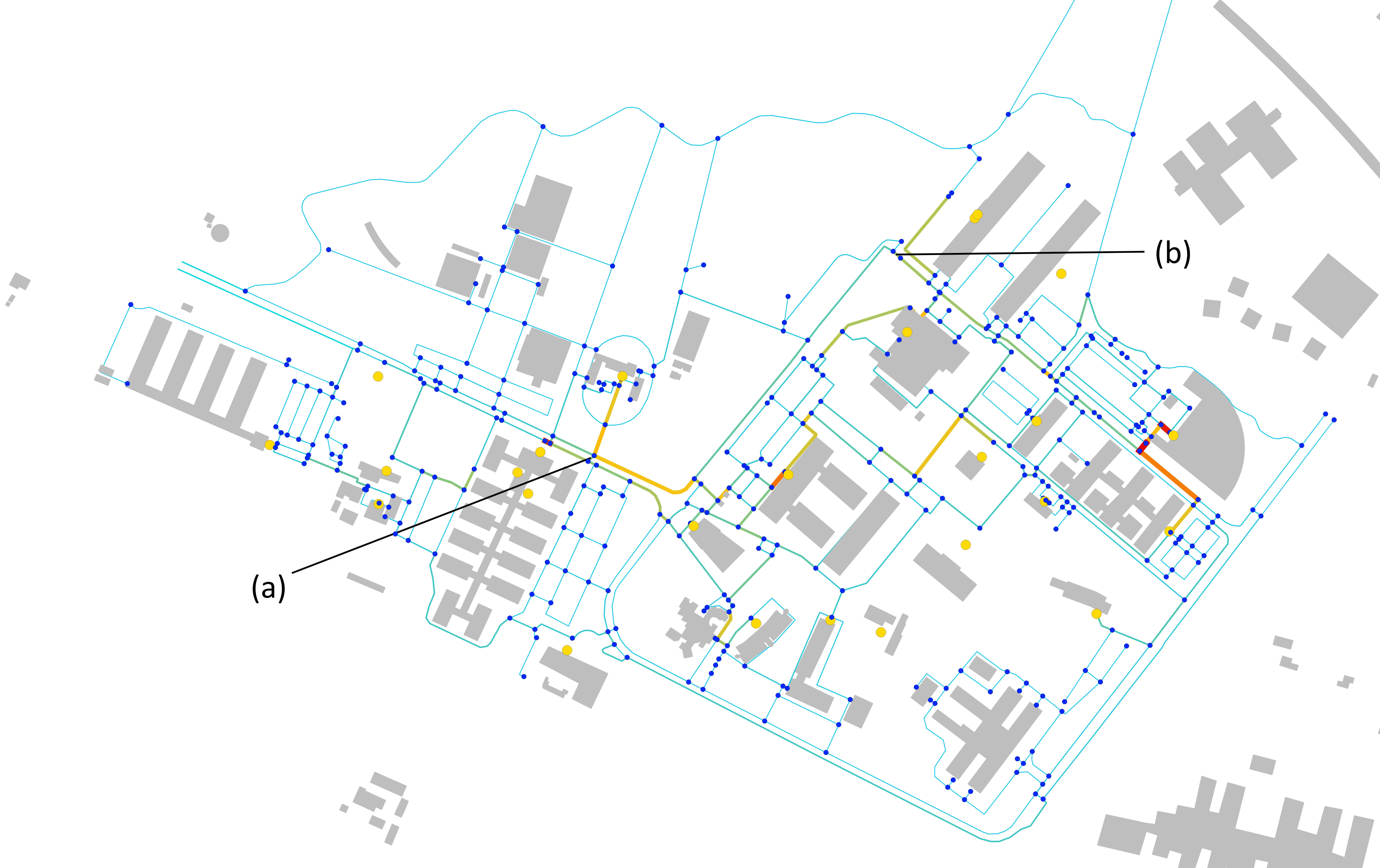}};

\draw (219,468) node [anchor=north west][inner sep=0.75pt, font=\footnotesize] {$\Delta p_{u}$ : Sum of the traffic variations (daily average number of people)};

\draw (348.36,490.6) node  {\includegraphics[width=329.31pt,height=8.75pt]{scala_cromatica.jpeg}};

\draw    (128.82,496.1) -- (567.91,496.1) 
          (201.82,490.1) -- (201.82,502.1)
          (274.82,490.1) -- (274.82,502.1)
          (347.82,490.1) -- (347.82,502.1)
          (420.82,490.1) -- (420.82,502.1)
          (493.82,490.1) -- (493.82,502.1)
          (566.82,490.1) -- (566.82,502.1);
\draw    (129.55,488.5) -- (129.2,503.1) ;

\draw (190.83,505.47) node [anchor=north west][inner sep=0.75pt, font=\footnotesize]  {$1235$};
\draw (124.27,505.47) node [anchor=north west][inner sep=0.75pt, font=\footnotesize]  {$0$};
\draw (265.7,505.47)  node [anchor=north west][inner sep=0.75pt, font=\footnotesize]  {$2469$};
\draw (336.88,505.47) node [anchor=north west][inner sep=0.75pt, font=\footnotesize]  {$3704$};
\draw (410.78,505.47) node [anchor=north west][inner sep=0.75pt, font=\footnotesize]  {$4939$};
\draw (478.23,505.47) node [anchor=north west][inner sep=0.75pt, font=\footnotesize]  {$6173$};
\draw (548.69,505.47) node [anchor=north west][inner sep=0.75pt, font=\footnotesize]  {$7408$};

\draw (210,630) node  {\includegraphics[width=185pt,height=110pt]{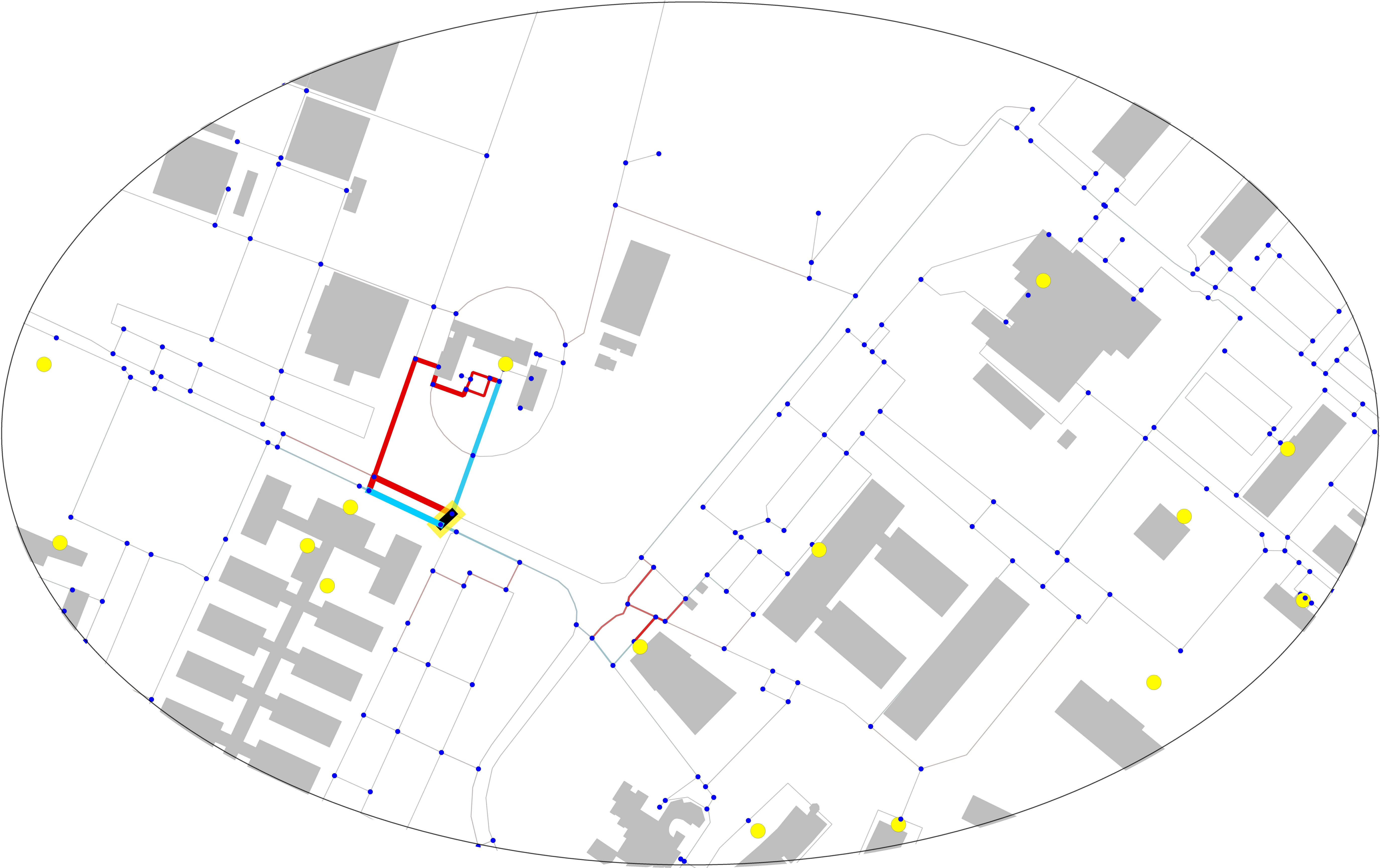}};
\draw (500,630) node  {\includegraphics[width=185pt,height=110pt]{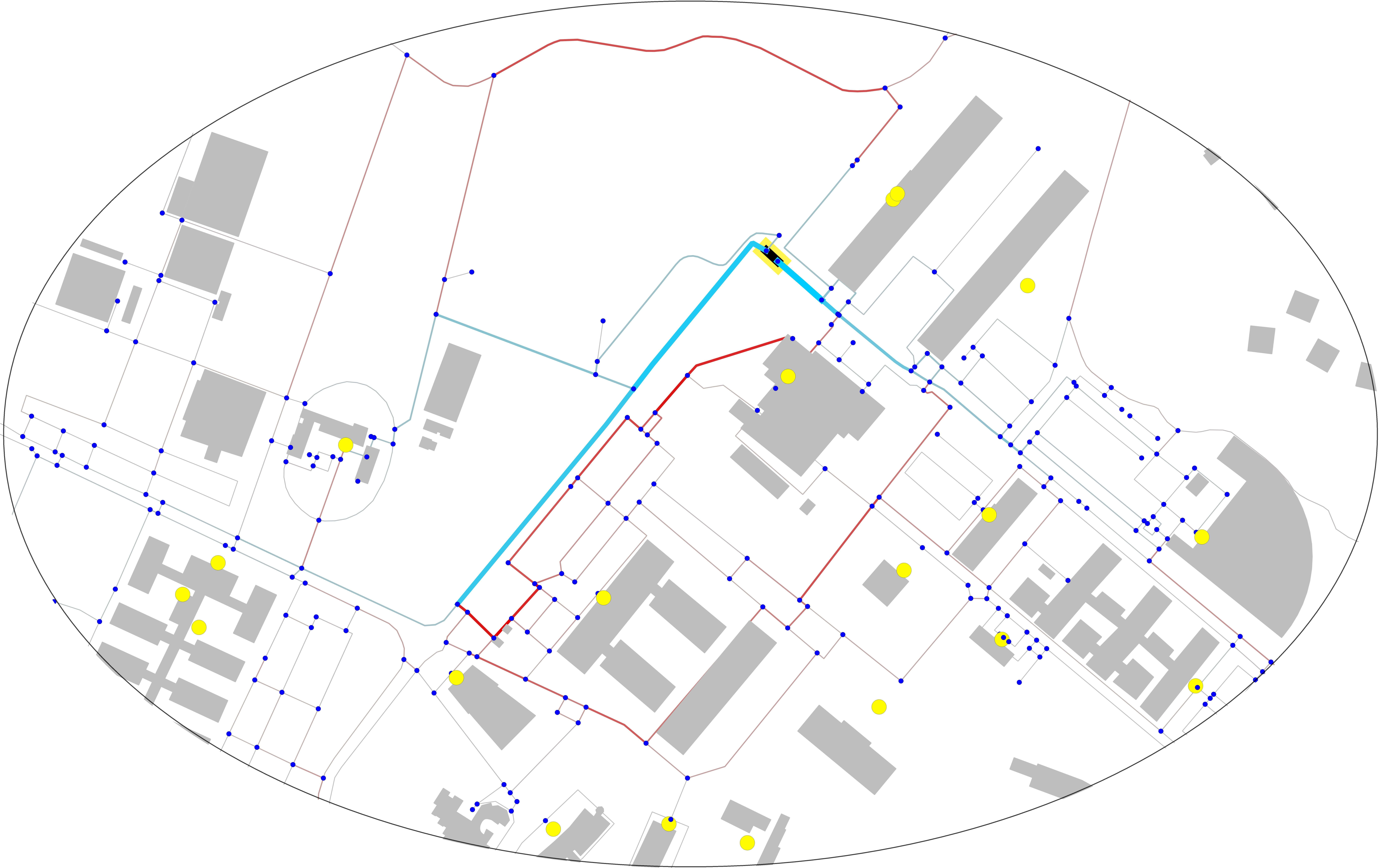}};

\draw (170,730) node [anchor=north west][inner sep=0.75pt]    {$( a) \ \Delta p_{u} = 2020.9$};
\draw (460,730) node [anchor=north west][inner sep=0.75pt]    {$( b) \ \Delta p_{u} = 2778.9$};

\draw (240,766) node [anchor=north west][inner sep=0.75pt, font=\footnotesize] {$p_{j | u} - p_{j}$: Traffic variation (daily average number of people)};

\draw (348.36,790) node  {\includegraphics[width=329.31pt,height=8.75pt]{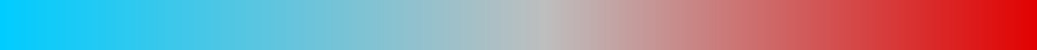}};

\draw    (128.82,795.5) -- (567.91,795.5) 
          (201.82,789.5) -- (201.82,801.5)
          (274.82,789.5) -- (274.82,801.5)
          (347.82,789.5) -- (347.82,801.5)
          (420.82,789.5) -- (420.82,801.5)
          (493.82,789.5) -- (493.82,801.5)
          (566.82,789.5) -- (566.82,801.5) ;
\draw    (129.55,787.9) -- (129.2,802.5) ;

\draw (190.83,814.47) node [anchor=north west][inner sep=0.75pt, font=\footnotesize]  {${-}165.06$};
\draw (124.27,814.47) node [anchor=north west][inner sep=0.75pt, font=\footnotesize]  {${-}235.77$};
\draw (265.7,814.47)  node [anchor=north west][inner sep=0.75pt, font=\footnotesize]  {${-}94.35$};
\draw (336.88,814.47) node [anchor=north west][inner sep=0.75pt, font=\footnotesize]  {${-}23.64$};
\draw (410.78,814.47) node [anchor=north west][inner sep=0.75pt, font=\footnotesize]  {$47.07$};
\draw (478.23,814.47) node [anchor=north west][inner sep=0.75pt, font=\footnotesize]  {$117.78$};
\draw (548.69,814.47) node [anchor=north west][inner sep=0.75pt, font=\footnotesize]  {$188.50$};

\end{tikzpicture}

%% file: scala_robust_2.tikz
\tikzset{every picture/.style={line width=0.75pt}}        

\begin{tikzpicture}[x=0.75pt,y=0.75pt,yscale=-1,xscale=1]

\draw (348.36,330) node  {\includegraphics[width=246.98pt,height=162.67pt]{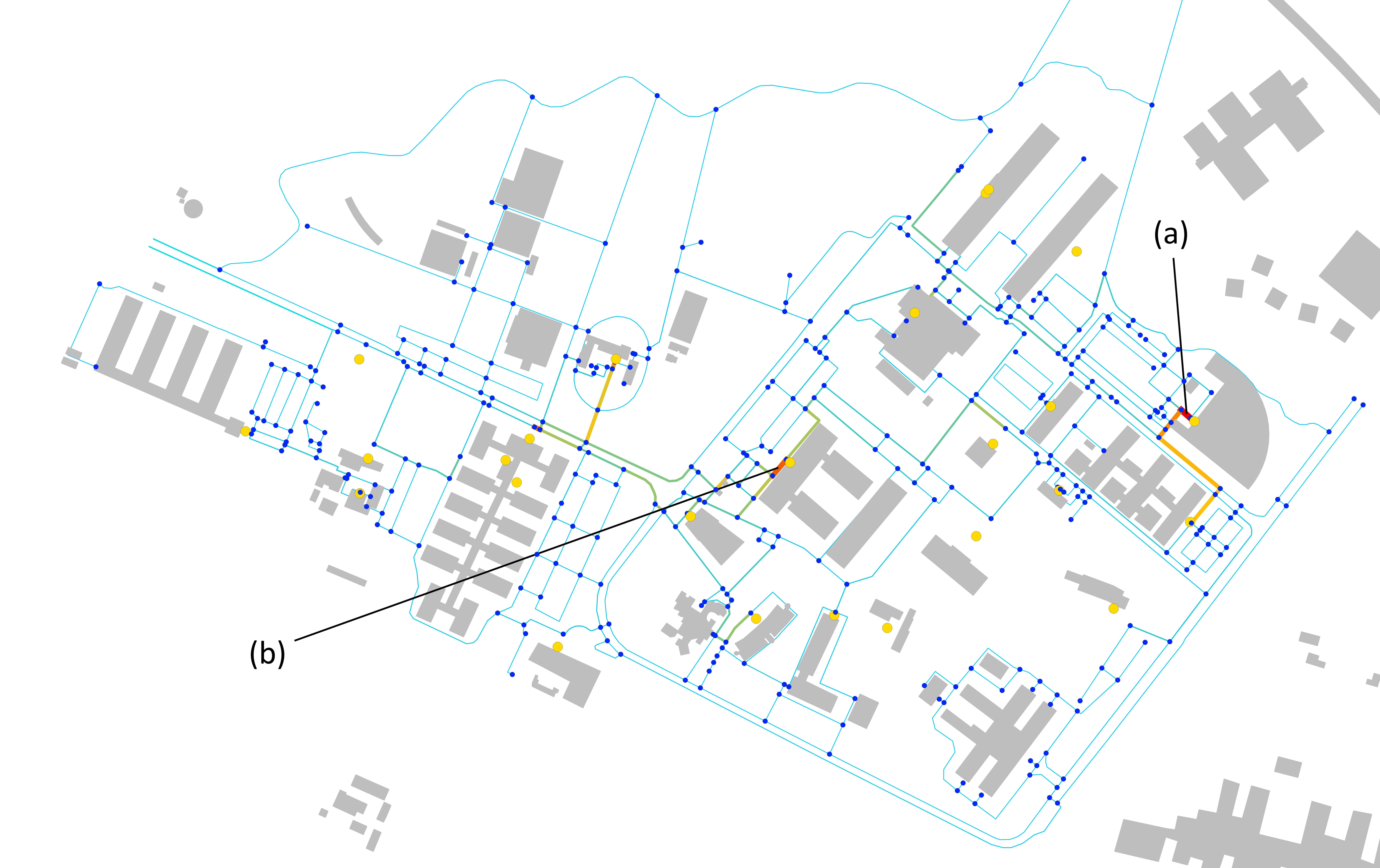}};

\draw (190,468) node [anchor=north west][inner sep=0.75pt, font=\footnotesize] 
{$\Delta p^{\max}_{u}$ : Maximum of the traffic variations (daily average number of people)};

\draw (348.36,490.6) node  {\includegraphics[width=329.31pt,height=8.75pt]{scala_cromatica.jpeg}};

\draw    (128.82,496.1) -- (567.91,496.1)
          (201.82,490.1) -- (201.82,502.1)
          (274.82,490.1) -- (274.82,502.1)
          (347.82,490.1) -- (347.82,502.1)
          (420.82,490.1) -- (420.82,502.1)
          (493.82,490.1) -- (493.82,502.1)
          (566.82,490.1) -- (566.82,502.1);
\draw    (129.55,488.5) -- (129.2,503.1);

\draw (190.83,505.47) node [anchor=north west][inner sep=0.75pt, font=\footnotesize]  {$132.31$};
\draw (124.27,505.47) node [anchor=north west][inner sep=0.75pt, font=\footnotesize]  {$0$};
\draw (265.7,505.47)  node [anchor=north west][inner sep=0.75pt, font=\footnotesize]  {$264.62$};
\draw (336.88,505.47) node [anchor=north west][inner sep=0.75pt, font=\footnotesize]  {$396.92$};
\draw (410.78,505.47) node [anchor=north west][inner sep=0.75pt, font=\footnotesize]  {$529.23$};
\draw (478.23,505.47) node [anchor=north west][inner sep=0.75pt, font=\footnotesize]  {$661.54$};
\draw (548.69,505.47) node [anchor=north west][inner sep=0.75pt, font=\footnotesize]  {$793.85$};

\draw (210,630) node  {\includegraphics[width=185pt,height=110pt]{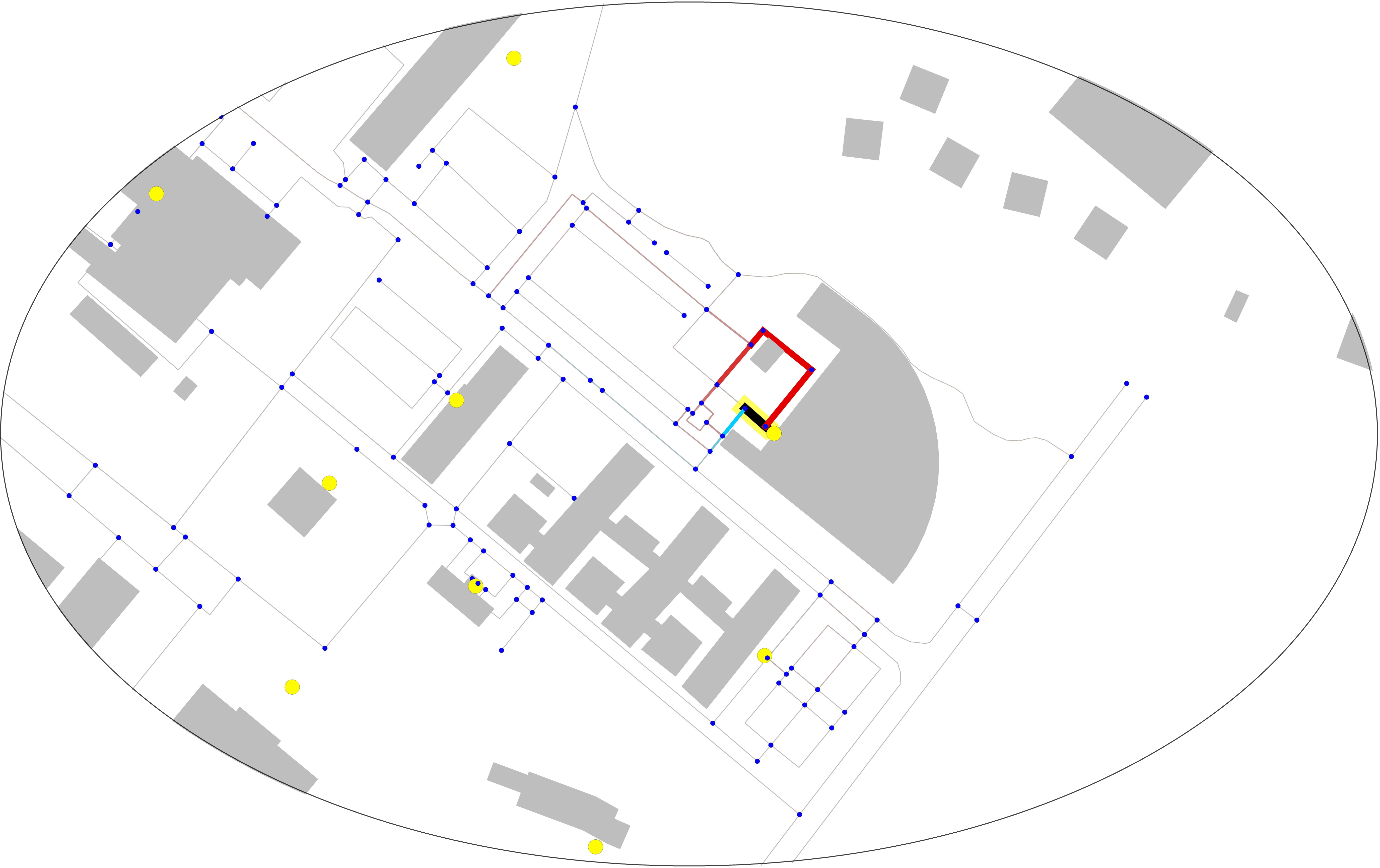}};
\draw (500,630) node  {\includegraphics[width=185pt,height=110pt]{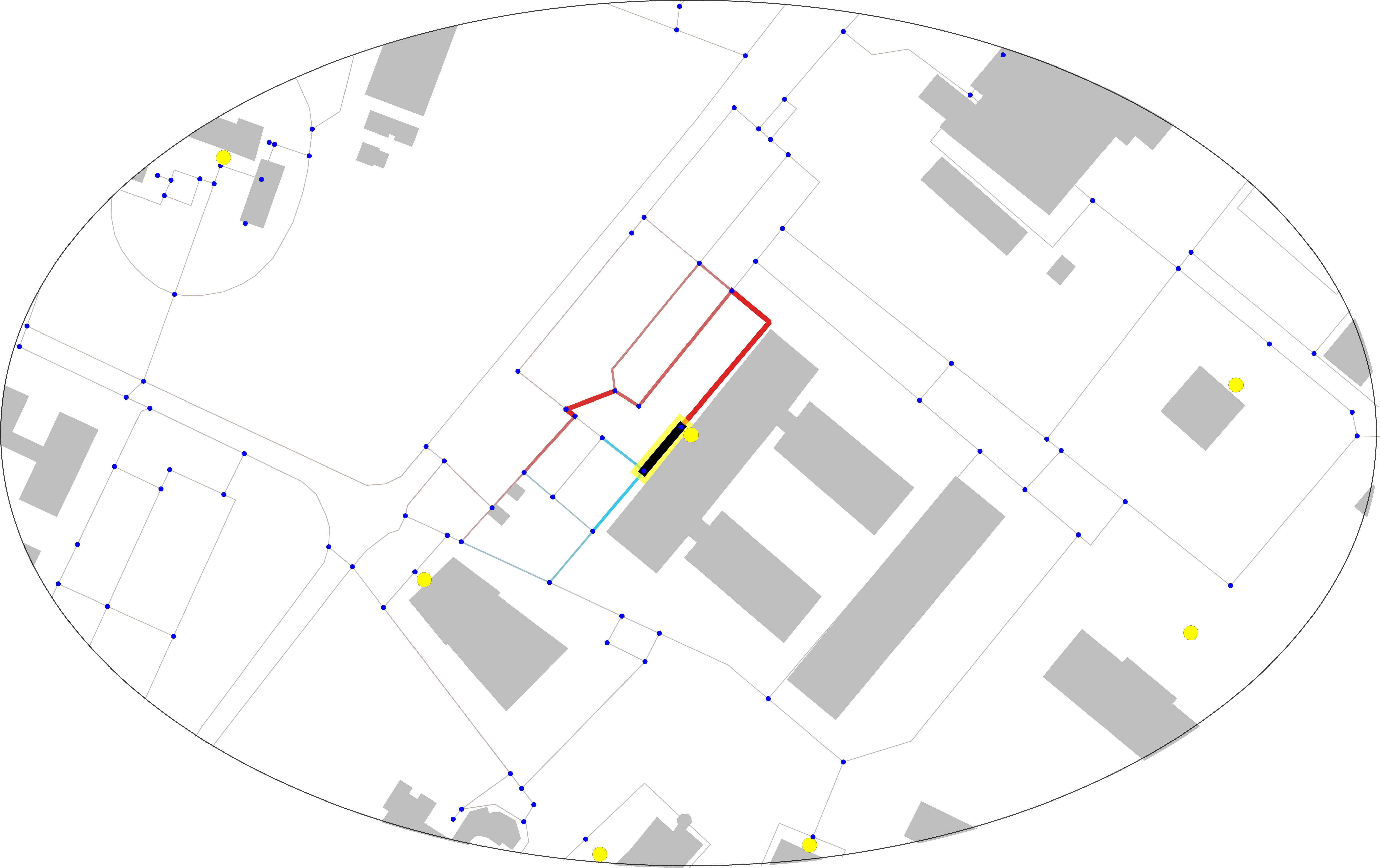}};

\draw (170,730) node [anchor=north west][inner sep=0.75pt]    {$( a) \ \Delta p_{u} = 7408$};
\draw (460,730) node [anchor=north west][inner sep=0.75pt]    {$( b) \ \Delta p_{u} = 6969$};

\draw (240,766) node [anchor=north west][inner sep=0.75pt, font=\footnotesize] {$p_{j | u} - p_{j}$: Traffic variation (daily average number of people)};

\draw (348.36,790) node  {\includegraphics[width=329.31pt,height=8.75pt]{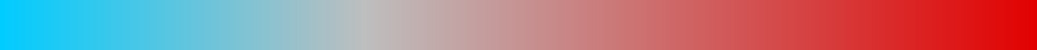}};

\draw    (128.82,795.5) -- (567.91,795.5)
          (201.82,789.5) -- (201.82,801.5)
          (274.82,789.5) -- (274.82,801.5)
          (347.82,789.5) -- (347.82,801.5)
          (420.82,789.5) -- (420.82,801.5)
          (493.82,789.5) -- (493.82,801.5)
          (566.82,789.5) -- (566.82,801.5);
\draw    (129.55,787.9) -- (129.2,802.5);

\draw (190.83,814.47) node [anchor=north west][inner sep=0.75pt, font=\footnotesize]  {${-}237.0$};
\draw (124.27,814.47) node [anchor=north west][inner sep=0.75pt, font=\footnotesize]  {${-}443.2$};
\draw (265.7,814.47) node [anchor=north west][inner sep=0.75pt, font=\footnotesize]  {${-}30.9$};
\draw (336.88,814.47) node [anchor=north west][inner sep=0.75pt, font=\footnotesize]  {$175.3$};
\draw (410.78,814.47) node [anchor=north west][inner sep=0.75pt, font=\footnotesize]  {$381.5$};
\draw (478.23,814.47) node [anchor=north west][inner sep=0.75pt, font=\footnotesize]  {$587.6$};
\draw (548.69,814.47) node [anchor=north west][inner sep=0.75pt, font=\footnotesize]  {$793.8$};

\end{tikzpicture}

%% file: scala_walk.tikz
\tikzset{every picture/.style={line width=0.75pt}} 

\begin{tikzpicture}[x=0.75pt,y=0.75pt,yscale=-1,xscale=1]

\draw (348.36,330) node {\includegraphics[width=246.98pt,height=173.06pt]{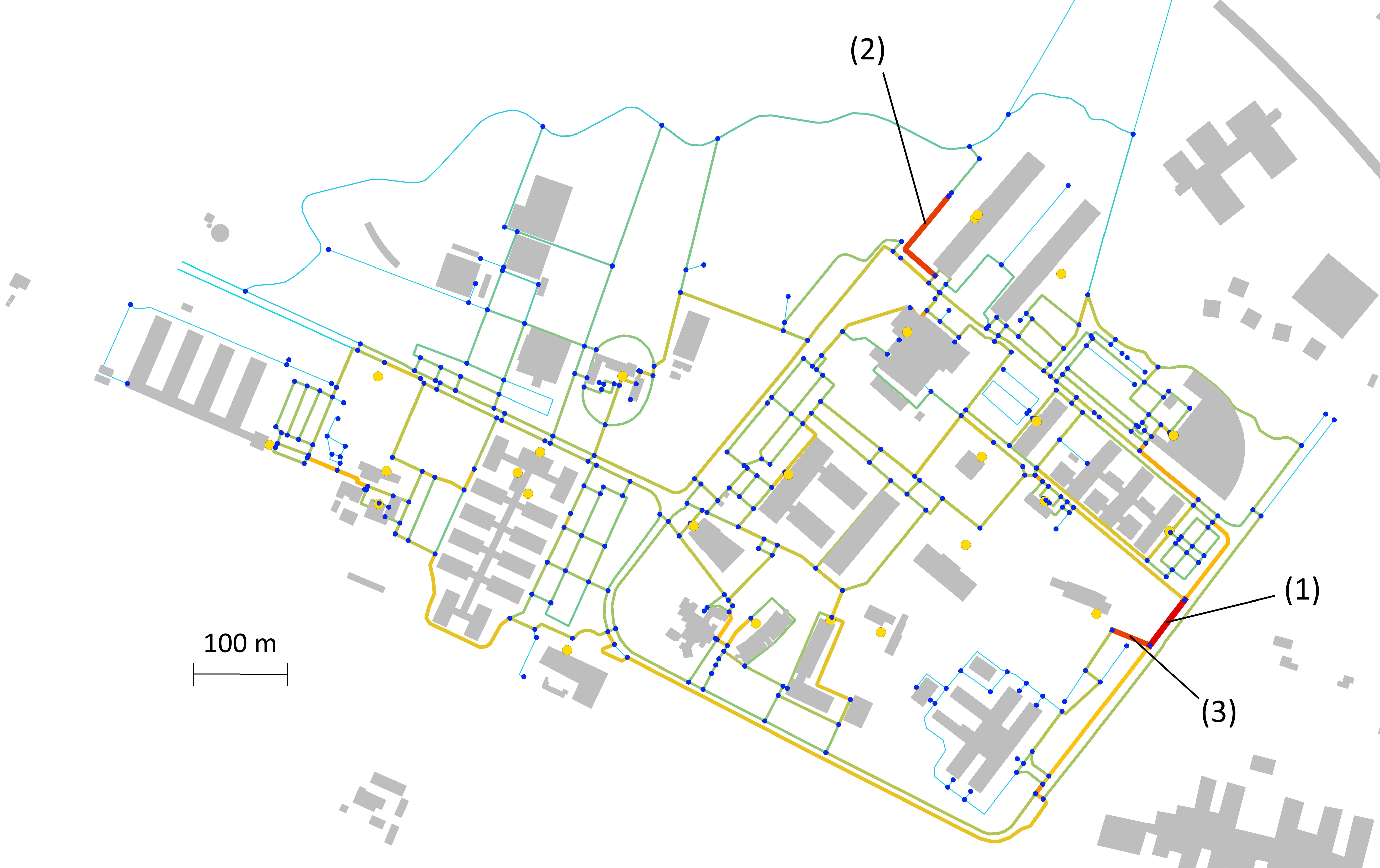}};

\draw (260,470) node [anchor=north west][inner sep=0.75pt, font=\footnotesize] 
    {$\Delta L_{u}$: Extra meters per displaced person};

\draw (348.36,492) node {\includegraphics[width=329.31pt,height=8.75pt]{scala_cromatica.jpeg}};

\draw (128.82,497.5) -- (567.91,497.5) 
      (201.82,491.5) -- (201.82,503.5)
      (274.82,491.5) -- (274.82,503.5)
      (347.82,491.5) -- (347.82,503.5)
      (420.82,491.5) -- (420.82,503.5)
      (493.82,491.5) -- (493.82,503.5)
      (566.82,491.5) -- (566.82,503.5);

\draw (129.55,489.9) -- (129.2,504.5);

\draw (124.27,514.47) node [anchor=north west][inner sep=0.75pt, font=\footnotesize] {$-282.29$};
\draw (190.83,514.47) node [anchor=north west][inner sep=0.75pt, font=\footnotesize] {$-158.94$};
\draw (265.7,514.47)  node [anchor=north west][inner sep=0.75pt, font=\footnotesize] {$-35.58$};
\draw (336.88,514.47) node [anchor=north west][inner sep=0.75pt, font=\footnotesize] {$87.77$};
\draw (410.78,514.47) node [anchor=north west][inner sep=0.75pt, font=\footnotesize] {$211.12$};
\draw (478.23,514.47) node [anchor=north west][inner sep=0.75pt, font=\footnotesize] {$334.48$};
\draw (548.69,514.47) node [anchor=north west][inner sep=0.75pt, font=\footnotesize] {$457.83$};

\end{tikzpicture}

%% file: scala_k_WiFi.tikz
\tikzset{every picture/.style={line width=0.75pt}} 

\begin{tikzpicture}[x=0.75pt,y=0.75pt,yscale=-1,xscale=1]

\draw (245,427) node [anchor=north west][inner sep=0.75pt, font=\footnotesize] 
    {Fraction of the total pedestrian traffic};

\draw (348.36,454.6) node {\includegraphics[width=329.31pt,height=15.75pt]{scala_cromatica.jpeg}};

\draw (128.82,465.1) -- (567.91,465.1)
      (201.82,459.1) -- (201.82,471.1)
      (274.82,459.1) -- (274.82,471.1)
      (347.82,459.1) -- (347.82,471.1)
      (420.82,459.1) -- (420.82,471.1)
      (493.82,459.1) -- (493.82,471.1)
      (566.82,459.1) -- (566.82,471.1);
\draw (129.55,457.5) -- (129.2,472.1);

\draw (124.27,482.07) node [anchor=north west][inner sep=0.75pt, font=\footnotesize] {$0 \%$};
\draw (190.83,482.07) node [anchor=north west][inner sep=0.75pt, font=\footnotesize] {$1.18 \%$};
\draw (265.7,482.07)  node [anchor=north west][inner sep=0.75pt, font=\footnotesize] {$4.22 \%$};
\draw (336.88,482.07) node [anchor=north west][inner sep=0.75pt, font=\footnotesize] {$11.68 \%$};
\draw (410.78,482.07) node [anchor=north west][inner sep=0.75pt, font=\footnotesize] {$27.96 \%$};
\draw (478.23,482.07) node [anchor=north west][inner sep=0.75pt, font=\footnotesize] {$59.12 \%$};
\draw (548.69,482.07) node [anchor=north west][inner sep=0.75pt, font=\footnotesize] {$100 \%$};

\draw (192.7,88.94) node  {\includegraphics[width=215.25pt,height=127.27pt]{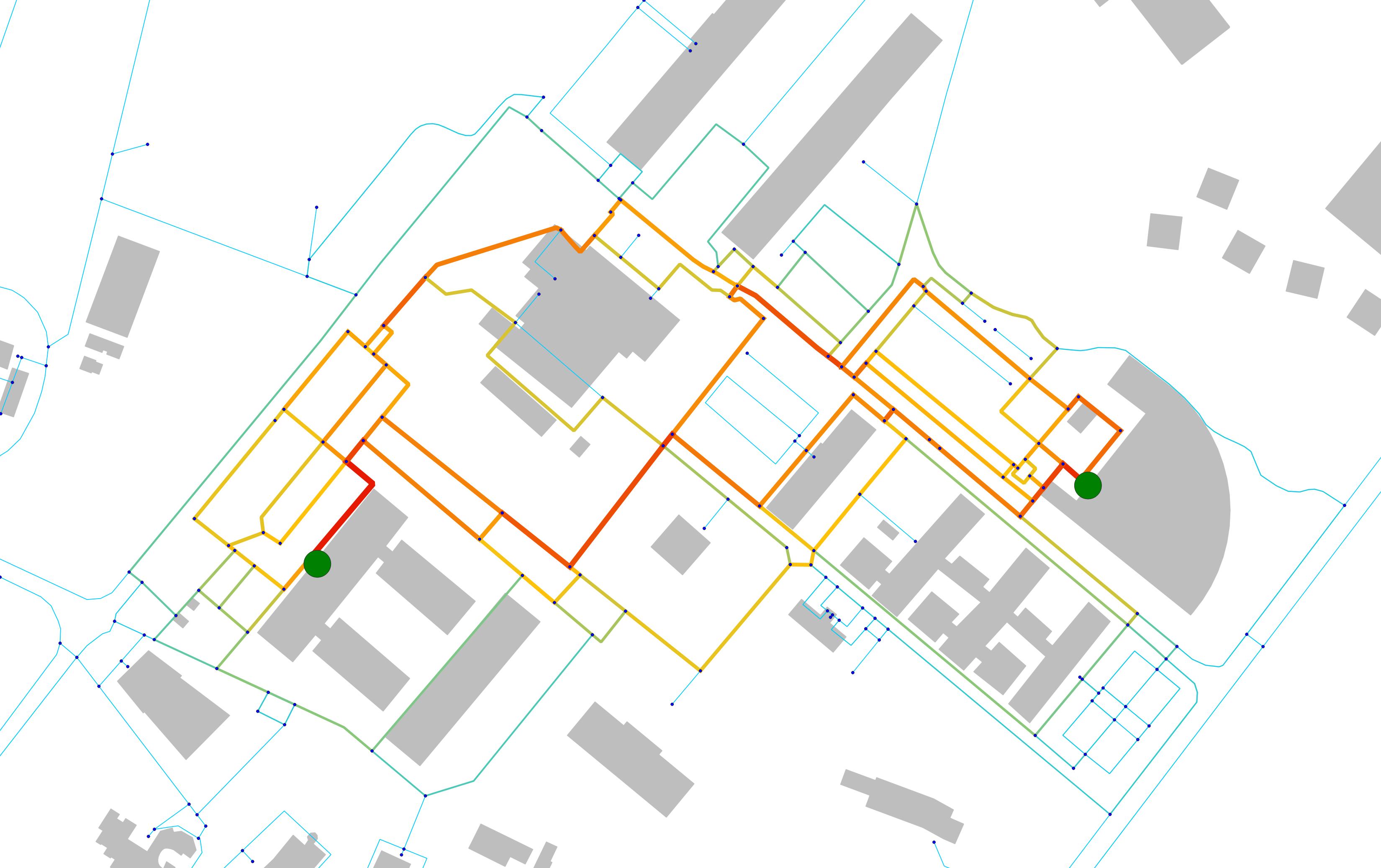}};
\draw (507.7,91.31) node  {\includegraphics[width=221.25pt,height=130.81pt]{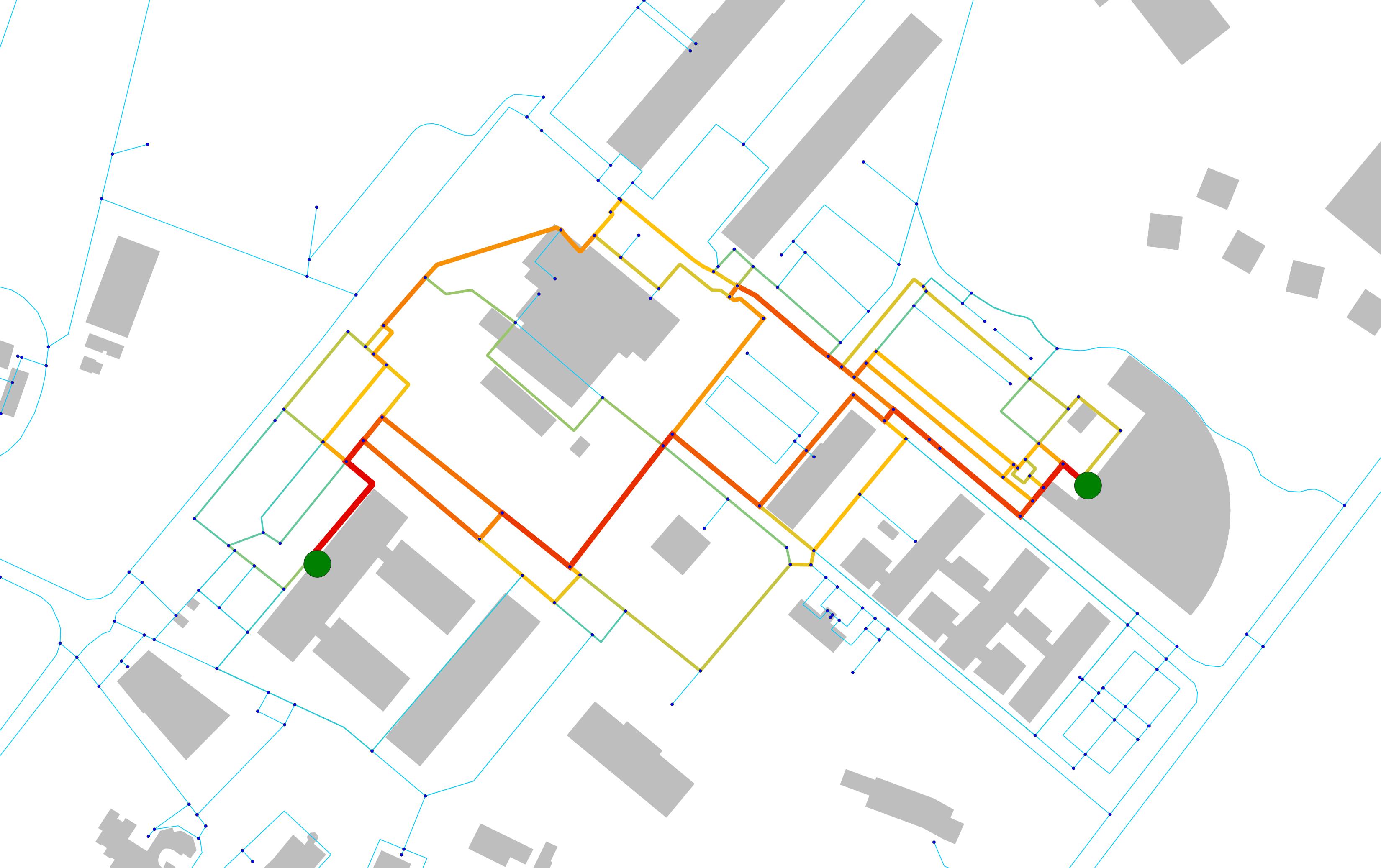}};
\draw (351.2,300.52) node  {\includegraphics[width=237pt,height=140.13pt]{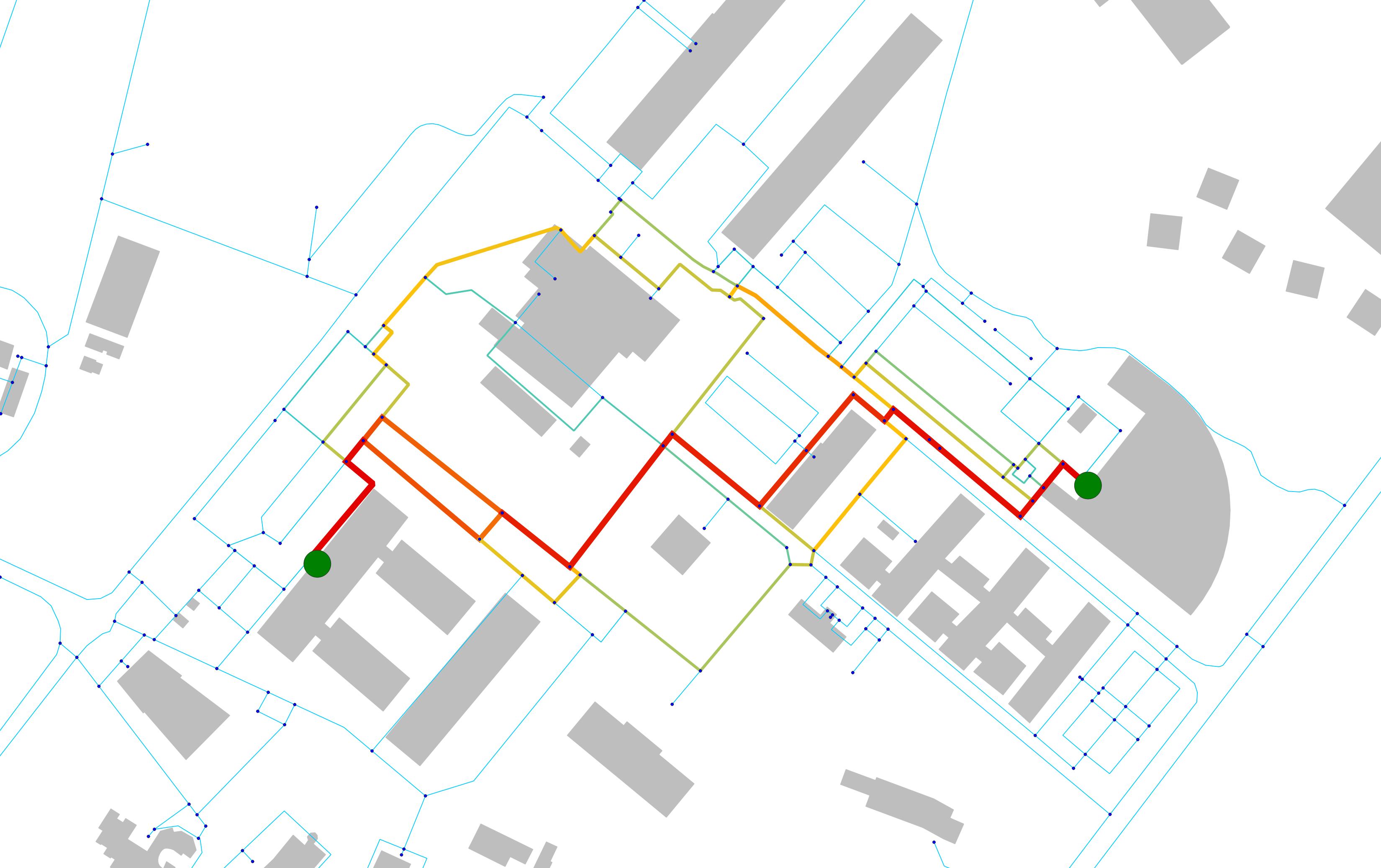}};

\draw (144,184.6) node [anchor=north west][inner sep=0.75pt] {$( a) \ k=5$};
\draw (470,183.6) node [anchor=north west][inner sep=0.75pt] {$( b) \ k=20$};
\draw (304,400.6) node [anchor=north west][inner sep=0.75pt] {$( c) \ k=50$};

\end{tikzpicture}

%% file: scala_no_WiFi.tikz
\tikzset{every picture/.style={line width=0.75pt}} 

\begin{tikzpicture}[x=0.75pt,y=0.75pt,yscale=-1,xscale=1]

\draw (245,427) node [anchor=north west][inner sep=0.75pt, font=\footnotesize] 
    {Pedestrian traffic (daily average number of people)};

\draw (348.36,454.6) node {\includegraphics[width=329.31pt,height=15.75pt]{scala_cromatica.jpeg}};

\draw (128.82,465.1) -- (567.91,465.1)
      (201.82,459.1) -- (201.82,471.1)
      (274.82,459.1) -- (274.82,471.1)
      (347.82,459.1) -- (347.82,471.1)
      (420.82,459.1) -- (420.82,471.1)
      (493.82,459.1) -- (493.82,471.1)
      (566.82,459.1) -- (566.82,471.1);
\draw (129.55,457.5) -- (129.2,472.1);

\draw (124.27,482.07) node [anchor=north west][inner sep=0.75pt, font=\footnotesize] {$0$};
\draw (190.83,482.07) node [anchor=north west][inner sep=0.75pt, font=\footnotesize] {$1.96$};
\draw (265.7,482.07)  node [anchor=north west][inner sep=0.75pt, font=\footnotesize] {$7.74$};
\draw (336.88,482.07) node [anchor=north west][inner sep=0.75pt, font=\footnotesize] {$29.62$};
\draw (410.78,482.07) node [anchor=north west][inner sep=0.75pt, font=\footnotesize] {$113.47$};
\draw (478.23,482.07) node [anchor=north west][inner sep=0.75pt, font=\footnotesize] {$433.86$};
\draw (548.69,482.07) node [anchor=north west][inner sep=0.75pt, font=\footnotesize] {$669.04$};

\draw (192.7,88.94) node  {\includegraphics[width=215.25pt,height=127.27pt]{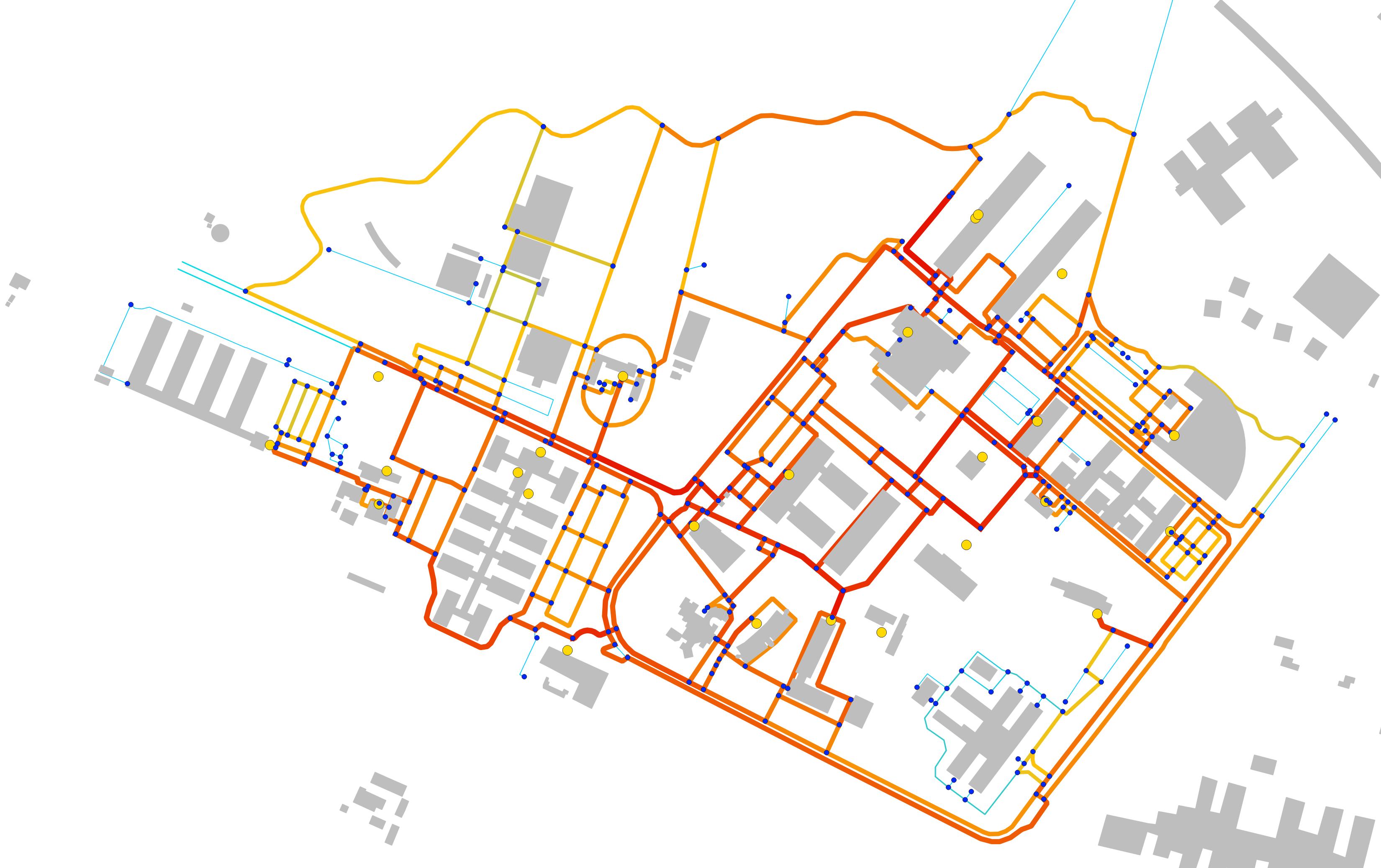}};
\draw (507.7,91.31) node  {\includegraphics[width=221.25pt,height=130.81pt]{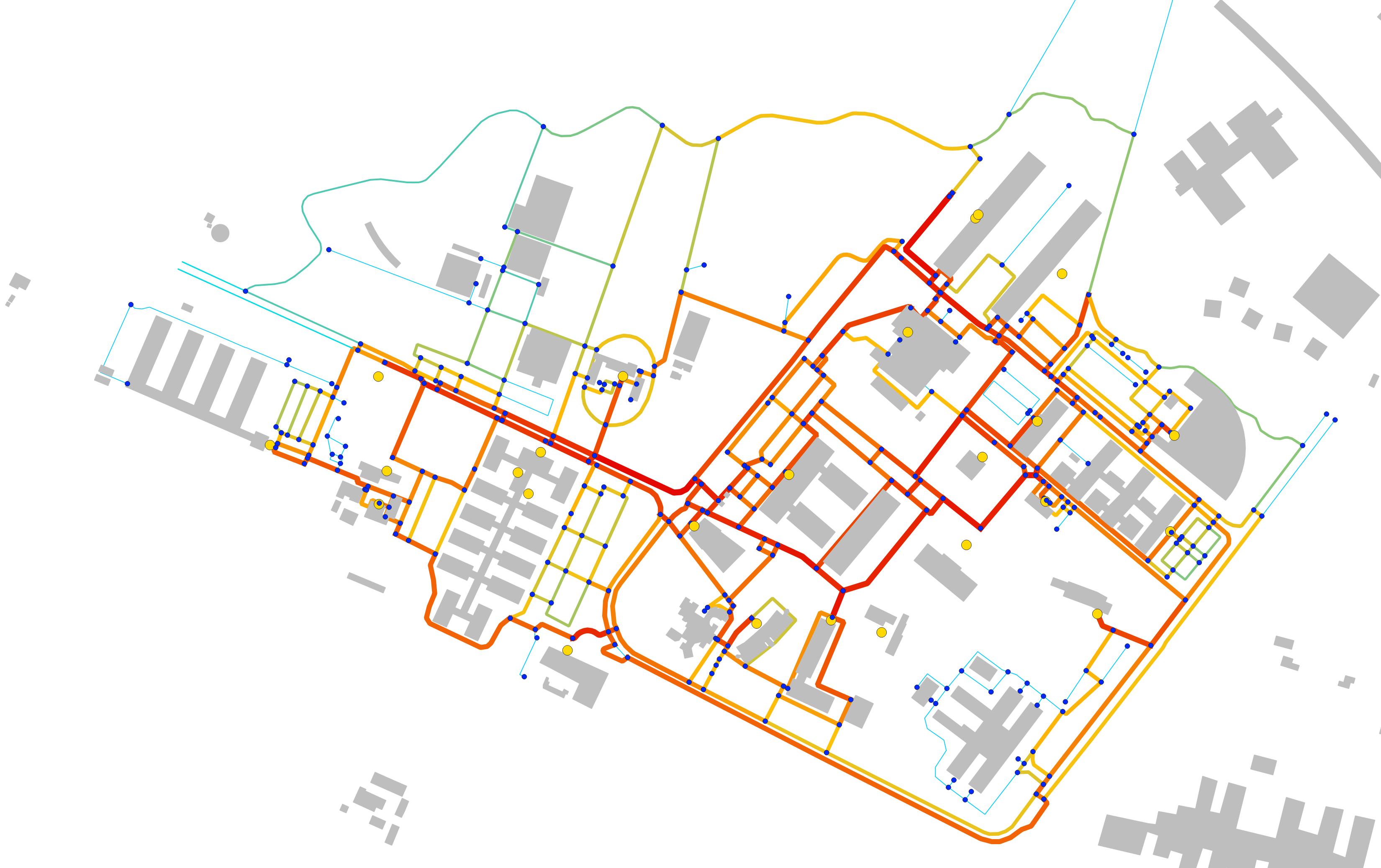}};
\draw (351.2,300.52) node  {\includegraphics[width=237pt,height=140.13pt]{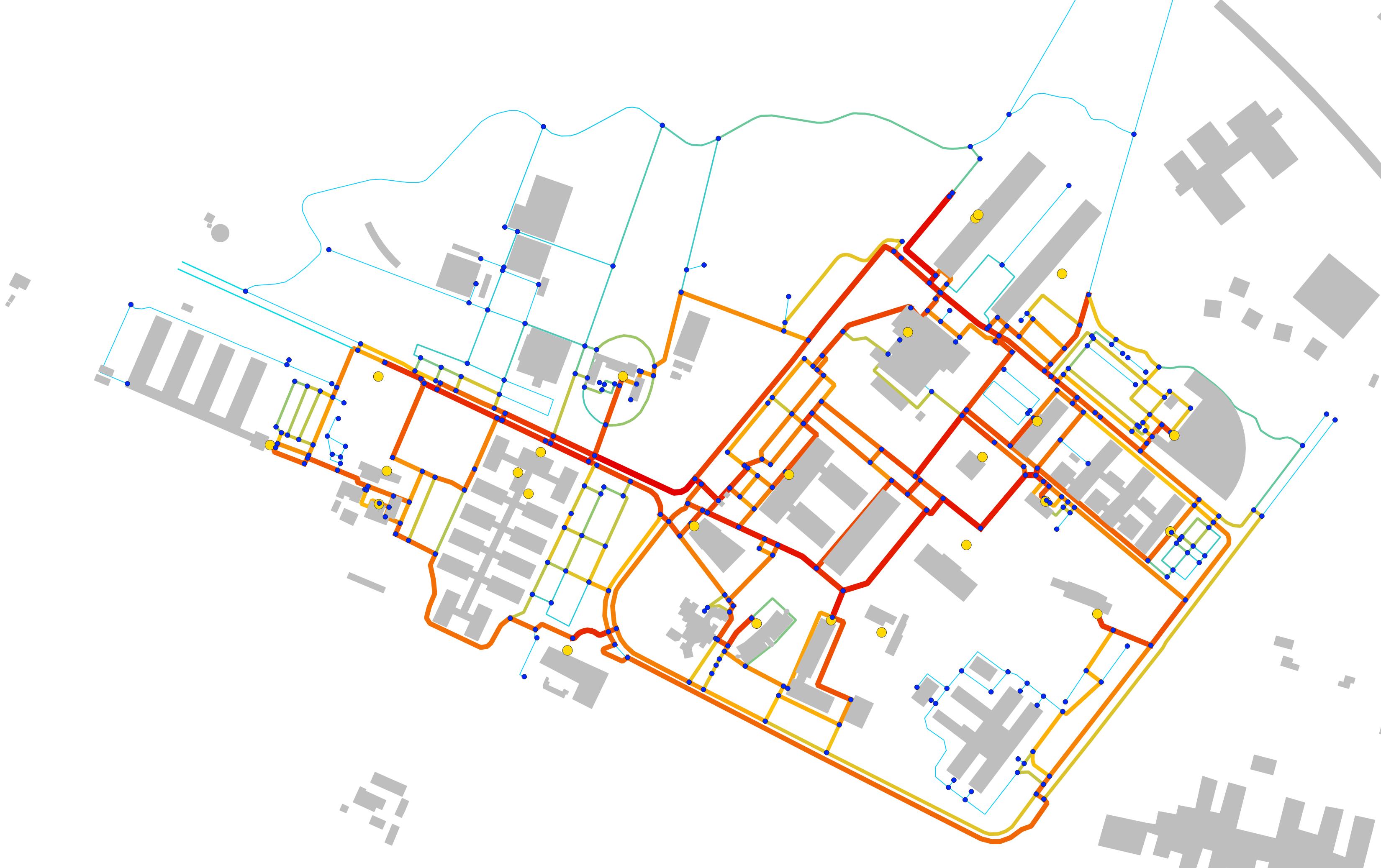}};

\draw (144,184.6) node [anchor=north west][inner sep=0.75pt] {$( a) \ k=5$};
\draw (470,183.6) node [anchor=north west][inner sep=0.75pt] {$( b) \ k=20$};
\draw (304,400.6) node [anchor=north west][inner sep=0.75pt] {$( c) \ k=50$};

\end{tikzpicture}